\documentclass[prd,preprintnumbers,nofootinbib,twocolumn]{revtex4-1}

\usepackage{amsmath}
\usepackage{amssymb}
\usepackage{booktabs}
\usepackage{graphicx}
\usepackage{multirow}
\usepackage{listings}
\usepackage{xcolor}

\usepackage{textcomp}

\usepackage[hypcap]{caption, subcaption}
\usepackage{braket}

\allowdisplaybreaks[1]

\renewcommand{\theequation}{\thesection.\arabic{equation}}
\numberwithin{equation}{section}
\newcommand{\black}{\color{black}}
\newcommand{\green}{\color{green}}

\newcommand{\refeq}[1]{Eq.~(\ref{eq:#1})}

\newcommand{\reffig}[1]{Fig.~\ref{fig:#1}}

\newcommand{\refsec}[1]{Sec.~\ref{sec:#1}}
\newcommand{\refapp}[1]{App.~\ref{app:#1}}
\newcommand{\reftab}[1]{Tab.~\ref{tab:#1}}

\newcommand{\gev}{\, \rm GeV}
\newcommand{\mev}{\, \rm MeV}

\def\simgt{\rlap{\lower 3.5 pt \hbox{$\mathchar \sim$}} \raise 1pt \hbox {$>$}}
\def\simlt{\rlap{\lower 3.5 pt \hbox{$\mathchar \sim$}} \raise 1pt \hbox {$<$}}

\def\3half{\textstyle\frac32}

\newcommand{\Br}{{\cal B}}
\newcommand{\BrCP}[1]{\overline{\cal B} {\left[ #1 \right]}}
\newcommand{\ACParg}[1]{{\cal A}_{\rm CP} {\left[ #1 \right]}}
\newcommand{\ACP}{{{\cal A}_{\rm CP}}}
\newcommand{\ACPnull}{{{\cal A}^0_{\rm CP}}}
\newcommand{\bbbar}{{$B_D$-$\overline{B}_D$}} 

\begin{document}

\title{
  Weak annihilation and new physics in charmless $\boldsymbol{B\to M M}$
  decays
}

\author{Christoph Bobeth}
\affiliation{
  Technische Universit\"at M\"unchen,
  Institute for Advanced Study,
  Lichtenbergstra{\ss}e 2a,
  D-85748 Garching,
  Germany
}

\author{Martin Gorbahn}
\affiliation{
  Department of Mathematical Sciences,
  University of Liverpool,
  L69 3BX Liverpool,
  United Kingdom
}

\author{Stefan Vickers}
\affiliation{
  Technische Universit\"at M\"unchen,
  Excellence Cluster Universe,
  Boltzmannstra{\ss}e 2,
  D-85748 Garching,
  Germany
}

\date{\today}

\preprint{FLAVOUR(267104)-ERC-78, LTH 1022}

\begin{abstract}
  We use currently available data of nonleptonic charmless 2-body $B\to MM$
  decays ($MM = PP,\, PV,\, VV$) that are mediated by $b\to (d, s)$ QCD- and
  QED-penguin operators to study weak annihilation and new-physics effects in
  the framework of QCD factorization. In particular we introduce one weak
  annihilation parameter for decays related by $(u\leftrightarrow d)$ quark
  interchange and test this universality assumption. Within the
  standard model, the data supports this assumption with the only exceptions in
  the $B\to K \pi$ system, which exhibits the well-known ``$\Delta \ACP$
  puzzle'', and some tensions in $B \to K^* \phi$.  Beyond the standard model,
  we simultaneously determine weak-annihilation and new-physics parameters from
  data, employing model-independent scenarios that address the ``$\Delta \ACP$
  puzzle'', such as QED-penguins and $b\to s\, \bar{u}u$ current-current
  operators.  We discuss also possibilities that allow further tests of our
  assumption once improved measurements from LHCb and Belle II become available.
\end{abstract}

\maketitle

%


%
%
%

\section{Introduction}
\label{sec:introduction}

Nonleptonic charmless 2-body decays $B\to M M$, with final state mesons $MM =
(PP,\, PV,\, VV)$, form a large class of decays that allow to test in principle
the underlying tree and penguin topologies at the parton level, as predicted by
the standard model (SM). Further, the subclass of QCD- and QED-penguin dominated
decays are sensitive to new physics (NP) beyond the SM, as any other $b \to
(d,s)$ flavor-changing neutral-current (FCNC) process, which makes them valuable
probes of the according short-distance couplings.

The major obstacle to constraining the short-distance couplings with data is the
evaluation of hadronic matrix elements in 2-body $B$-meson decays beyond naive
factorization. In view of this, strategies have been developed to construct
tests of the weak phases of the Cabibbo-Kobayashi-Maskawa (CKM) quark-mixing
matrix of the SM where the hadronic matrix elements are determined from data,
usually involving additional assumptions of $SU(2)$ and/or $SU(3)$ flavor
symmetries. Although this allows to test the consistency of weak phases
extracted in tree- and loop-induced processes in the framework of the SM, no
other detailed information can be obtained on particular short-distance
couplings of the involved QCD- and QED-penguin operators.

In this respect, systematic expansions in the heavy bottom quark mass, $m_b$,
yield at leading order in $1/m_b$ a simplified representation of hadronic 2-body
matrix elements in terms of rather well known heavy-to-light form factors and
distribution amplitudes (DA) of the involved mesons. These approaches, QCD
factorization (QCDF) \cite{Beneke:1999br, Beneke:2000ry, Beneke:2001ev,
  Beneke:2002jn, Beneke:2003zv, Beneke:2004bn}, soft-collinear
effective theory (SCET) \cite{Chay:2003ju, Bauer:2004tj, Bauer:2005kd,
  Williamson:2006hb} or perturbative QCD (pQCD) \cite{Keum:2000ph, Keum:2000wi, 
  Lu:2000em, Ali:2007ff}, 
provide predictions at leading order in $1/m_b$ that allow in principle to
test short-distance couplings with data.

Weak annihilation (WA) contributions are formally of subleading order in
$1/m_b$, but an additional chiral enhancement makes them phenomenologically
relevant for a consistent description of experimental data in the SM and
scenarios beyond. In QCDF and SCET, they are plagued by nonfactorizable divergences, which are
present in endpoint regions of convolutions of meson DAs. In QCDF, these divergences are
frequently parameterized by a phenomenological complex parameter
\cite{Beneke:2001ev} and hence are model-dependent.  In particular, the
associated strong phase governs the size of CP asymmetries. In practice this
leads to large theoretical uncertainties in the prediction of observables
\cite{Beneke:2003zv, Hofer:2010ee}. Although branching fractions and CP
asymmetries are sensitive to new physics effects, the model-dependence and the
arising uncertainties due to the involved strong phases raise the question how
reliable information can be extracted on the short-distance couplings.

Here we determine the model-dependence in the framework of QCDF from data,
admitting one phenomenological parameter for decays $B \to M_a M_b$ that are
related by $(u \leftrightarrow d$) quark exchange. The theoretical uncertainties
of all other input parameters (see \refapp{numeric:input}) are treated as
uncorrelated and have been included into the likelihood as explained in
\refapp{statistics}.  We use data of mostly QCD-penguin dominated $B_{u,d}$
decays into $PP = K\pi,\, K\eta^{(')},\, KK$ or $PV = K\rho,\, K\phi,\,
K\omega,\, K^*\pi ,\, K^*\eta^{(')}$ or $VV = K^*\rho,\, K^*\phi,\, K^*\omega,\,
K^*K^*$, and further $B_s$ decays into $PP = \pi\pi,\, KK,\, K\pi$ or $VV =
\phi\phi,\, K^*\phi,\, K^* K^*$ final states.  We determine also the relative
magnitude of subleading WA amplitudes compared to the relevant leading order
amplitudes. The results within the context of the SM are presented in
\refsec{SM-fit-results}.  Given the current data, a simultaneous fit of the WA
parameters and the short-distance couplings is pursued in
\refsec{NP-fit-results} for generic NP extensions of the SM in order to explore
the constraining power of these decays. Before presenting our results of the
fit, we review the observables and collect the experimental input of charmless
2-body decays in \refsec{observables}.  The relevant details of QCDF and the
definition of the phenomenological parameter are summarized in
\refsec{QCDF}. Various appendices collect additional material on numerical input
in \refapp{numeric:input} and the statistical treatment of experimental and
theoretical uncertainties as well as determination of pull values and $p$ values
in \refapp{statistics}.

%
%
%

\section{$B \to MM$ observables and data}
\label{sec:observables}

The 2-body decays of $B$ mesons into final states $f = PP,\, PV,\, V_h V_h$ with
light charmless pseudo-scalar ($P$) and/or vector ($V_h$) mesons with
polarization mode $h = L, \perp, \parallel$ provide various observables in
time-integrated, time-dependent, and also angular analyses. These are reviewed
in the first part of this section, whereas in the second part the according
available experimental data is listed that has been used in the fits.

%
%
\subsection{Observables}

The most important observables for decays of charged $B_u$ mesons into a final
state $f$ are the CP-averaged branching fraction and the (direct) CP-asymmetry
\begin{equation}
\begin{aligned}
  \BrCP{B_u \to f} &
  = \frac{\tau_{B_u}}{2} \left( \Gamma[\bar{B}_u \to \bar{f}] + \Gamma[B_u \to f] \right)\,,
\\[0.2cm]
  \ACParg{B_u \to f} & 
  = \frac{\Gamma[\bar{B}_u \to \bar{f}] - \Gamma[B_u \to f]}
         {\Gamma[\bar{B}_u \to \bar{f}] + \Gamma[B_u \to f]} \,.
\end{aligned}
\end{equation}
 
Concerning decays of neutral $B_D$ mesons ($D = d, s$) into a common final state
$f$ for both flavor eigenstates $\bar{B}_D$ and $B_D$, the simplest measurements
are untagged rates. The decay is governed by the decay rates $R_f^{\rm H,L}
\equiv \Gamma[B_D^{\rm H,L} \to f]$ of the heavy and light mass eigenstates
$B_D^{\rm H, L}$, which yield the averaged and time-integrated branching
fraction
\begin{equation}
\begin{aligned}
  \BrCP{B_D \to f} &
  = \frac{1}{2} \int dt \left( R_f^{\rm L} \, e^{- \Gamma^{\rm L}_D t}
                             + R_f^{\rm H} \, e^{- \Gamma^{\rm H}_D t} \right)
\\
  & = \frac{1}{2} \left( \frac{R_f^{\rm L}}{\Gamma^{\rm L}_D}
                       + \frac{R_f^{\rm H}}{\Gamma^{\rm H}_D} \right) \,,
\end{aligned}
\end{equation}
with their respective lifetimes $\Gamma_D^{\rm H,L}$ in the two
exponentials. The mass eigenstates are related to the flavor eigenstates
$|B_D^{\rm L} \rangle = p |B_D \rangle + q |\bar{B}_D\rangle$ and $|B_D^{\rm H}
\rangle = p |B_D \rangle - q |\bar{B}_D\rangle$ defining $q$ and $p$. On the
other hand, theoretical predictions are made for the flavor eigenstates,
implying $t = 0$,
\begin{align}
  \Br[B_D \to f] &
   = \frac{\frac{1}{2} \left(\Gamma[\bar{B}_D \to f] + \Gamma[B_D \to f]\right) }
          {\frac{1}{2} \left(\Gamma^{\rm L}_D + \Gamma^{\rm H}_D \right)}
\end{align}
with the average lifetime $\tau_{B_D} = (\Gamma_D)^{-1} \equiv 2/(\Gamma^{\rm
  L}_D + \Gamma^{\rm H}_D)$. Both branching fractions are related via
\cite{DeBruyn:2012wj}
\begin{align}
  \label{eq:BR:exp:theo}
  \BrCP{B_D \to f} &
  = \Br[B_D \to f] \, \frac{1 + y_D \, H_f}{1 - y_D^2}
\end{align}
where $y_D$ is proportional to the width difference $\Delta \Gamma_D$
\begin{align}
  y_D &
  = \frac{\Delta\Gamma_D}{2 \Gamma_D}
  \equiv \frac{\Gamma^{\rm L}_D - \Gamma^{\rm H}_D}{\Gamma^{\rm L}_D + \Gamma^{\rm H}_D} \,. 
\end{align}
The CP asymmetry due to nonvanishing width difference,
\begin{align}
  H_f & 
  = \frac{R_f^{\rm H} - R_f^{\rm L}}
         {R_f^{\rm H} + R_f^{\rm L}}\,,
\end{align}
is an independent observable and provides complementary tests of physics beyond
the SM.  It can be obtained from measurements of effective lifetimes in
untagged, but time-dependent rate measurements \cite{Fleischer:2010ib} or
together with the mixing-induced CP asymmetry $S_f$ and the direct CP asymmetry
$C_f = - \ACP$ of a time-dependent analysis
\begin{align}
  \ACParg{B_D\to f}(t) & =
    \frac{S_f \sin(\Delta m_D\, t) - C_f \cos(\Delta m_D\, t)}
         {\cosh\left(\frac{\Delta \Gamma_D}{2} t\right) -
      H_f \sinh\left(\frac{\Delta \Gamma_D}{2} t\right)}\,,
\end{align}
where the mass difference of the heavy and light mass eigenstates is denoted as
$\Delta m_D = m_D^{\rm H} - m_D^{\rm L} > 0$. The three CP asymmetries are not
independent of each other, $|S_f|^2 +|C_f|^2+|H_f|^2 = 1$, and are given in
terms of one complex quantity
\begin{align}
   \label{eq:lambda_f}
   \lambda_f & = \frac{q}{p} \frac{\overline{A}_f}{A_f}\,,
\end{align}
as follows
\begin{equation}
  \label{eq:def:CP-asymmetries:S-H-C}
\begin{gathered}
\begin{aligned}
  S_f & = \frac{2\, \mbox{Im}(\lambda_f)}{1+\left|\lambda_f\right|^2}\,, \qquad &
  H_f & = \frac{2\, \mbox{Re}(\lambda_f)}{1+\left|\lambda_f\right|^2}\,, &
\end{aligned}
\\
\begin{aligned}
  C_f & = \frac{1-\left|\lambda_f\right|^2}{1+\left|\lambda_f\right|^2}\,.
\end{aligned}
\end{gathered}
\end{equation}
In \refeq{lambda_f} $\overline{A}_f = A[\bar{B}_D\to f]$ and $A_f = A[B_D\to f]$
denote the decay amplitudes and in \refsec{QCDF} we review their calculation
in QCDF.

In the limit $\Delta \Gamma_D \to 0$ one obtains $\BrCP{B_D \to f} = \Br[B_D \to
f]$.  This is the case for $D = d$ to a very good approximation in the SM where
$y_d|_{\rm SM}Ê= (0.21 \pm 0.04) \cdot 10^{-2} \ll 1$
\cite{Lenz:2011ti}. Currently, the precision of experimental results does not
yet allow to test the SM prediction.  Measurements are available from
$B$-factories $|y_d| = (0.7 \pm 0.9) \cdot 10^{-2}$ \cite{Amhis:2012bh} and a
recent determination of LHCb from effective lifetimes $y_d = (-2.2 \pm 1.4)\cdot
10^{-2}$ \cite{Aaij:2014owa}, which assumes the SM result for
$H_f$. Model-independent analysis of effects of NP in $\Delta \Gamma_d$ show
that there is still room for huge nonstandard contributions
\cite{Bobeth:2014rda}. We use the approximation $y_d = 0$ in all our
predictions, which is well justified in the SM and also the considered NP
scenarios.

On the other hand, $\Delta \Gamma_s$ is not negligible and the current world
average from $B_s \to J/\psi \phi$ analyses alone is~\cite{Amhis:2012bh}
\begin{align}
  \label{eq:exp:value:y_s}
  y_s & = (5.8\pm 1.0) \cdot 10^{-2} \,,
\end{align}
which will be used in our analysis. In general $-1 \leq H_f \leq 1$, and
therefore the correction factor on the r.h.s. of \refeq{BR:exp:theo} can become
of ${\cal O}(10\%)$ for final states $f$ that are CP eigenstates, as has been
found for some cases~\cite{DeBruyn:2012wj}. Other averages take into account
$B_s \to J/\psi \pi\pi$ angular analysis, the effective lifetime measurement of
$B_s \to K^+K^-$ and flavor-specific $B_s$ lifetime averages, which involve
additional assumptions in the potential presence of new physics. They yield a
slightly larger value then in \refeq{exp:value:y_s}, $y_s = (6.2 \pm 0.9) \cdot
10^{-2}$ \cite{Amhis:2012bh}, being consistent within the uncertainties.

Besides branching fractions and CP-asymmetries, 2-body decays $B \to VV$ with
subsequent decays $V \to PP$ provide additional observables in the full angular
analysis of the 4-body final state \cite{Korner:1979ci}.  The decay can be
described in terms of three amplitudes, which can be chosen to correspond to
definite helicities of the final-state vector mesons $V_{a,h_a} V_{b,h_b}$ with
$h_a = h_b = (L, +, -)$ or, as in the following, transversity amplitudes
$A_{\parallel, \perp} = (A_+ \pm A_-)/\sqrt{2}$.  The three magnitudes and two
relative phases of the $A_h$ can be measured in a three-fold angular decay
distribution, where we follow the definitions \cite{Beneke:2006hg}. Hence, five
CP-averaged and CP-asymmetric observables can be measured in the case where
tagging of the initial $B$-flavor is possible.  There are polarization fractions
and relative phases for $\bar{B}$ decays
\begin{align}
  f_{h}^{\bar{B}} & 
  = \frac{|\bar{A}_h|^2}{\sum_h |\bar{A}_h|^2}\,, &
  \phi^{\bar{B}}_{\parallel, \perp} & 
  = \arg \frac{\bar{A}_{\parallel,\perp}}{\bar{A}_L}\,.
\end{align}
In view of the normalisation condition $f_L^{\bar{B}} + f_\parallel^{\bar{B}} +
f_\perp^{\bar{B}} = 1$ one uses the branching fraction and two of the
polarization fractions. In combination with the same quantities from $B$ decays,
replacing $\bar{A}_h \to A_h$, one has three CP-averaged polarization fractions
and three CP-asymmetries
\begin{align}
  f_h &
  = \frac{1}{2} \left( f_h^{\bar{B}} + f_h^B \right)\,, &
  \ACP_h &
  = \frac{f_h^{\bar{B}} - f_h^B}{f_h^{\bar{B}} + f_h^B}\,.
\end{align}
Concerning the phases, the following two CP averaged and CP violating
observables can be constructed for $h = (\parallel, \perp)$
\begin{equation}
\begin{aligned}
  \phi_h &
  = \frac{1}{2} \left(\phi^{\bar{B}}_h + \phi^B_h \right)
\\ 
  & \quad 
  - \pi \, \mbox{sgn}\left(\phi^{\bar{B}}_h + \phi^B_h \right)
     \theta\left(\phi^{\bar{B}}_h - \phi^B_h - \pi\right) \,,
\\
  \Delta \phi_h &
  = \frac{1}{2} \left(\phi^{\bar{B}}_h - \phi^B_h \right)
  - \pi \, \theta\left(\phi^{\bar{B}}_h + \phi^B_h \right) \,.
\end{aligned}
\end{equation}
This convention implies $\phi_h = \Delta \phi_h = 0$ at leading order in QCDF,
where all strong phases are zero \cite{Beneke:2006hg} and might differ for the
sign of $A_L$ relative to $A_{\parallel,\perp}$ adopted by experimental
collaborations.

\begin{table*}
\begin{center}
\renewcommand{\arraystretch}{1.3}
\begin{tabular}{ll|ll|ll|ll|ll||ll|ll}
\hline\hline
  \multicolumn{10}{c||}{$\boldsymbol{b \to s}$}
& \multicolumn{4}{c}{$\boldsymbol{b \to d}$}
\\
\hline
  \multicolumn{2}{c|}{$B \to K \pi$}
& \multicolumn{2}{c|}{$B \to K \eta$}
& \multicolumn{2}{c|}{$B \to K \eta'$}
& \multicolumn{2}{c|}{$B_s \to KK$ \cite{Aaij:2013tna}}
& \multicolumn{2}{c||}{$B_s \to \pi\pi$}
& \multicolumn{2}{c|}{$B \to K K$}
& \multicolumn{2}{c}{$B_s \to K \pi$}
\\
  $K^0 \pi^0 \,:$   & $\Br,\, C,\, S$
& $K^0 {\eta} :$    & $\Br$
& $K^0 {\eta'} :$   & $\Br,\, C,\, S$
& $K^+ K^- :$       & $\Br,\, C,\, S$
& $\pi^+ \pi^- :$   & $\Br \qquad$
& $K^0 \bar{K}^0 \,:$  & $\Br$
& $K^+ \pi^- :$        & $\Br,\, C$
\\
  $K^+ \pi^- \!:$   & $\Br,\, C$
& $K^+ {\eta} \!:$  & $\Br,\, C \quad$
& $K^+ {\eta'} \!:$ & $\Br,\, C$
& & & & & 
  $K^+ K^- \!:$     & $\Br$
& &
\\
  $K^+ \pi^0 :$     & $\Br,\, C$
& & & & & & & & &
  $K^+ K^0 :$       & $\Br,\, C  \quad$
& &
\\
  $K^0 \pi^+ :$     & $\Br,\, C$
& & & & & & & & & &
\\
\hline\hline
\end{tabular}
\caption{Observables of $B\to PP$ decays mediated by $b\to s$ and $b\to d$
  transitions that are used in the fit.}
\label{tab:exp:input:obs:PP}
\renewcommand{\arraystretch}{1.0}
\end{center}
\end{table*}

\begin{table*}
\begin{center}
\renewcommand{\arraystretch}{1.3}
\begin{tabular}{ll|ll|ll|ll|ll|ll}
\hline\hline
  \multicolumn{12}{c}{$\boldsymbol{b \to s}$}
\\
\hline\hline
  \multicolumn{2}{c|}{$B \to K^* \pi$}
& \multicolumn{2}{c|}{$B \to K \rho$}
& \multicolumn{2}{c|}{$B \to K^* \eta$}
& \multicolumn{2}{c|}{$B \to K^* \eta'$}
& \multicolumn{2}{c|}{$B \to K \phi$ \cite{Chen:2003jfa, Acosta:2005eu, Lees:2012kxa, Aaij:2013lja}}
& \multicolumn{2}{c}{$B \to K \omega$ \cite{Chobanova:2013ddr, Aubert:2007si}}
\\
  $K^{*0} \pi^0 \,:$   & $\Br,\, C$
& $K^{0} \rho^0 \,:$   & $\Br,\, C,\, S$
& $K^{*0} {\eta} :$    & $\Br,\, C$
& $K^{*0} {\eta'} :$   & $\Br, \, C$
& $K^{0} \phi :$       & $\Br,\, C,\, S$
& $K^{0} \omega :$     & $\Br,\, C,\, S$
\\
  $K^{*+} \pi^- \!:$   & $\Br,\, C$
& $K^{+} \rho^- \!:$   & $\Br,\, C$
& $K^{*+} {\eta} :$    & $\Br,\, C$
& $K^{*+} {\eta'} :$   & $\Br,\, C$
& $K^{+} \phi \!:$     & $\Br,\, C$
& $K^{+} \omega \!:$   & $\Br,\, C$
\\
  $K^{*+} \pi^0 :$     & $\Br,\, C$
& $K^{+} \rho^0 :$     & $\Br,\, C$
 & &  & &  & &  & &
\\
  $K^{*0} \pi^+ :$     & $\Br,\, C$
& $K^{0} \rho^+ :$     & $\Br,\, C$
& &  & &  & &  & &
\\
\hline\hline
\end{tabular}
\caption{Observables of $B\to PV$ decays mediated by $b\to s$ transitions that
  are used in the fit.}
\label{tab:exp:input:obs:PV}
\renewcommand{\arraystretch}{1.0}
\end{center}
\end{table*}

In the case of $B_s \to VV$ decays, again a correction factor
\refeq{BR:exp:theo} due to $y_s \neq 0$ applies, however now
\begin{equation}
\begin{aligned}
  \BrCP{B_D \to f} &
  = \sum_{h = L, \parallel, \perp} \!\!\!\!\!
    \Br[B_D \to f_h] \, \frac{1 + y_D \, H_{f_h}}{1 - y_D^2} \,,
\\[0.2cm]
  \bar{f}_h &
  = \frac{\Br[B_D \to f_h]}{\displaystyle \sum_{h = L,\parallel, \perp} \!\!\!\!\! \Br[B_D \to f_h]} 
    \, \frac{1 + y_D \, H_{f_h}}{1 - y_D^2}\,.
\end{aligned}
\end{equation}
Here $H_{f_h}$ is defined as in \refeq{def:CP-asymmetries:S-H-C} and the
quantity $\lambda_f$ is evaluated with $A_f \to A_{f_h}$.

Besides these observables, further combinations are considered that involve
different types of charged and neutral $B$, $M_a$ and $M_b$ mesons. They are
either ratios of branching fractions or differences of direct CP-asymmetries.
The complete set of ratios \cite{Fleischer:1997um, Hofer:2010ee} is
\begin{widetext}
\begin{equation}
  \label{eq:DefOfRatios}
\begin{aligned}
  R_c^B     & = 2\, \frac{\BrCP{B^-\to M_a^- M_b^0}}{\BrCP{B^- \to M_a^0 M_b^-}}\,,       \qquad &
  R_c^{M_a} & = 2\, \frac{\BrCP{B^-\to M_a^- M_b^0}}{\BrCP{\bar{B}^0 \to M_a^- M_b^+}}\,, \qquad &
  R_c^{M_b} & =     \frac{\BrCP{B^-\to M_a^0 M_b^-}}{\BrCP{\bar{B}^0 \to M_a^- M_b^+}}\,,
\\[0.2cm]
  R_n^B     & = \frac{1}{2}\, \frac{\BrCP{\bar{B}^0\to M_a^- M_b^+}}{\BrCP{\bar{B}^0 \to M_a^0 M_b^0}}\,, \qquad &
  R_n^{M_a} & = \frac{1}{2}\, \frac{\BrCP{B^-\to M_a^0 M_b^-}}{\BrCP{\bar{B}^0 \to M_a^0 M_b^0}}\,, \qquad &
  R_n^{M_b} & =               \frac{\BrCP{B^-\to M_a^- M_b^0}}{\BrCP{\bar{B}^0 \to M_a^0 M_b^0}}\,,
\end{aligned}
\end{equation}
where factors of $\tau_{B^0}/\tau_{B^-}$ are not included in the definition of
$R_{c,n}^{M_a,M_b}$, contrary to \cite{Hofer:2010ee}. It is anticipated that
these ratios are measured directly in experimental analyses, such that common
experimental systematic errors cancel. Further, the following two differences of
direct CP asymmetries are frequently considered
\begin{equation}
  \label{eq:def:DeltaACP}
\begin{aligned}
  -\Delta C = \Delta \ACP   & = 
    \ACParg{B^- \to M_a^- M_b^0} - \ACParg{\bar{B}^0 \to M_a^- M_b^+} \,, &
\\[0.2cm]
  \Delta \ACPnull & =
    \ACParg{B^- \to M_a^0 M_b^-} - \ACParg{\bar{B}^0 \to  \, M_a^0 \,M_b^0 \hspace{0.25mm} }\,,
\end{aligned}
\end{equation}
\end{widetext}
in which in QCDF a cancellation of uncertainties takes place
\cite{Lunghi:2007ak}.

In order to separate NP effects in decays from those in {\bbbar} mixing in
$S_f$, we define the observables \cite{Beneke:2005pu, Buchalla:2005us}
\begin{align}
\Delta S_f &= - \eta_f S_f - \left\{
  \begin{array}{ll}
     S(\bar{B}_d \to J/\psi\, \bar{K}_S) & D = d  \\[0.2cm]
     S(\bar{B}_s \to J/\psi\, \phi)      & D = s
  \end{array} \right.
\end{align}
with $\eta_f = \pm 1$ the CP eigenvalue of the final state $f$.  The decays
$\bar{B}_d \to J/\psi\, \bar{K}_S$ and $\bar{B}_s \to J/\psi\, \phi$ are
dominated by contributions from charm tree-level operators and CP violation in
the decay is both parametrically (CKM) and topologically (loop) suppressed.  We
expect that the CP-violating phase in {\bbbar} mixing, $\phi_{B_d}$ and
$\phi_{B_s}$, can clearly be extract from those decays, even in the presence of
most NP scenarios \cite{Nir:2005js}
\begin{equation} 
\begin{aligned}
  \lambda_{J/\psi\, \bar{K}_S} & \simeq e^{-i\phi_{B_d}}, & \quad 
  S(\bar{B}_d \to J/\psi\, \bar{K}_S) & \simeq \sin 2\beta,
\\
  \lambda_{J/\psi\, \phi}      & \simeq e^{-i\phi_{B_s}}, & \quad 
  S(\bar{B}_s \to J/\psi\, \phi) & \simeq \sin 2\beta_s,
\end{aligned}
\end{equation}
in which the angles of the CKM unitarity triangle are defined as $\beta =
\arg\left(\lambda_c^d/\lambda_t^d \right) $ and $\beta_s =
\arg\left(\lambda_t^s/\lambda_c^s \right)$.  This source of CP violation enters
the mixing-induced CP asymmetry of most decays that are triggered by $b \to s$
transition in the same way and can be eliminated by the construction of $\Delta
S_f$, which therefore exclusively measures the interference of CP violation in
the decay and in mixing.

%
%
\subsection{Data \label{sec:data}}

\begin{table*}
\begin{center}
\renewcommand{\arraystretch}{1.3}
\begin{tabular}{ll|ll|ll|ll|ll||ll|ll}
\hline\hline
  \multicolumn{10}{c||}{$\boldsymbol{b \to s}$}
& \multicolumn{4}{c}{$\boldsymbol{b \to d}$}
\\
\hline\hline
  \multicolumn{2}{c|}{ $B   \to K^*  \rho$}
& \multicolumn{2}{c|}{ $B   \to K^*  \phi$ \cite{Aubert:2008zza, Prim:2013nmy, Aaij:2014tpa}}
& \multicolumn{2}{c|}{ $B   \to K^*  \omega$}
& \multicolumn{2}{c|}{ $B_s \to \phi \phi$}
& \multicolumn{2}{c||}{$B_s \to K^*  K^*$ \cite{Aaij:2011aj}}
& \multicolumn{2}{c|}{ $B_s \to K^* \phi$}
& \multicolumn{2}{c}{  $B   \to K^* K^*$}
\\
  $K^{*0} \rho^0 \,:$   & $\Br,\, C,\, f_L$
& $K^{*0} \phi   \,:$   & $\Br,\, C,\, C_{L, \perp},$
& $K^{*0} \omega \,:$   & $\Br,\, C,\, f_L$
&  $\phi   \phi  \,:$   & $\Br,\, f_L$
& $K^{*0} K^{*0} \,:$   & $\Br,\, f_L$
& &
& $K^{*0} K^{*0} :$    & $\Br,\, f_L$
\\
  $K^{*+} \rho^- \!:$   & $\Br,\, C,\, f_L$
&                       & $f_{L, \perp},\, \phi_{\parallel, \perp}$
& $K^{*+} \omega \,:$   & $\Br,\, C,\, f_L$
& &  & &  & 
  $K^{*+} \phi \,:$    & $\Br,\, f_L$
& $K^{*+} K^{*0} :$    & $\Br,\, f_L$
\\
  $K^{*+} \rho^0 :$     & $\Br,\, C,\, f_L$
& $K^{*+} \phi \,:$     & $\Br,\, C,\, C_{L, \perp},$
& &  & &  & &  & &  & &
\\
  $K^{*0} \rho^+ :$     & $\Br,\, C,\, f_L$
&                       & $f_{L, \perp},\, \phi_{\parallel, \perp}$
& &  & &  & &  & &  & &
\\
\hline\hline
\end{tabular}
\caption{Observables of $B\to VV$ decays mediated by $b\to s$ and $b \to d$
  transitions that are used in the fit.}
  \label{tab:exp:input:obs:VV}
\renewcommand{\arraystretch}{1.0}
\end{center}
\end{table*}

We investigate mainly $B\to MM$ decays mediated by $b\to s$ transitions but will
consider also some $b\to d$ examples. The final 2-meson state $MM$ consists
either out of two pseudo-scalars ($MM = PP$) or one pseudo-scalar and one vector
($MM = PV$)\footnote{Here $MM = PV$ stands for both, $MM = PV$ and $MM = VP$.}
or two vectors ($MM = VV$), which are listed in \reftab{exp:input:obs:PP},
\reftab{exp:input:obs:PV}, and \reftab{exp:input:obs:VV}, respectively, together
with the observables that have been measured. We use the most recent values of
branching ratios $\Br$ as well as direct and mixing-induced CP asymmetries $\ACP
= - C$ and $S$ from the Heavy Flavor Averaging Group (HFAG) 2012 compilation and
updates from 2013/2014 on the website~\cite{Amhis:2012bh}. For decays into
$VV$-final states we include also the data of polarization fractions $f_{L,
  \perp}$, the relative phases $\phi_{\parallel,\perp}$ and CP-asymmetries
$C_{L,\perp}$.  Meanwhile, some observables had been updated or measured for the
first time from individual experiments and not yet included in the HFAG
averages. In these cases we do not make use of HFAG averages, but instead all
measurements from individual experiments enter the likelihood function in
\refeq{def:likelihood} as single measurements. The according references are
given explicitly in the tables for such cases.

In addition we investigate the complementarity of composed observables, the
ratios $R_{c,n}^{B, M_a, M_b}$ \eqref{eq:DefOfRatios} of branching fractions and
differences of CP asymmetries $\Delta C$ \eqref{eq:def:DeltaACP}.  In the
future, it is desirable to have direct experimental determinations of the
uncertainties for these ``composed'' observables that already account for the
cancellation of common experimental systematic uncertainties, which are only
accessible to the experimental collaborations themselves. This is important,
since usually outsiders are not in the position to account retroactively for
cancellations of systematic errors and are restricted to the application of
rules of error propagation to the uncertainties of the measurements of the
involved components, which then might result in too conservative estimates. Of
course such a procedure on the experimental side requires that the according
decay modes with charged and neutral initial/final states can be analyzed
simultaneously, which is the case for Babar, Belle and also Belle II. In this
context it should be noted that ratios of gaussian distributed quantities are
not gaussian distributed, although the differences are small as long as the tail
regions of the distribution do not contribute. The details of the treatment of
Gaussian and ratio of Gaussian distributed experimental probability
distributions of the measurements are given in \refapp{statistics}.

The tables \reftab{exp:input:obs:PP}, \reftab{exp:input:obs:PV}, and
\reftab{exp:input:obs:VV} show that the decay systems $B\to K\pi,\, K^*\pi,\,
K\rho,\, K^*\rho$ (and $K^*\phi$) are the ones with the most measured
observables, allowing to investigate the complementarity of the constraints
imposed on the phenomenological parameter of WA by branching fractions versus CP
asymmetries versus other observables in $VV$-final states. In these cases we can
also form the ratios of branching fractions \refeq{DefOfRatios} and differences
of CP asymmetries \refeq{def:DeltaACP}.  We will perform fits using two
different sets of observables for these systems.  In the first, called ``Set
I'', we will use four branching fractions and four direct CP asymmetries. In the
lack of precise experimental data on the mixing-induced CP-asymmetry $S$, we
rather prefer to predict them from the results of the fit then including them in
the fit, see \refapp{predictions} for details on the procedure. Such predictions
can be tested with measurements of $S$ by Belle II and LHCb in the near future
\cite{Abe:2010sj, Meadows:2011bk, Bediaga:2012py} and are given for the SM and
some NP fits in \reftab{Chadr:pred} and \reftab{tree-bsuu:pred}.  The second
``Set II'' contains the fully independent observables of one branching fraction,
three ratios $R_{c,n}^{B,M_a,M_b}$, three direct CP asymmetries $C$ and the
difference of CP asymmetries $\Delta C$ -- see \reftab{SM-pulls} for the
explicit list of observables. In summary:
\begin{equation}
  \label{eq:def:obs:sets}
\begin{aligned}
  \mbox{Set I}  & : 
  (4 \times \Br) + (4 \times C)
\\[0.2cm]
  \mbox{Set II} & :  
  (1 \times \Br) + (3 \times R_{c,n})
\\ 
  & \qquad \qquad + (3 \times C) + \Delta C \,.
\end{aligned}
\end{equation}

%
%
%

\section{$B \to MM$ in QCD factorization}
\label{sec:QCDF}

Here we revisit the building blocks that arise in QCDF to calculate the final
state-dependent corrections needed for a suitable prediction of the decay
amplitudes $\lambda_f$ \refeq{lambda_f} and details of their treatment in our
analysis. Most importantly, the parametrization of the endpoint divergences
arising in weak-annihilation (WA) and hard-scattering (HS) contributions are
given, which will be determined from experimental data in
\refsec{SM-fit-results} and \refsec{NP-fit-results} in the framework of the SM
and scenarios of NP, respectively. Further, we describe in
\refsec{def:sub-leading:ratios} the determination of the relative magnitude of
WA amplitudes compared to the leading ones in the SM and NP scenarios, as they
are formally of subleading order in $1/m_b$, but chirally enhanced.

In our analysis all decay modes are driven by the same flavor transition $b
\rightarrow D$ ($D = d, s$), which is described by the effective Hamiltonian of
electroweak interactions. In the SM \cite{Beneke:2000ry, Beneke:2001ev}
\begin{equation}
  \label{eq:heff-b-one}
\begin{aligned}
  {\cal H}_{\rm eff}^D &
  = \frac{G_F}{\sqrt2} \sum_{p=u,c} \lambda_p^{(D)} \Big(
        C_1 O_1^p + C_2 O_2^p
\\ 
  & \quad
  + \sum_{i=3}^{10} C_i O_i + C_{7\gamma} O_{7\gamma} + C_{8g} O_{8g} \Big)
  + \mbox{h.c.},
\end{aligned}
\end{equation}
where $G_F$ denotes the Fermi constant and $\lambda_p^{(D)} \equiv V_{pb}^{}
V_{pD}^*$ are products of elements of the CKM matrix. The flavor-changing
operators are
\begin{align}
  O_1^p & = (\bar Dp)_{V-A} (\bar pb)_{V-A},
\nonumber \\[0.2cm]
  O_2^p & = (\bar D_{\alpha}p_{\beta})_{V-A} (\bar p_{\beta}b_{\alpha})_{V-A},
\nonumber \\[0.1cm]
  O_{3(5)} & = (\bar Db)_{V-A}\sum_{q}(\bar qq)_{V \mp A},
\nonumber \\
  O_{4(6)} & = (\bar D_\alpha b_{\beta})_{V-A}\sum_{q}(\bar q_{\beta}q_{\alpha})_{V \mp A},
\nonumber \\
  O_{7(9)} & = \frac{3}{2}(\bar Db)_{V-A}\sum_{q}e_q(\bar qq)_{V\pm A},
\nonumber \\
  O_{8(10)} & = \frac{3}{2}(\bar D_{\alpha}b_{\beta})_{V-A}\sum_{q}e_q(\bar{q}_{\beta}q_{\alpha})_{V \pm A},
\nonumber \\[0.2cm]
  O_{7\gamma}& = -\frac{e\, m_b}{8\pi^2}\, (\bar D\,\sigma^{\mu\nu}(1+\gamma_5)\, b) F_{\mu\nu} ,
\nonumber \\[0.1cm]
  O_{8g}     & = -\frac{g_s m_b}{8\pi^2}
                 (\bar D_\alpha \sigma^{\mu\nu}(1+\gamma_5) T^a_{\alpha\beta}  b_\beta) G_{\mu\nu}^a,
\label{eq:curcur}
\end{align}
where $(\bar q_1 q_2)_{V\pm A}=\bar q_1\gamma_\mu(1\pm\gamma_5)q_2$, the sum is
over active quarks $q = (u,d,s,c,b)$, with $e_q$ denoting their electric charge
in fractions of $|e|$ and $\alpha,\beta$ denoting color indices.  The Wilson
coefficients, $C_i$, are obtained by a matching calculation at a scale typically
of the order $\mathcal{O}(M_W)$ of the $W$-boson mass and are evolved down to
the low energy scale of $\mathcal{O}(m_b)$ of the bottom-quark mass by means of
the renormalization group (RG) equation. Here, we will use the modified counting
scheme, as described in \cite{Beneke:2001ev}, in which the dominant part of the
electroweak penguin Wilson coefficients $C_i$ ($i = 7,\ldots,10$) are treated as
a leading order effect. The electroweak penguin operators belong to the
isospin-violating part of ${\cal H}_{\rm eff}^D$ and affect in principle
isospin-asymmetric observables, as for example \refeq{DefOfRatios} and
\refeq{def:DeltaACP}.

%
%
%

\subsection{Weak annihilation in QCDF}

As was established by Beneke, Buchalla, Neubert and Sachrajda
\cite{Beneke:2000ry, Beneke:2001ev}, the matrix elements of the involved
operators can be treated systematically in a $1/m_b$ expansion that has become
known as QCD factorization (QCDF). At leading order this yields
\begin{widetext}
\begin{equation}
  \label{eq:QCDF-theorem}
\begin{aligned}
  \bra{M_1 M_2}O_i\ket{B} & =
    \sum_j F_j^{B\rightarrow M_1}(m_2^2) \int_0^1 du \, T^{\rm I}_{ij}(u) \Phi_{M_2}(u)
  + (M_1 \leftrightarrow M_2 )
\\
  &  \quad + \int_0^1 d\xi dv du \, T^{\rm II}_{i}(u,v,w) \Phi_{B}(\xi) \Phi_{M_1}(v) \Phi_{M_2}(u)
  + \mathcal{O}\left(\frac{\Lambda_{\rm QCD}}{m_b}\right)
\end{aligned}
\end{equation}
\end{widetext}
two terms with hard-scattering kernels $T^{\rm I,II}$, which are calculable in
perturbation theory to higher orders in the QCD coupling $\alpha_s$. They are
convoluted with light-cone distribution amplitudes (DA) of the light mesons,
denoted as $\Phi_{M_i}$, and are multiplied by the corresponding heavy-to-light
form factors $F_j^{B\rightarrow M_i}$ in the case of $T^{\rm I}$ and involve an
additional convolution with the $B$-meson distribution amplitude $\Phi_B$ in the
case of $T^{\rm II}$. In \refeq{QCDF-theorem}, the meson $M_1$ inherits the
spectator quark of the decaying $B$ meson, and depending on the final state, the
decay amplitude might depend also on matrix elements with $M_1 \leftrightarrow
M_2$, see \cite{Beneke:2001ev, Beneke:2003zv} for details.

At leading order in $1/m_b$, the perturbative kernels $T^{\rm I,II}$ have been
calculated up to NLO in strong coupling $\alpha_s$ \cite{Beneke:2000ry,
  Beneke:2001ev} and throughout we will stay within this approximation.
Contrary to previous works \cite{Beneke:2001ev, Beneke:2003zv, Hofer:2010ee}, we
employ Wilson coefficients of the weak Hamiltonian evaluated at the scale $m_b$
even in WA and HS contributions, only the strong coupling $\alpha_s$ is
evaluated at the semi-hard scale $\mu_h = \sqrt{\Lambda_{\rm QCD} m_b}$. In the
SCET approach this is apparent as a subsequent matching step from QCD to
SCET${}_{\rm I}$ taken at $\mu \sim m_b$ such that the Wilson coefficients of
the weak Hamiltonian do not run below $m_b$, whereas $\alpha_s$ does. Equivalent
arguments in the framework of QCDF can be found in \cite{Beneke:2004dp}.

The NNLO $\alpha_s$ corrections to $T^{\rm I,II}$ are work in progress and by
now the only lacking part are corrections to $T^{\rm I}$ from QCD- and
QED-penguin operators $i =3,\dots, 10$ as well as the dipole operators $i =
7\gamma, 8g$. These NNLO corrections are especially important for decays under
consideration here because strong phases are generated in QCDF only at NLO and
higher order corrections might be large, apart from the reduction of
renormalization scheme dependences. In the case of the color-allowed and
color-suppressed current-current contributions due to $O^p_{1,2}$, the NNLO
contributions to $T^{\rm I}$ cancel in large parts for both, real and imaginary
parts, \cite{Bell:2007tv, Bell:2009nk, Beneke:2009ek} with the ones to $T^{\rm
  II}$ \cite{Beneke:2005vv, Kivel:2006xc, Pilipp:2007mg} in the corresponding
amplitudes $\alpha_{1,2}(MM)$ \cite{Bell:2009fm, Beneke:2009ek} for $MM =
\pi\pi,\, \rho\pi$, leaving them close to NLO predictions. This might not be the
case for final states considered here.

As it is discussed in detail in the literature \cite{Beneke:2001ev,
  Beneke:2003zv}, contributions from HS and WA topologies, which are subleading
in $1/m_b$, elude so far from a systematic treatment in QCDF. However, they can
be chirally enhanced and contribute sizable corrections in predictions. Due to
the ignorance of the respective QCD mechanisms, additional phenomenological
parameters were introduced
\begin{equation}
  \label{Hard_scattering}
\begin{aligned}
  X_k & = \left(1 + \rho_k\right) \ln\frac{m_b}{\Lambda_{\rm QCD}} \,,
\\[0.2cm]
  \rho_k & \equiv |\rho_k| e^{i\phi_k}
\end{aligned}
\end{equation}
with the complex parameters $\rho_k$ for $k = A, H$.

In the HS they originate from terms involving twist-3 light-cone DAs
$\Phi_{m1}(y)$ with $\Phi_{m1}(y) \neq 0$ for $y\to1$ in convolutions
\begin{equation}
\begin{aligned}
  \int_0^1  \frac{\Phi_{m1}(y) dy}{1 - y}   &  \equiv \Phi_{m1}(1) X_H
+ \int_0^1 \frac{\Phi_{m1}(y) dy}{[1 - y]_+}\, , 
\end{aligned}
\end{equation}
which are regulated by the introduction of the phenomenological parameter
$X_H$\footnote{ In principle one might introduce a separate $X_H$ for each meson
  $M_1$ and $M_2$ as well as for each operator insertion.}, representing a
soft-gluon interaction with the spectator quark. As indicated above, it is
expected that $X_H \sim \ln(m_b/\Lambda_{\rm QCD})$ because it arises in a
perturbative calculation of these soft interactions that are regulated in
principle latest by a physical scale of order $\Lambda_{\rm QCD}$.  Neither the
adequate degrees of freedom nor their interactions, which should be used in an
effective theory below this scale are known. It is also conceivable that
factorization might be achieved at some intermediate scale between $m_b$ and
$\Lambda_{\rm QCD}$. The factor $(1 + \rho_H)$ summarises the remainder of an
unknown nonperturbative matrix element, including the possibility of a strong
phase, which affects especially the predictions of CP asymmetries. The numerical
size of the complex parameter $\rho_H$ is unknown, however too large values will
give rise to numerically enhanced subleading $1/m_b$ contributions compared to
the formally leading terms putting to question the validity of the $1/m_b$
expansion of QCDF.

WA is entirely subleading in $1/m_b$ and consists in principle of six different
building blocks $A_{k}^{i,f}$ ($k = 1,2,3$), which are characterized by gluon
emission from the initial ($i$) and final ($f$) states and the three possible
Dirac structures that are involved: $k = 1$ for $(V-A)\otimes (V-A)$, $k = 2$
for $(V-A)\otimes (V+A)$ and $k = 3$ for $(-2)(S-P)\otimes(S+P)$. They
contribute to non-singlet annihilation amplitudes with specific combinations of
Wilson coefficients of the 4-quark operators \cite{Beneke:2001ev, Beneke:2003zv}
\begin{align}
  b_1 & = \frac{C_F}{N_c^2} C_1 A_1^i \,,
\nonumber \\
  b_2 & = \frac{C_F}{N_c^2} C_2 A_1^i \,,
\nonumber \\
  b_3^p & = \frac{C_F}{N_c^2}
  \left[ C_3 A_1^i + C_5 (A_3^i + A_3^f) + N_c C_6 A_3^f \right] \,,
\nonumber \\
  b_4^p & = \frac{C_F}{N_c^2}
  \left[ C_4 A_1^i + C_6 A_2^i \right] \,,
\nonumber \\
  b_{3, {\rm EW}}^p & = \frac{C_F}{N_c^2}
  \left[ C_9 A_1^i + C_7 (A_3^i + A_3^f) + N_c C_8 A_3^f \right] \,,
\nonumber \\
  b_{4, {\rm EW}}^p & = \frac{C_F}{N_c^2}
  \left[ C_{10} A_1^i + C_8 A_2^i \right] \,,
\label{eq:WAamp:WC:Aif}
\end{align}
and depend on $M_1$ and $M_2$. Here, $N_c = 3$ denotes the number of colors and
the color factor $C_F = 4/3$. In particular, they correspond to the amplitudes
due to current-current ($b_1,\, b_2$), QCD-penguin ($b_3^p,\, b_4^p$) and
electroweak penguin ($b_{3, {\rm EW}}^p,\, b_{4, {\rm EW}}^p$) annihilation.
Below we will frequently refer to WA amplitudes with the normalization
\cite{Beneke:2003zv}
\begin{align}
  \label{eq:def:beta_i}
  \beta_i^{(p)} & =
    {\cal N}_\beta \,
    \left\{ \begin{array}{cl} \displaystyle
      \frac{b_i^{(p)}}{m_{M_2}}  & \quad \mbox{for } M_1 M_2 = V^\pm V^\pm
    \\[0.4cm] \displaystyle
      \frac{b_i^{(p)}}{m_{B}}    & \quad \mbox{for all others}
    \end{array}Ê\right.
\end{align}
where the argument $M_1 M_2$ has been suppressed and ${\cal N}_\beta \equiv f_B
f_{M_1}/(m_B F^{B\to M_1})$ is independent on $M_2$.  As in the case of HS, the
endpoint singularities in WA amplitudes are regulated in a model-dependent
fashion. The results are expressed in terms of convolutions of hard-scattering
kernels with DAs of twist-2 and chirally enhanced twist-3, involving
phenomenological parameters $X_A$
\begin{equation}
  \label{eq:def:X_A}
\begin{aligned}
  \int_0^1 \frac{dy}{y} & \to X_A \,,
\\[0.2cm]
  \int_0^1 \frac{\ln y \, dy }{y} & \to -\frac{1}{2} (X_A)^2 \,,
\end{aligned}
\end{equation}
which in principle are different for each meson and each building block
$A_k^{i,f}$.  Explicit expressions for $A_k^{i,f}$ in terms of $X_A$ are given
for $MM = PP,\, PV,\, VP,\, VV$ in the literature \cite{Beneke:2003zv,
  Beneke:2006hg, Bartsch:2008ps}, but independently one has $A^f_{1,2} = 0$. As
a further simplification, it is assumed in the literature that there is only one
phenomenological parameter, independent of meson type and Dirac structure, such
that $A^{i,f}_k(X_A)$ are functions of the same parameter. In this context we
would like to note that in the most relevant WA amplitude $\beta_3^c$ the
building block $A^f_3$ is parametrically enhanced by $N_c$ and in the SM a large
Wilson coefficient. In consequence its contribution dominates over the ones of
$A^i_{1,3}$.  It should be noted that the WA amplitudes \eqref{eq:WAamp:WC:Aif}
in the light-cone sum rule (LCSR) approach exhibit the same dependence on the
products of Wilson coefficients and building blocks \cite{Khodjamirian:2005wn},
however in this approach the calculation of $A^{i,f}_k$ does not suffer from
endpoint singularities due to different assumptions and approximations.  With
the latter in mind, a more general approach would be to interpret the building
blocks themselves as phenomenological parameters, or equivalently introduce one
$X_A$ for each of them. When investigating new-physics effects, it is desirable
to keep the explicit dependence on the Wilson coefficients in
\eqref{eq:WAamp:WC:Aif} since they depend on NP parameters, including new weak
phases. In the case of non-negligible WA contributions, the CP asymmetries and
branching fractions will be sensitive to the interference of the new physics
phases and the strong phases from $X_A$.

As already indicated, the phenomenological parameters $X_{A,H}$ are unknown and
their size is conventionally adjusted within some range $|\rho_{A,H}| \lesssim
2$ to reproduce data whereas the phase $\phi_{A,H}$ is kept arbitrary and varied
freely to estimate the uncertainty in theoretical predictions of observables
within QCDF due to WA and HS. This procedure showed the phenomenological
importance of WA and constitutes a major source of theoretical uncertainty in
predictions within the SM \cite{Beneke:2003zv} and searches beyond
\cite{Hofer:2010ee} and below we will refer to it as ``conventional QCDF''.

In this work, we are going to fit $\rho_A$ --- and for $B\to K \pi$ also
$\rho_H$ --- from data. As a consequence, no predictions will be possible for
those observables that are used in the fit, while the fitted values of $\rho_A$
depend on the short-distance model under consideration. Yet, the consistency of
the underlying short-distance model can be tested. We perform our fits in the
framework of the SM, and further in new physics scenarios simultaneously with
the additional NP parameters. In the latter case, the determination of the NP
parameters will take into account the uncertainty of the WA contribution when
marginalizing over $\rho_A$.

This procedure is different to conventional QCDF in as much as it assumes one
universal parameter $\rho_A$ for all observables in one specific decay
mode. Indeed, in conventional QCDF the independent variation of $\rho_{A,H}$ for
each observable in a specific decay corresponds to a different WA (and HS)
parameter for each observable. However, since in QCDF the parameters
$\rho_{A,H}$ are introduced at the level of decay amplitudes one would expect
that they are the same for all observables of a specific decay
mode. Consequently, conventional QCDF allows for situations where experimental
measurements and theory predictions for two observables are in agreement,
although for the first observable the agreement is reached for values of
$\phi_{A,H}$ that might be much different from those where the agreement is
reached for the second observable.

In the lack of precise data for most of the decays, we make the further
assumption of a WA parameter that is even universal for decay modes that are
related by the exchange of $(u \leftrightarrow d)$ quarks. As an example, this
allows to combine observables of the four decay channels $\bar{B}^0\to \bar{K}^0
\pi^0,\, K^-\pi^+$ and $B^-\to K^-\pi^0,\, \bar{K}^0\pi^-$, to which we refer as
``decay system'' $B\to K\pi$.  All considered decay systems and the according
observables have been listed in \reftab{exp:input:obs:PP},
\reftab{exp:input:obs:PV} and \reftab{exp:input:obs:VV}.  This assumption is
motivated by the circumstance that the dominant contributions to the amplitude
in all considered decays come actually from the linear combination
$\hat{\alpha}_4(M_1 M_2) = \alpha_4(M_1 M_2) + \beta_3(M_1 M_2)$, which is due
to isospin-conserving QCD penguin operators $O_{3,\ldots,6}$ \eqref{eq:curcur}.
The definition of all $\alpha_i$'s can be found in \cite{Beneke:2003zv}, whereas
$\beta_i$'s are given in \refeq{def:beta_i}. Other assumptions have been tested
in the literature as for example universal weak annihilation among $B_s$ and
$B_d$ decays into final states containing kaons and pions \cite{Chang:2012xv}.

The procedure reflects the general idea inherent to $1/m_b$ expansions, which
aim at a factorization into short-distance and universal nonperturbative
quantities, where the latter are determined from data in the lack of first
principle determinations.  Presently, however, factorization theorems are not
yet established at subleading order that would support the existence of such
universal quantities. In view of this, our study can affirm at most experimental
evidence against the assumption of one universal parameter per decay
system. Therefore a positive affirmation may not be over interpreted. Finally,
it must also be noted that contributions of not included NNLO corrections could
be sizeable and in our fits they are interpreted as part of the phenomenological
WA parameter.

%
%
%

\subsection{Size of power suppressed corrections}
\label{sec:def:sub-leading:ratios}

In this work, we determine the size of subleading WA (and HS) contributions from
data in the framework of the SM and NP scenarios. Due to the chiral enhancement,
WA contributions are not necessarily $1/m_b$ suppressed numerically with respect
to the leading order amplitudes. Therefore, it is of interest to know the
relative magnitude of WA to leading amplitudes for the best fit regions of
$\rho_{A(H)}$. For this purpose we introduce the quantities
\begin{align}
  \label{eq:def:xi_A}
  \xi_i^A(\rho_A) & 
  = \left| \frac{\beta_i(\rho_A)}{\alpha_{(i +\delta_{i3}), \rm I}} \right|\,, &
\intertext{for WA and}
  \xi_i^H(\rho_H) & 
  = \left| \frac{\alpha_{i, \rm II}^{\rm tw-3}(\rho_H)}
                {\alpha_{i,\rm I} + \alpha_{i,\rm II}^{\rm tw-2}} \right|\,,
\end{align}
for HS amplitudes. For the latter, $\alpha_{i, \rm II}^{\rm tw-3}$ denotes the
subleading, but chirally enhanced, twist-3 contribution, whereas $\alpha_{i,\rm
  II}^{\rm tw-2}$ is the leading HS contribution from twist-2 DAs, which is free
of endpoint divergences.  The leading amplitudes are splitted into $\alpha_i =
\alpha_{i, \rm I} + \alpha_{i, \rm II}$, with the two contributions from kernel
I and II introduced in \refeq{QCDF-theorem}. Note that $\xi^A_3 =
|\beta_3/\alpha_{4, \rm I}|$.  This definition is generalized for the case $MM =
VV$, where $\xi^A$ is defined as the mean value of the corresponding ratios for
the longitudinal and negative polarized amplitudes $\xi_{i,VV}^A \equiv
(\xi_{i,L}^A + \xi_{i,-}^A)/2$.

The most important contribution from power-suppressed corrections are clearly
obtained from HS in $\alpha_2$, which is enhanced by the large Wilson
Coefficient $C_1$ and from the WA correction $\beta_3$ in QCD-penguin dominated
decays. Therefore $\xi_2^H$ and $\xi_3^A$ will play an important role in the
phenomenological part of this work.

In the SM, the $\xi^A$-ratios depend exclusively on $\rho_A$ and contour lines
of constant $\xi^A$ can be easily obtained in the complex $\rho_A$-plane.
Concerning fits in new-physics scenarios, the $\xi^A$-ratios depend in addition
on new-physics parameters $\boldsymbol{x}^{\rm NP}$, where
dim($\boldsymbol{x}^{\rm NP}$) corresponds to their number. The dependence is
both, explicit in the Wilson coefficients and implicit on data via the
likelihood. In this case one would be interested in the minimal value of
$\xi^A_i(\boldsymbol{x}^{\rm NP})$ in the 68\% credibility regions (CR) of all
NP parameters, but marginalized over $\rho_A$. Since the determination of this
CRs requires huge computational efforts when dim$(\boldsymbol{x}^{\rm NP}) > 2$,
we proceed differently. In the course of the fit, we histogram in all
2-dimensional subspaces of NP parameters $(x_a^{\rm NP},\, x_b^{\rm NP})$, with
$a \neq b$, those values $\xi^A$ in each bin $(x_a^{\rm NP},\, x_b^{\rm NP})$
that belong to the largest likelihood value when sampling the complementary
subspace of the remaining NP parameters $x_c^{\rm NP}$ with $c\neq a$ and $c\neq
b$. As a result, in each of the 2D-marginalized planes, ``labeled'' by $(a, b)$,
the 68\% CR will contain a smallest and a largest $\xi^A_i$ in one of the bins
in $(x_a^{\rm NP},\, x_b^{\rm NP})$, which all belong to the minimal $\chi^2$ in
the subspace of $x_c$. As the final range we choose the minimum of the smallest
values and the maximum of the largest values from all the pairs $(a, b)$ in our
NP analyses in \refsec{NP-fit-results}.

%
%
%

\section{Weak annihilation in the Standard Model}
\label{sec:SM-fit-results}

\begin{table*}
\begin{center}
\renewcommand{\arraystretch}{1.3}
\begin{tabular}{ll|ccccccc}
\hline\hline
  \multicolumn{9}{l}{$\boldsymbol{MM = PP}$} 
\\
  &
  &   $B \to K \pi$
  &   $B \to K \eta$			               
  &   $B \to K \eta'$			               
  &   $B \to K K$			               
  &   $B_s \to K K$			               
  &   $B_s \to \pi \pi$
  &   $B_s \to K \pi$
\\
  \multirow{3}{*}{$\xi_3^A$}
  & BFP
  & $0.39$
  & $0.08$                         
  & $1.83$
  & $0.58$
  & $1.83$                         
  & --
  & $0.96$
\\
  & $68\%$ CR
  & $[0.37;\, 0.54]$ 		    
  & $[0.00;\, -]$                         
  & $[0.18;\,3.25]$
  & $[0.00;\,2.07]$
  & $[0.02;\,2.09]$                         
  & --
  & $[0.56;\,1.54]$
\\
  & $95\%$ CR
  & $[0.34;\, 0.69]$
  & $[0.00;\, -]$
  & $[0.16;\, 3.34]$
  & $[0.00;\, 2.10]$
  & $[0.00;\, 2.13]$
  & --
  & $[0.44;\, 1.83]$
\\
  \cline{2-9}
  \multirow{2}{*}{$|\rho_A|$}
  & lower
  & $>1.8$                     
  & $>0$                         
  & $>0.9$
  & $>0\, (0.9)$
  & $>0$                         
  & $>3.4$
  & $>2.3$
\\
  & upper
  & $<3.9$                     
  &  --                         
  & $<7.7$
  & $<6.1\, (8.6)$
  & $<5.5$                         
  & $<10.9$
  & $<4.8$
\\
\hline\hline
  \multicolumn{9}{l}{$\boldsymbol{MM = PV}$} 
\\  
  & 
  &   $B \to K^* \pi$			               
  &   $B \to K \rho$
  &   $B \to K^* \eta$			               
  &   $B \to K^* \eta'$			               
  &   $B \to K \phi$			               
  &   $B \to K \omega$  
  &   
\\
   \multirow{3}{*}{$\xi_3^A$}
  & BFP
  & $0.89$
  & $0.78$                         
  & $2.74$
  & $0.48$
  & $0.50$                         
  & $2.7$
  & 
\\
  & $68\%$ CR
  & $[0.75;\, 1.40]$ 		    
  & $[0.39;\, 1.55]$                         
  & $[0.71;\, 3.77]$
  & $[0.02;\, 7.84]$
  & $[0.40;\, 2.41 ]$                         
  & $[0.63;\, 2.88]$
  & 
\\
  & $95\%$ CR
  & $[0.69;\, 1.56]$ 		    
  & $[0.16;\, 2.18]$                         
  & $[0.64;\, 5.06]$
  & $[0.02;\, 8.41]$
  & $[0.32;\, 2.54]$                         
  & $[0.57;\, 2.88]$
  & 
\\
  \cline{2-9}
  \multirow{2}{*}{$|\rho_A|$}
  & lower
  & $>1.4$                     
  & $>0.8$                         
  & $>1.1$
  & $>0$
  & $>0.8$                         
  & $>1.3$
  & 
\\
  & upper
  & $<3.4$                     
  & $<3.4$                         
  & $<4.4$
  & $<6.1$
  & $<3.6$                         
  & $<4.3$
  & 
\\
\hline\hline
  \multicolumn{9}{l}{$\boldsymbol{MM = VV}$} 
\\
  & 
  & $B \to K^* \rho$			               
  & $B \to K^* \phi$
  & $B_s \to K^* \phi$			               
  & $B \to K^* K^*$			               
  & $B_s \to K^* K^*$			               
  & $B_s \to \phi \phi$			               
  & $B \to K^* \omega$  
\\
    \multirow{3}{*}{$\xi_3^A$}
  & BFP
  & $1.33$
  & $0.38$                         
  & $1.53$
  & $1.84$
  & $3.01$                         
  & $0.50$
  & $0.91$
\\
  & $68\%$ CR
  & $[0.84;\, 1.94]$ 		    
  & $[0.31;\, 0.43]$                         
  & $[0.30;\, 2.05]$
  & $[0.85;\, 2.68]$
  & $[1.94;\, 3.79]$                         
  & $[0.49;\, 1.11]$
  & $[0.20;\, 1.39]$
\\
  & $95\%$ CR
  & $[0.56;\, 2.33]$ 		    
  & $[0.25;\, 0.50]$                         
  & $[0.10;\, 2.15]$
  & $[0.09;\, 2.90]$
  & $[0.96;\, 4.17]$                         
  & $[0.41;\, 1.38]$
  & $[0.09;\, 1.46]$
\\
  \cline{2-9}
    \multirow{2}{*}{$|\rho_A|$}
  & lower
  & $>1.0$                     
  & $>0.6$                         
  & $>0.3$
  & $>1.2$
  & $>1.6$                         
  & $>0.7$
  & $>0.3$
\\
  & upper
  & $<2.9$                     
  & $<1.8$                         
  & $<3.2$
  & $<3.0$
  & $<3.6$                         
  & $<2.3$
  & $<2.4$
\\
\hline\hline
\end{tabular}
\caption{ Compilation of the power suppressed ratio $\xi_3^A$ at the best-fit
  point (BFP) and in the $68\%$ and $95\%$ CRs, as well as lower and upper
  bounds on the fit parameter $|\rho_A|$ in the 68\% CR for all relevant decay
  systems.  For $B \to (K\pi,\, K^*\pi,\, K\rho,\, K^*\rho)$, values correspond
  to the fit with the observable Set II. The pure WA decay $B^0 \to K^+ K^-$ is
  not included in the decay system $B \to KK$ and $\rho_A$-bounds are given
  separately in parenthesis.  }
\label{tab:SM-xi-values}
\renewcommand{\arraystretch}{1.0}
\end{center}
\end{table*}

\begin{table*}
\begin{center}
\renewcommand{\arraystretch}{1.3}
\begin{tabular}{l||cc|cc|cc|cc}
\hline\hline
    $M_a M_b$
  & \multicolumn{2}{c}{$K \pi$}
  & \multicolumn{2}{c}{$K^* \pi$}
  & \multicolumn{2}{c}{$K \rho$}
  & \multicolumn{2}{c}{$ K^* \rho$} 
\\
    set
  & SI                            & SII
  & SI                            & SII
  & SI                            & SII
  & SI                            & SII
\\
\hline
    $p$ value
  & 0.44                          & 0.04
  & 0.95                          & 0.90
  & 1                             & 1
  & 1                             & 0.97
\\
    best-fit point
  & 3.39; 2.73                    & 3.34; 2.71
  & 1.79; 5.85                    & 1.61; 5.84
  & 2.57; 2.79                    & 2.69; 2.68
  & 2.31; 2.74                    & 1.56; 5.66
\\
\hline
    $\Br(\bar{B}^0 \to M_a^0 M_b^0)$
  & $+0.3 \, \sigma$              &  --
  & $-0.3 \, \sigma$              &  --
  & $\hphantom{+}0.0 \, \sigma$   &  --
  & $\hphantom{+}0.0 \, \sigma$   &  --
\\ 
    $\Br(\bar{B}^0 \to M_a^- M_b^+)$
  & $\hphantom{+}0.0 \, \sigma$   &  --
  & $\hphantom{+}0.0 \, \sigma$   &  --
  & $\hphantom{+}0.0 \, \sigma$   &  --
  & $+0.3 \, \sigma$              &  $+0.1 \, \sigma$
\\ 
    $\Br(     B^- \to M_a^- M_b^0)$
  & $\hphantom{+}0.0 \, \sigma$   &  --
  & $+0.6 \, \sigma$              &  --
  & $\hphantom{+}0.0 \, \sigma$   &  --
  & $\hphantom{+}0.0 \, \sigma$   &  --
\\  
    $\Br(     B^- \to M_a^0 M_b^-)$
  & $\hphantom{+}0.0 \, \sigma$   & $+0.2 \, \sigma$
  & $\hphantom{+}0.0 \, \sigma$   & $+0.1 \, \sigma$
  & $\hphantom{+}0.0 \, \sigma$   & $+0.1 \, \sigma$
  & $\hphantom{+}0.0 \, \sigma$   &  --
\\  
\hline
    $R^B_c$
  &  --                           &  --
  &  --                           &  --
  &  --                           &  --
  &  --                           & $-0.5 \, \sigma$
\\ 
    $R^B_n$
  &  --                           & $\bf{-1.9 \, \sigma}$
  &  --                           & $+0.6 \, \sigma$
  &  --                           & $\hphantom{+}0.0 \, \sigma$
  &  --                           & $+0.6 \, \sigma$
\\
    $R^{M_a}_c$
  &  --                           & $ \hphantom{+}0.0 \, \sigma$
  &  --                           & $+0.8 \, \sigma$
  &  --                           & $+0.7 \, \sigma$
  &  --                           & $-0.8 \, \sigma$
\\ 
    $R^{M_b}_c$
  &  --                           & $+0.9 \, \sigma$
  &  --                           & $\hphantom{+}0.0\, \sigma$
  &  --                           & $-0.2 \, \sigma$
  &  --                           &  --
\\ 
\hline
    $C(\bar{B}^0 \to M_a^0 M_b^0)$
  & $\hphantom{+}0.0 \, \sigma$   & $\hphantom{+}0.0 \, \sigma$
  & $+0.5 \, \sigma$              & $+0.4 \, \sigma$
  & $\hphantom{+}0.0 \, \sigma$   & $\hphantom{+}0.0 \, \sigma$
  & $\hphantom{+}0.0 \, \sigma$   & $\hphantom{+}0.0 \, \sigma$
\\ 
    $C(\bar{B}^0 \to M_a^- M_b^+)$
  & $+0.7 \, \sigma$              & $+0.1 \, \sigma$
  & $+0.1 \, \sigma$              & $+0.1\, \sigma$
  & $ \hphantom{+}0.0 \, \sigma$  & $+0.1 \, \sigma$
  & $+0.5 \, \sigma$              & $+0.6 \, \sigma$
\\ 
    $C(     B^- \to M_a^- M_b^0)$
  & $\bf{-2.1 \, \sigma}$         &  --
  & $\hphantom{+}0.0 \, \sigma$   &  --
  & $ \hphantom{+}0.0 \, \sigma$  &  --
  & $+0.3 \, \sigma$              &  --
\\  
    $C(     B^- \to M_a^0 M_b^-)$
  & $\bf{+1.0 \, \sigma}$         & $\bf{+1.0 \, \sigma}$
  & $+0.9 \, \sigma$              & $\bf{+1.0\,\sigma}$
  & $+0.7 \, \sigma$              & $+0.7 \, \sigma$
  & $+0.1 \, \sigma$              & $ \hphantom{+}0.0 \, \sigma$
\\ 
\hline
    $\Delta C$
  &  --                           & $\bf{-2.8 \, \sigma}$
  &  --                           & $-0.1 \, \sigma$
  &  --                           & $ \hphantom{+}0.0 \, \sigma$
  &  --                           & $\hphantom{+}0.0 \, \sigma$
\\ 
\hline
    $f_L(\bar{B}^0 \to M_a^0 M_b^0)$
  &  --                           &  --
  &  --                           &  --
  &  --                           &  --
  & $ \hphantom{+}0.0 \, \sigma$  & $ \hphantom{+}0.0 \, \sigma$
\\ 
    $f_L(\bar{B}^0 \to M_a^- M_b^+)$
  &  --                           &  --
  &  --                           &  --
  &  --                           &  --
  & $-0.6 \, \sigma$              & $-0.5 \, \sigma$
\\ 
    $f_L(     B^- \to M_a^- M_b^0)$
  &  --                           &  --
  &  --                           &  --
  &  --                           &  --
  & $+0.7 \, \sigma$              & $+0.9 \, \sigma$
\\ 
    $f_L(     B^- \to M_a^0 M_b^-)$
  &  --                           &  --
  &  --                           &  --
  &  --                           &  --
  & $ \hphantom{+}0.0 \, \sigma$  & $ \hphantom{+}0.0 \, \sigma  $
\\ 
\hline\hline
\end{tabular}
\caption{ Compilation of $p$ values and pulls of the SM fit, evaluated at the
  best-fit point of $\rho_A^{M_a M_b}$ for the two different sets of observables
  Set I and Set II for the decays $B \to K\pi,\, K^*\pi,\, K\rho,\, K^*\rho$.  }
\label{tab:SM-pulls} 
\renewcommand{\arraystretch}{1.0}
\end{center}
\end{table*}

In this section we present the results of the determination of the WA parameter
$\rho_A$ from data of various QCD-penguin\green-\black and WA\green-\black
dominated nonleptonic charmless $B\to MM$ decays in the framework of the SM.
This includes characteristics of the best-fit regions, the $p$ values at the
best-fit point and pull values for observables, as well as the relative amount
of the subleading WA contribution needed to explain the data, which we quantify
by the ratio $\xi_3^A$ defined in \refeq{def:xi_A}.

We start with an extensive discussion of the $B \to K\pi$ system, which shows
the largest deviations from SM predictions for the difference of CP asymmetries
$\Delta C(K\pi)$ (see \refeq{def:DeltaACP}), commonly known in the literature as
the ``$\Delta \ACP$ puzzle'' and to a lesser extent in the ratio
$R_n^B(K\pi)$. We investigate the ``$\Delta \ACP$ puzzle'' further in a
simultaneous fit of the parameter of WA, $\rho_A$, and HS, $\rho_H$, and discuss
the implications on other CP asymmetries in $B_{d,s}\to K\pi$.

We turn then to the discussion of the decays $B\to K\rho,\, K^*\rho,\, K^*\pi$,
which allow also for studies of different sets of observables in Set I and Set
II due to the rather numerous and quite precise measurements.  Subsequently, we
discuss shortly the results for other decays listed in
\reftab{exp:input:obs:PP}, \reftab{exp:input:obs:PV} and
\reftab{exp:input:obs:VV} with some special comments on $B\to K \omega$ and
$B\to K^*\phi$. For each decay system we present separate constraints from
branching fractions, CP asymmetries, polarization fractions and relative phases
on the WA parameter $\rho_A$, besides the combined ones.

Apart from the above listed penguin dominated decays, we also study decays
mediated solely by weak annihilation, such as $B \to K^+ K^-$ and $B_s \to \pi^+
\pi^-$. Being independent of $\beta_3$ and hence $A_3^f$, these decay modes are
sensitive to a WA contribution from $A_{1,2}^i$ and provide access to different
building blocks.

Based on our previous fit results, we discuss finally the assumption of a
universal WA for $B_d$ and $B_s$ decays into the same final states and
investigate in particular consequences for CP asymmetries in $B_s \to K\pi$ in
view of the ``$\Delta \ACP$ puzzle'' in $B\to K\pi$.

The statistical procedure used in all fits is described in \refapp{statistics}.
In the SM, we deal mostly with the fit of one complex-valued parameter $\rho_A$
except for the $B\to K \pi$ system, where we also perform a simultaneous fit of
$\rho_A$ and $\rho_H$. When fitted, for both parameters a uniform prior
\begin{align}
  \label{eq:rho:prior:ranges}
  0 & \leq |\rho_{A,H}| \leq 8\,, &
  0 & \leq  \phi_{A,H} \leq 2\pi\,,    
\end{align} 
is assumed and no restriction is imposed on the phases. In comparison, in
conventional QCDF the magnitude $|\rho_{A,H}| \leq 2$ is used for uncertainty
estimates of theoretical predictions.  In the case that $\rho_H$ is not fitted,
but treated as a nuisance parameter instead, we use $|\rho_H| = 1$ and vary $0
\leq \phi_H \leq 2 \pi$.

Our findings for lower and upper bounds on $\rho_A$ in the 68\% CR are
summarized in \reftab{SM-xi-values} for all considered decay systems. It can be
seen that data requires non-zero values of $|\rho_A|$ to be in agreement with
QCDF predictions in the SM. In some cases they are much larger compared to the
conventionally adapted ranges, allowing thus in principle for a better agreement
of theoretical predictions with data. Since we use Wilson coefficients at the
scale $\mu \sim m_b$ in WA (and HS) contributions, contrary to
\cite{Beneke:2003zv, Hofer:2010ee}, our numerical values of $|\rho_A|$ are in
general a bit larger compared to the ones known in the literature. Representing
the size of a nonperturbative quantity, $|\rho_A|$ is expected naively to be of
order one, whereas too large values would put in doubt the convergence of the
$1/m_b$ expansion.

Further we list the ratio $\xi_3^A$ of WA amplitudes to leading ones as a
measure of the numerical relevance of these formally subleading but chirality
enhanced contributions. At the best-fit point of $\rho_A$ the according value is
indeed $\xi_3^A < 1$ for many decay systems. Although at the best-fit point
$\xi_3^A$ might reach values up to 2 or even 3 for some decay systems, once
considering the 68\% CR in $\rho_A$, it is possible to have again $\xi_3^A < 1$
(except for $B_s \to K^*K^*$) for the price of some tension among data and
prediction. Bearing in mind the chirality enhancement, our fits of the data thus
do not indicate anomalously huge WA contributions, which put QCDF into question
in principle. By definition, there is no $\xi_3^A$ for the two pure WA modes
$B_d \to K^+K^-$ and $B_s \to \pi^+\pi^-$.

%
%

\subsection{Results for $B \to K \pi$}
\label{sec:SM:BKpi}

\begin{figure*}
  \begin{subfigure}[t]{0.32\textwidth}
    \centering
    \includegraphics[width=\textwidth]{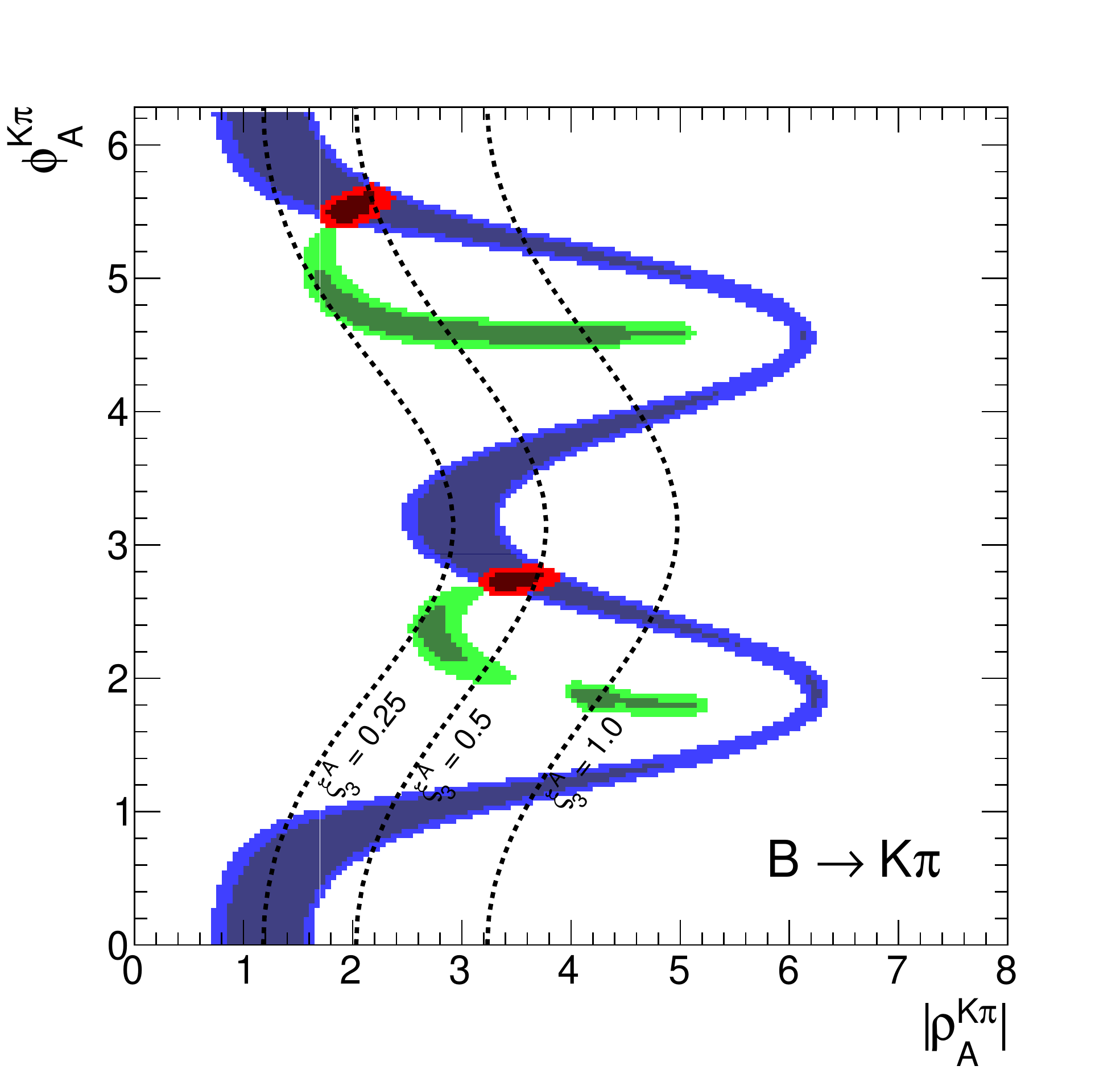}
    \caption{}
    \label{fig:BAT_SM_Kpi_a}
  \end{subfigure}
  \begin{subfigure}[t]{0.32\textwidth}
    \centering
    \includegraphics[width=\textwidth]{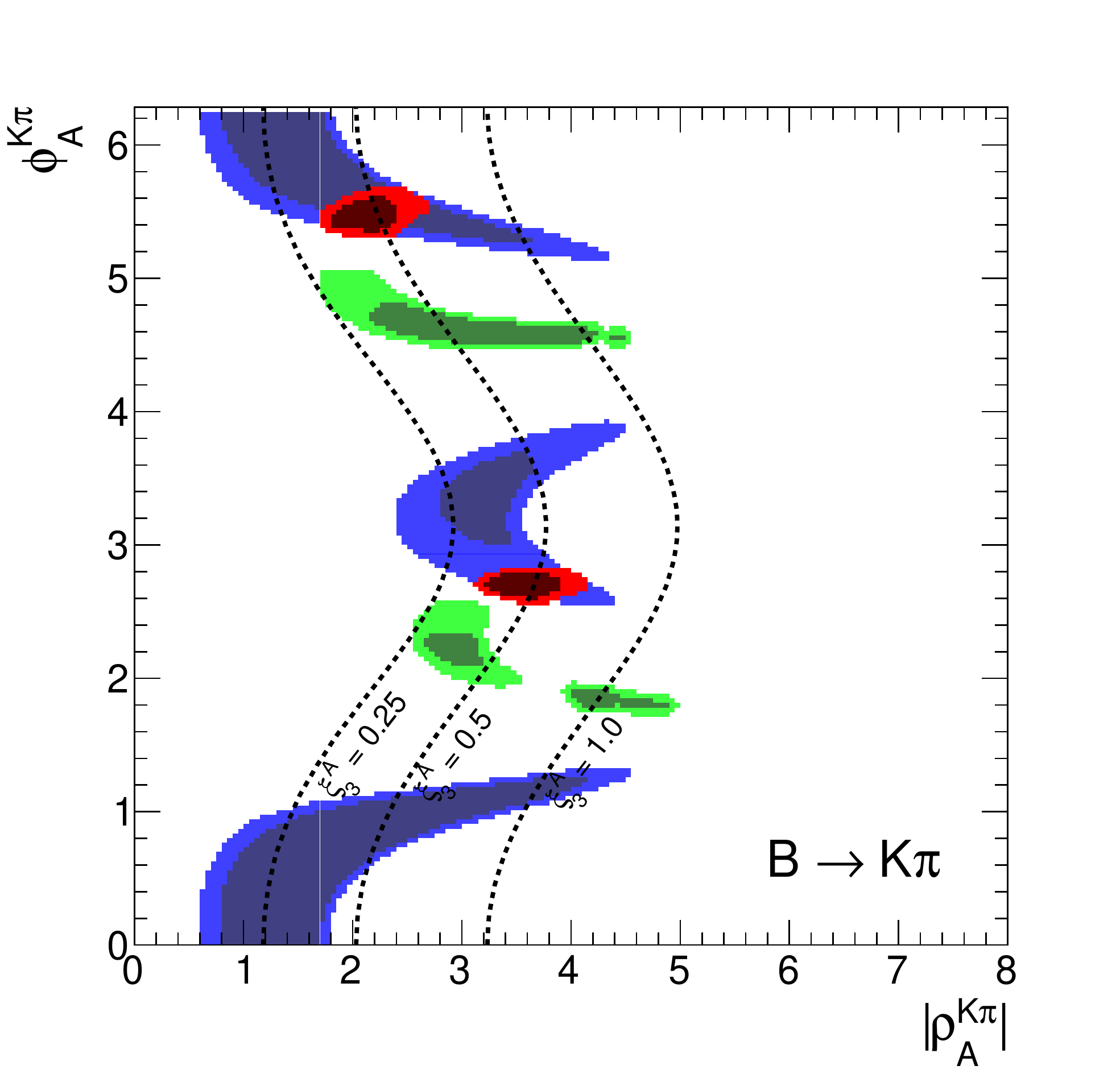}
    \caption{}
    \label{fig:BAT_SM_Kpi_b}
  \end{subfigure}
  \begin{subfigure}[t]{0.32\textwidth}
    \centering
    \includegraphics[width=\textwidth]{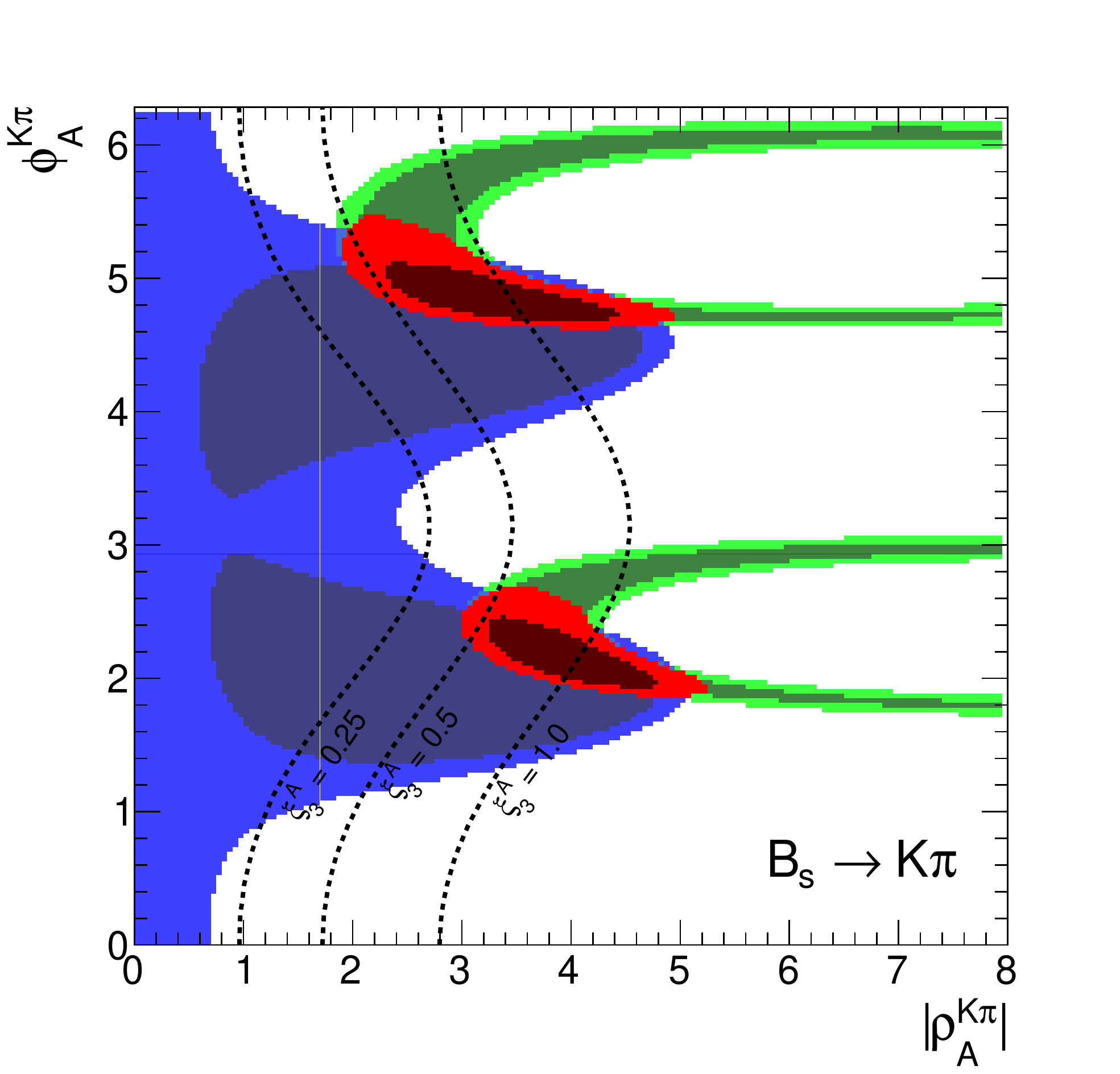}
    \caption{}
    \label{fig:BAT_SM_BsKpi}
  \end{subfigure}  

  \caption{ The 68\% (dark) and 95\% (bright) CRs of $\rho_A^{M_1 M_2}$ from a
    fit of observables in the ($B \to K \pi$)-system in (a) Set I and (b) Set
    II, and for comparison in the (c) ($B_s \to K \pi$)-system.  Allowed regions
    are shown for $\Br$ and $R_{n,c}$ (blue), $C$ and $\Delta C$ (green) and
    their combination (red).  The dashed lines correspond to constant $\xi^A_3 =
    (0.25,\, 0.5,\, 1.0)$ from left to right.  }
\label{fig:BAT_SM_Kpi}
\end{figure*}

The $B\to K \pi$ system offers the most precise measured branching fractions and
CP asymmetries (see \reftab{exp:input:obs:PP}) among the decays considered here.
In consequence, we find stringent bounds on the WA parameter $\rho_A^{K\pi}$.
This can be seen in \reffig{BAT_SM_Kpi_a} when using the observables in Set~I
and \reffig{BAT_SM_Kpi_b} for Set~II. The allowed regions from both, $\Br$
and/or $R_{c,n}^{B,M_a,M_b}$ (blue) as well as $C$ and/or $\Delta C$ (green) are
very distinct leaving two tiny overlap regions (red) at 68\% probability around
$\rho_A^{K\pi} \approx 2.1 \exp(i \, 5.5)$ and $\rho_A^{K\pi}\approx 3.4
\exp(i\, 2.7)$, which differ only slightly for both Sets I and II. The best-fit
points listed in \reftab{SM-pulls} fall into the solution with larger
$|\rho_A^{K\pi}|\approx 3.4$, but it must be noted that the other solution
provides almost equally good fits in terms of $\chi^2$.

As shown in \reftab{SM-pulls}, the $p$ values at the best-fit points of the fits
of Set~I and Set~II are very different: 0.44 versus 0.04, respectively.  The
reason are large pull values of composed observables in Set~II at the best-fit
point: $-2.8\sigma$ for $\Delta C(K\pi)$ and $-1.9\sigma$ for $R_n^B(K\pi)$,
compared to Set~ÊI: $-2.1\sigma$ for $C(B^- \to K^- \pi^0)$, showing the
importance and complementarity of composed observables. The large pull values in
CP asymmetries arise from the higher statistical weight of the preciselier
measured branching fractions in their combined fit, reflecting the ``$\Delta
\ACP$ puzzle'' in the $B\to K \pi$ sytem.  The individual pull values of $C(B^-
\to K^- \pi^0)$ and $C(\bar{B}^0 \to K^- \pi^+)$ in Set~I add up to the large
pull of $\Delta C(K\pi)$ in Set~II.

In the following we will elaborate on the constraints posed by individual
observables. For example, the general shape of the contour of branching
fractions can be easily understood as follows: The leading contribution of the
decay amplitude $\hat{\alpha}_4^c(K\pi) = \alpha_4^c(K\pi) + \beta_3^c(K\pi)$ is
given as the sum of the QCD-penguin and the $\rho_A$-dependent WA amplitudes
$\alpha_4^c$ and $\beta_3^c$, respectively. The experimental measurement of the
branching fraction restricts $\hat{\alpha}_4^c$ to a circle in its imaginary
plane
\begin{align}
  \label{eq:BKpi:BR}
  \sqrt{\Br} & \simeq \lambda_c^{(s)} F_0^{B\to \pi} f_K \,
  \big|\hat{\alpha}^c_4\big| \, \big(1 + {\cal O}(r_i)\big)
\end{align}
where the $r_i = (r_{\rm T}^{},\, r_{\rm T}^{\rm C},\, r_{\rm EW}^{},\, r_{\rm
  EW}^{\rm C},\, r_{\rm EW}^{\rm A})$'s \cite{Hofer:2010ee} are numerically
small, mode-dependent corrections, normalized to $\hat{\alpha}_4^c$.
Consequently, $\beta_3^c(\rho_A^2, \rho_A^{})$ can interfere constructively or
destructively with $\alpha_4^c$ depending on the phase of $\rho_A^{K\pi}$.  For
$\phi_A^{K\pi} \sim 0,\, \pi$, the WA contribution is mainly real and
contributes constructively to $\alpha_4^c$. However, the contributions to
$\beta_3^c$ that are linear and quadratic in $\rho_A$ also interfere with each
other either constructively ($\phi_A^{K\pi} \sim 0$) leading to small
$|\rho_A^{K\pi}| \sim 2.0$ or destructively ($\phi_A^{K\pi} \sim \pi$), leading
to larger $|\rho_A^{K\pi}| \sim 3.4$. On the other hand, large values
$|\rho_A^{K\pi}| \sim 6.0$ are required for $\phi_A^{K\pi} \sim \pi/2,\,
(3\pi/2)$ where $\beta_3^c$ becomes purely imaginary and interferes
destructively with $\alpha_4^c$. In summary, the four branching fraction
measurements in Set~ÊI of the $B\to K \pi$ system can be described by a single
universal $\rho_A$ and by themselves they do not exclude any value of the phase
and allow up to $|\rho_A^{K\pi}|\lesssim 6$.

Whereas the branching fractions fix the modulus of
\begin{align}
  \label{eq:alpha_4^c}
  \hat{\alpha}_4^c & = |\hat{\alpha}_4^c|\,Ê\exp (i\, \hat\phi_4^c)\,,
\end{align}
the ratios of branching fractions
(see \refeq{DefOfRatios}) depend strongly on the real part of $\hat{\alpha}_4^c$,
i.e., are sensitive to the phase $\hat\phi_4^c$
\begin{align}
  R_{c,n}^{B,K,\pi} & \simeq
  1 + \cos \hat\phi_4^c \sum_i c_i \mbox{Re}(r_i)  + \ldots
\end{align}
Here the $c_i$ denote proportionality factors and terms proportional to
$\mbox{Im}(r_i)$ are denoted by dots. The latter become numerically important
only in the vicinity of $\hat\phi_4^c \sim \pi/2,\, (3\pi/2)$, and are fully
included in the fits.  Hence these ratios are sensitive to flips of
$\hat\phi_4^c$ by $\pi$.  As can be seen in \reffig{BAT_SM_Kpi_b}, the data
disfavors and excludes to a large extent the scenario of large WA when using
observable Set~II, i.e., purely imaginary $\beta_3^c$ that would interfere
destructively with $\alpha_4^c$.  There is no need for anomalously large WA
contributions to describe $B\to K\pi$ data of branching fractions and their
ratios in QCDF within the SM. Moreover, at 68\% probability the largest portion
of allowed $\rho_A^{K\pi}$ parameter space is within $\xi^A_3(K\pi) < 0.5$.

\begin{figure*}
  \begin{subfigure}[t]{0.32\textwidth}
    \centering
    \includegraphics[width=\textwidth]{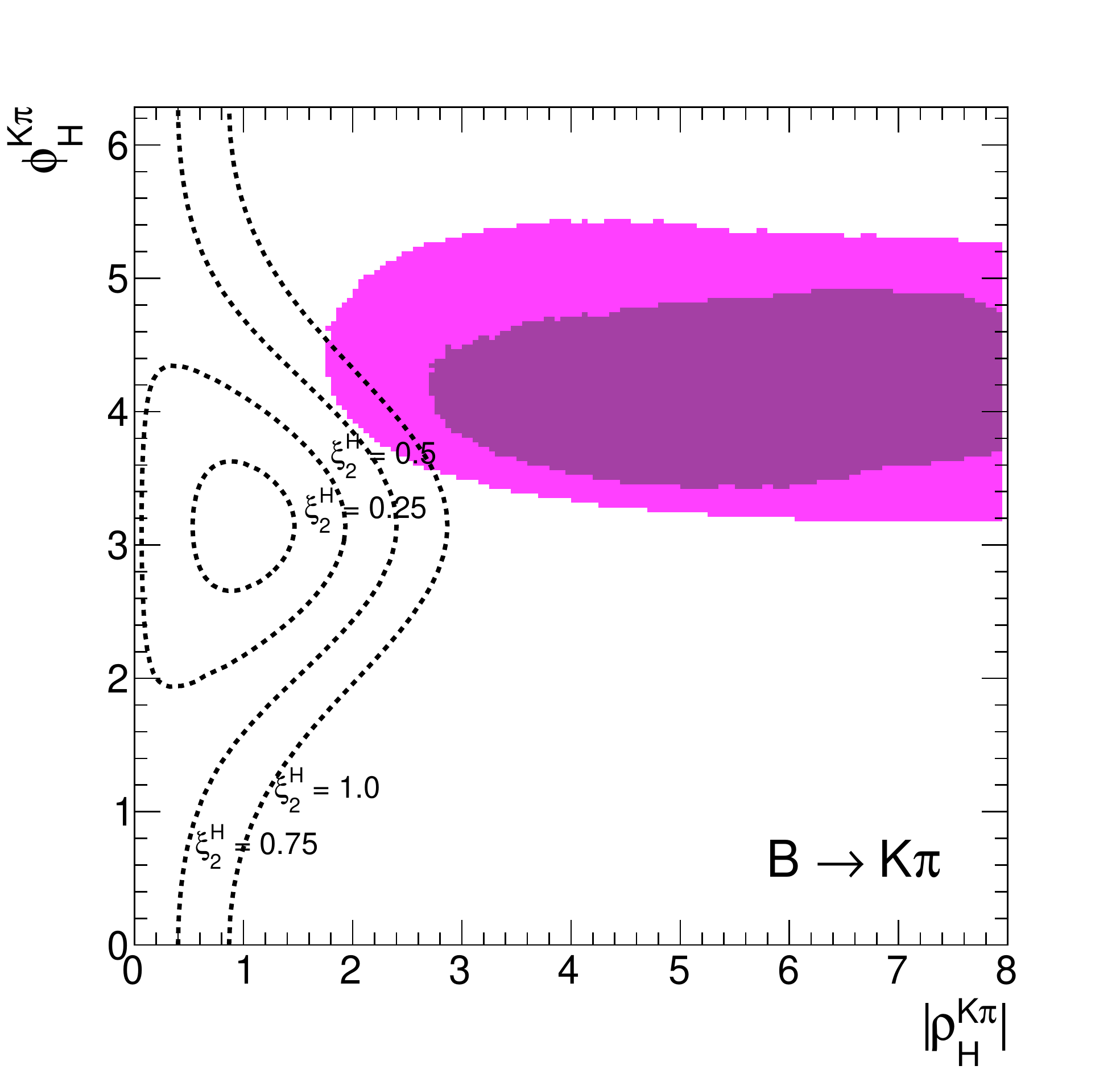}
    \caption{}
    \label{fig:BAT_SM_XH_BKpi}
  \end{subfigure}
  \begin{subfigure}[t]{0.32\textwidth}
    \centering
    \includegraphics[width=\textwidth]{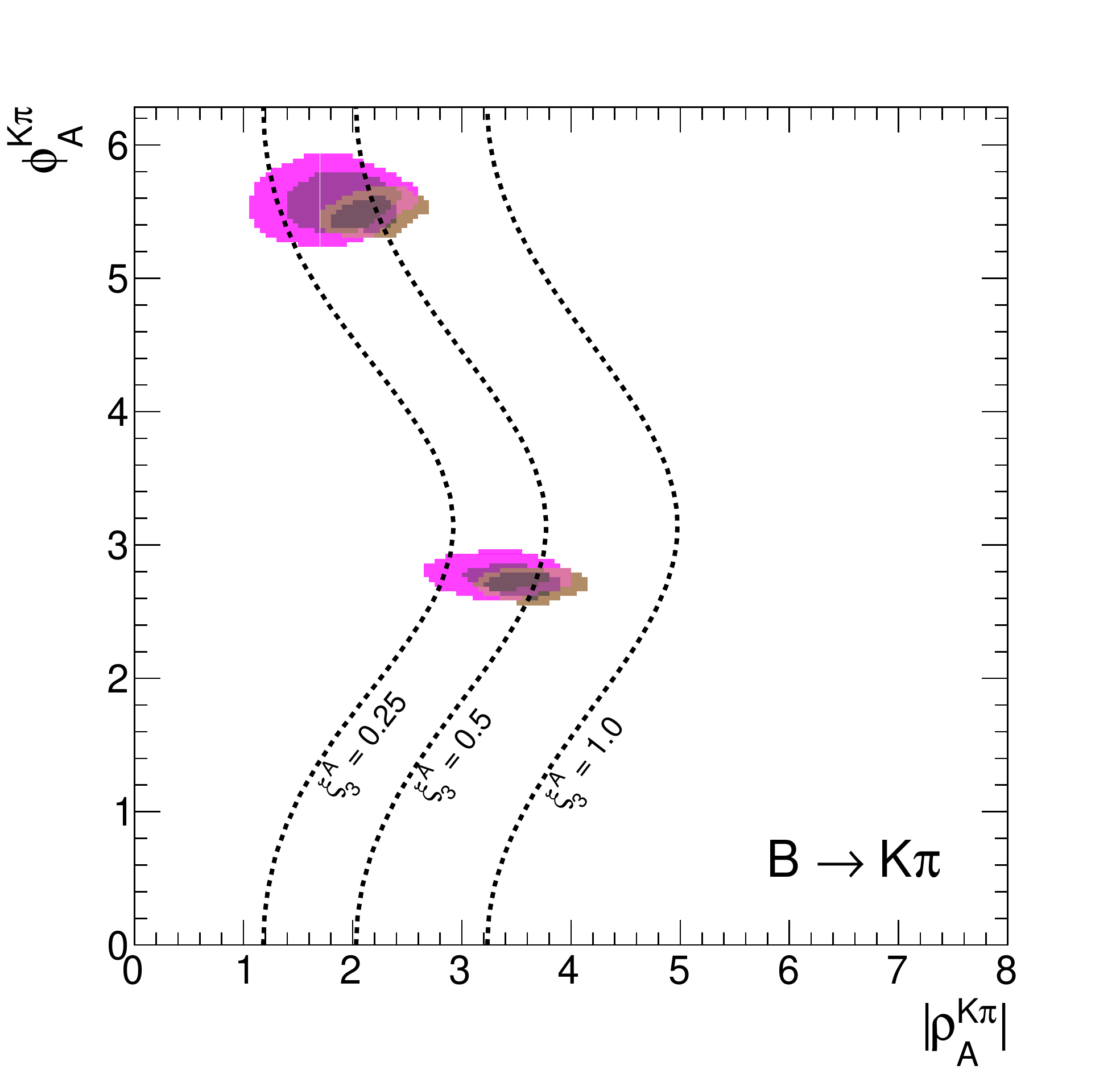}
    \caption{}
    \label{fig:BAT_SM_XAwithXH_BKpi}
  \end{subfigure}
\\
  \begin{subfigure}[t]{0.32\textwidth}
    \centering
    \includegraphics[width=\textwidth]{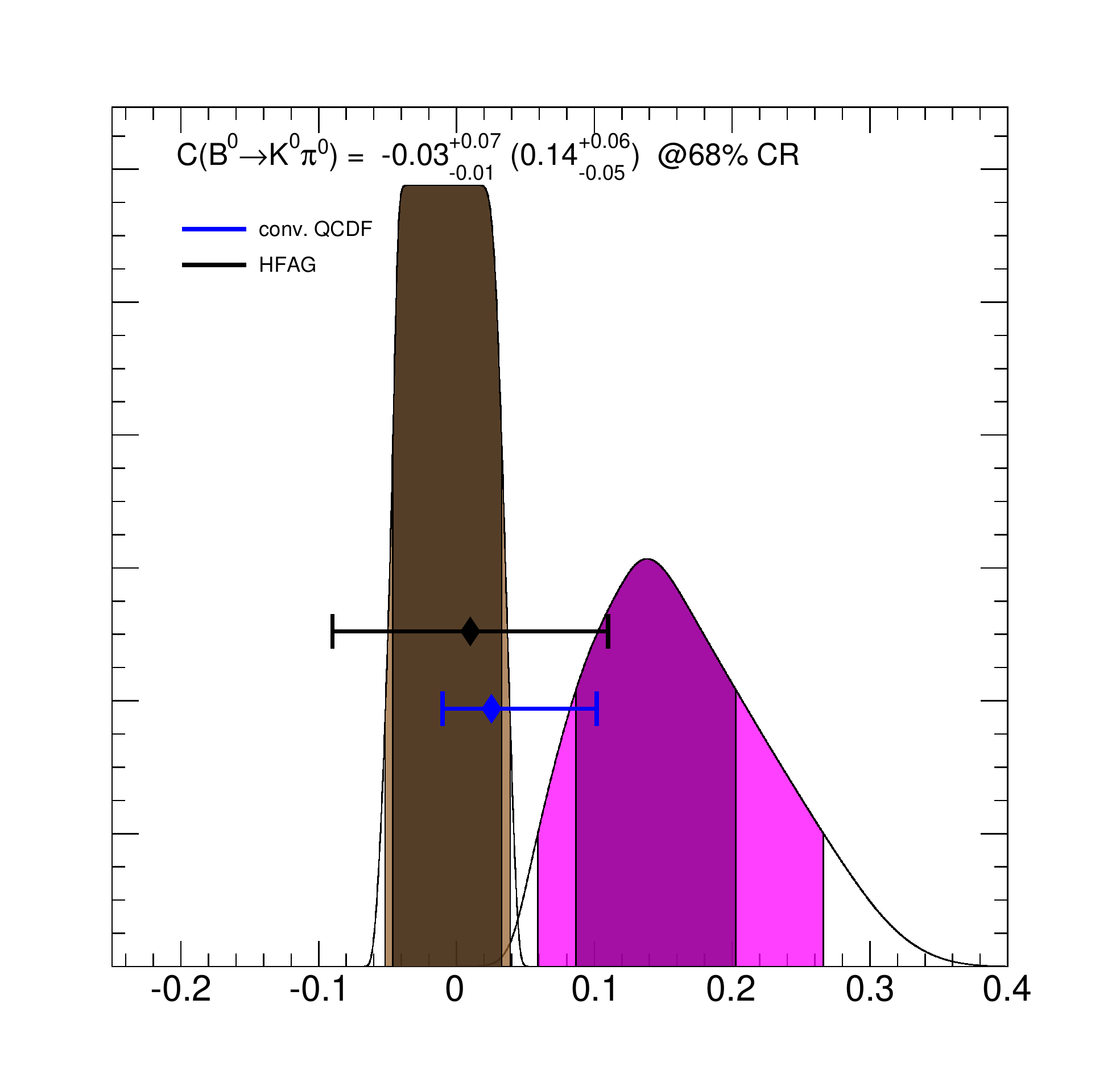}
    \caption{}
    \label{fig:BAT_SM_pred_CPA_BdK0pi0}
  \end{subfigure}
  \begin{subfigure}[t]{0.32\textwidth}
    \centering
    \includegraphics[width=\textwidth]{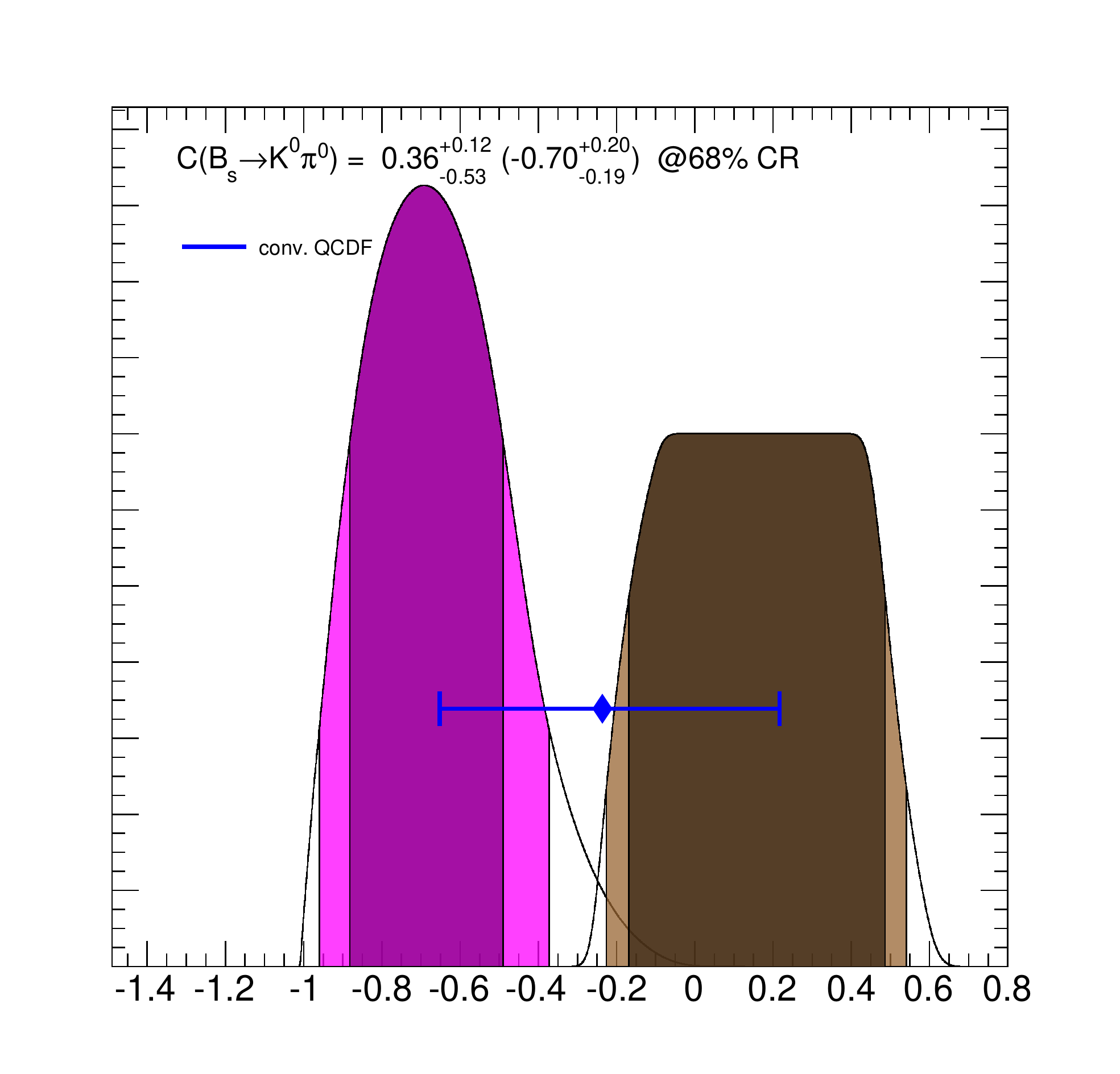}
    \caption{}
    \label{fig:BAT_SM_pred_CPA_BsK0pi0}
  \end{subfigure}

  \caption{ The 68\% (dark) and 95\% (bright) CRs of $\rho_H^{K\pi}$ (upper
    left) and $\rho_A^{K\pi}$ (upper right), obtained from a fit with partial
    (see text) observable Set II for $B\to K\pi$ decays, treating
    $\rho_H^{K\pi}$ either as a fit parameter (purple) or nuisance parameter as
    described in \refapp{prob:region} (brown).  The dashed lines correspond to
    constant $\xi^H_2(K\pi) = (0.25,\, 0.5,\, 0.75,\, 1.0)$ and $\xi^A_3(K\pi) =
    (0.25,\, 0.5,\, 1.0)$, respectively.  The lower panels show the predictions
    for $C(B_d\to K^0\pi^0)$ (left) and $C(B_s \to K^0\pi^0)$ (right). The
    available experimental results are shown with $1\sigma$ errors and a
    prediction from QCDF with the conventional uncertainty estimate is labeled
    ``conv. QCDF''. The 68\% credibility intervals for the predictions are given
    on the top of both panels for conventional $\rho_H$ (brown) and in brackets
    for fitted $\rho_H$ (purple).  }
\label{fig:BAT_SM_PP}
\end{figure*}

We provide also separately the constraints from direct CP asymmetries. In QCDF
the strong phase, necessary for CP violation, arises at ${\cal O}(\alpha_s)$,
respectively ${\cal O}(1/m_b)$, and is thus included only to leading order in
our numerical evaluations. Currently, CP asymmetries with neutral kaons in the
final state are measured to be small with large errors, whereas the ones with
charged kaons are observed to be large and with a relative opposite sign. For
the latter decays, the leading terms to the CP asymmetries are from
color-allowed, $r_{\rm T}$, and color-suppressed, $r_{\rm T}^{\rm C}$,
penguin-to-tree ratios \cite{Hofer:2010ee},
\begin{equation}
  \label{eq:CPA:BKpi:exp}
\begin{aligned}
  C(B^- \to K^- \pi^0) & 
  \approx 2\, \mbox{Im} (r_{\rm T} + r_{\rm T}^{\rm C}) \sin \gamma
\\
  & = (-4.0 \pm 2.1) \% \,,
\\[0.2cm]
  C(\bar{B}^0 \to K^- \pi^+) & 
  \approx 2\, \mbox{Im} (r_{\rm T}) \sin \gamma
\\
  & = (+8.2 \pm 0.6) \% \,, 
\end{aligned}
\end{equation}
where the measured values are taken from \cite{Amhis:2012bh}, and $\gamma$
denotes the angle of the CKM unitarity triangle.  Their difference is dominated
by the color-suppressed tree amplitude
\begin{align}
  \label{eq:DeltaACP:BKpi}
  \Delta C & 
  \simeq 2\, \mbox{Im} (r_{\rm T}^{\rm C}) \sin \gamma
  \simeq (-12.2 \pm 2.2) \% \,.
\end{align}
In QCDF, one has
\begin{equation}
  \label{eq:}
\begin{aligned}
  \mbox{Im} (r_{\rm T}) &
  \propto - \frac{\mbox{Re}(\alpha_1)}{|\hat\alpha_4^c|} \sin\hat\phi_4^c 
          + \frac{\mbox{Im}(\alpha_1)}{|\hat\alpha_4^c|} \cos\hat\phi_4^c \,,
\\
  \mbox{Im} (r_{\rm T}^{\rm C}) & 
  \propto - \frac{\mbox{Re}(\alpha_2)}{|\hat\alpha_4^c|} \sin\hat\phi_4^c 
          + \frac{\mbox{Im}(\alpha_2)}{|\hat\alpha_4^c|} \cos\hat\phi_4^c \,,
\end{aligned}
\end{equation}
where $\hat\alpha_4^c$ depends on $\rho_A$. The numerical values for 
\begin{equation}
\begin{aligned}
  100 \cdot \frac{\alpha_1}{|\hat\alpha_4^c|} & \approx -17.4^{+1.0}_{-0.9} - i \, 0.4^{+0.6}_{-0.5}\,,
\\ 
  100 \cdot \frac{\alpha_2}{|\hat\alpha_4^c|} & \approx  -5.8^{+1.1}_{-3.4} + i \, 1.5^{+0.3}_{-0.5}\, - 2.1^{+0.3}_{-1.2}\, \rho_H^{K\pi},
\end{aligned}
\end{equation} 
hold for the central values as well as the variation of theory parameters and
$\rho_A^{K\pi}$ at the best-fit point of Set II listed in \reftab{SM-pulls}.  We
kept explicitly the dependence of $r_{\rm T}^{\rm C}$ on the HS parameter
$\rho_H^{K\pi}$, which is numerically irrelevant for $r_{\rm T}$. It can be seen
that $r_{\rm T}^{\rm C}$ can be enhanced if $\mbox{Re}(\rho_H^{K\pi}) > 0$ and
$\mbox{Im}(\rho_H^{K\pi}) < 0$ --- see later discussion concerning
\reffig{BAT_SM_XH_BKpi}.

The allowed regions obtained from a fit of only CP asymmetries in
\reffig{BAT_SM_Kpi_a} and \reffig{BAT_SM_Kpi_b} (shown in green) are similar for
Set I and Set II. The best-fit point is at $\rho_A^{K\pi} \approx 4.1 \exp(i\,
1.8)$, along the branch of $\phi_A^{K\pi} \sim \pi/2$ that gives rise to strong
cancellations in destructive interference of $\alpha_4^c$ and $\beta_3^c$ and
leads to large theoretical uncertainties, which in turn allows for good
agreement with the data. Apart from the fact that branching fraction
measurements would become incompatible at more than $30\sigma$, neglected higher
order perturbative and power corrections would become important in these regions
of parameter space putting into doubt the reliability of the prediction.
However, there are substantial parts of the 68\% CRs with $\xi^A(K\pi) < 0.5$
and the size of WA contributions can be as low as $0.25$, for which these
comments do not apply.

The very same figures (\reffig{BAT_SM_Kpi_a} and \reffig{BAT_SM_Kpi_b}) show
also that there is no or hardly any overlap at 95\% probability of the allowed
regions from branching fractions (blue) and those from CP asymmetries
(green). In our approach, the so-called ``$\Delta \ACP$ puzzle'' manifests
itself only in the combined fit of branching fractions and CP asymmetries, where
then large pull values arise for $\Delta C$ (Set II), or equivalently also
$C(B^- \to K^- \pi^0)$ (Set I). These pull values are shown in \reftab{SM-pulls}
and caused by the higher statistical weight of the branching fraction
measurements. So one might wonder, how previous QCDF analyses, for example
\cite{Hofer:2010ee, Hofer:2012vc}, arrived at a ``$\Delta \ACP$ puzzle'' based
on the conventional approach, where $\rho_A$ is varied independently for each
observable? The answer is rather simple: there the uncertainty of an observable
is determined by the spread of values obtained in a scan of $\rho_A$ with
$|\rho_A^{K\pi}| \approx 1$ and arbitrary phase $\phi_A^{K\pi}$. Moreover, the
central value of the observable is usually assigned by definition to
$\phi_A^{K\pi} = 0$.  This corresponds in \reffig{BAT_SM_Kpi_a} and
\reffig{BAT_SM_Kpi_b} to a line of constant $|\rho_A^{K\pi}| = 1$, which yields
only consistent results for branching fractions, but never for the combination
of CP-asymmetries, i.e. those CP asymmetries which dominate statistically over
the ones with large experimental errors. In the conventional approach there will
be no ``$\Delta \ACP$ puzzle'' once larger values of $|\rho_A^{K\pi}| \lesssim2$
and a fine stepsize for the variation of $\phi_A^{K\pi}$ are permitted in the
scan, implying of course larger $\xi^A(K\pi) \lesssim 0.5$ and increased
theoretical uncertainties.  We note that $C(B^- \to K^0 \pi^-)$ almost vanishes
in QCDF and even in the presence of large power corrections it is difficult to
increase the predictions beyond $1\%$, such that the current pull of $1 \sigma$
(see \reftab{SM-pulls}) can hardly be reduced.  We emphasize again the different
assumptions underlying our approach, i.e., WA parameters are universal among
decays related by $(u\leftrightarrow d)$ quark exchange, but need not be small
and are determined from data, contrary to the conventional approach, i.e., WA
parameters are scanned over a rather small range and the resulting errors
correspond to non-universal parameters.

The results of the combined fit of branching fractions and CP asymmetries had
been already discussed at the beginning of this section. Whereas CP asymmetries
involved in the ``$\Delta \ACP$ puzzle'' exhibit larger pull values for both
sets of observables Set I and Set II, predictions of branching fractions are in
good agreement with the corresponding measurements, in part also due to large
form factor uncertainties. The latter parametric dependence cancels to a large
extent in ratios of branching fractions and yields a large pull value of
$-1.9\sigma$ for $R_n^B(K\pi)$ in Set~II, which contributes also to the
problematic $p$ value of 0.04. In a fit of Set~II without CP asymmetries we
obtain pull values of $-1.2\sigma$ for $R_n^B(K\pi)$, $0.4\sigma$ for
$R_c^K(K\pi)$ and $0.6\sigma$ for $R_c^\pi(K\pi)$, which can be compared to the
pull in \reftab{SM-pulls} when including CP asymmetries. This can be also seen
in \reffig{BAT_SM_Kpi_b} where the solution of the combined fit (red) at $\rho_A
\sim 3.3 \exp(2.7\,i)$ does neither overlap with the 68\% CRs from $\Br/R_{n,c}$
(blue) nor from $C/\Delta C$ (green).

Finally, we explore in more detail the discrepancy in $\Delta C$, departing from
the conventional error estimate of power corrections of HS contributions that
had been used until now, see \refapp{numeric:input}.  As previously mentioned in
\refeq{DeltaACP:BKpi}, the color-suppressed tree amplitude $\alpha_2^u$
determines the magnitude of $r_{\rm T}^{\rm C} \propto |\lambda_u^{(s)}/\lambda_c^{(s)}|
\, \alpha_2^u/\hat\alpha_4^c$. Possible large NNLO corrections might relax the
tension in the case of destructive interference to the real and constructive
interference to the imaginary part. For $\pi\pi$-final states, however, such
NNLO vertex corrections are cancelled by the NLO HS corrections
\cite{Bell:2007tv, Bell:2009nk, Bell:2009fm, Beneke:2009ek}, which might not
necessarily takes place to the same extent in $K\pi$ final states.
Nevertheless, in the following we will assume no large perturbative higher order
corrections and fit instead the phenomenological parameter $\rho_H^{K\pi}$ in
addition to $\rho_A^{K\pi}$. We point out that $\Delta C$ depends on
the sum $\alpha_{2,\rm{I}} + \alpha_{2,\rm{II}}^{{\rm tw}-2} 
+ \alpha_{2,\rm{II}}^{{\rm tw}-3}(\rho_H) + \beta_2(\rho_A^i)$, where $\beta_2$ is
dominated by building block $A_1^i$ and the corresponding WA parameter $\rho_A^i$
--- see \refeq{WAamp:WC:Aif}. The contribution of $\beta_2$ is always much smaller
than $\alpha_{2,\rm{II}}^{{\rm tw}-3}$ unless $\rho_H \ll \rho_A^i$. Hence we prefer
to fit $\rho_H$ and assume $\rho_A^i = \rho_A^{}$, the common WA parameter of all
building blocks $A_{1,2,3}^{i,f}$. Only for a rather large $\rho_A^i \gtrsim 4$
will the $\beta_2$ contribution be comparable to the theory 
uncertainties of the leading amplitudes in $\Delta C$.

The results of a fit to the partial observable Set II for the parameters
$\rho_H^{K\pi}$ and $\rho_A^{K\pi}$ is shown in purple in
\reffig{BAT_SM_XH_BKpi} and \reffig{BAT_SM_XAwithXH_BKpi}, respectively. We have
removed the CP asymmetry $C(B_d \to K^0\pi^0)$ that is very sensitive to HS, but
its current measurement does not provide any constraints, making it an ideal
candidate for a prediction. In contrast, $R_n^B(K\pi)$ is not very sensitive to
HS, but its large pull value would force $\rho_H^{K\pi}$ to large values, which
might not be necessary to explain $\Delta C$. For comparison we depict as brown
contours also the ones from \reffig{BAT_SM_Kpi_b} and provide contour lines of
constant $\xi_2^H(K\pi)$ and $\xi_3^A(K\pi)$.

We find that the prediction of $\Delta C = -0.11^{+0.04}_{-0.02}$ at the
best-fit point $\rho_H^{K\pi} = 3.3 \exp(3.7\, i)$ coincides, within
experimental uncertainties, with the measurement \refeq{DeltaACP:BKpi}.  The
preferred phase of $\phi_H^{K\pi} \sim (3\pi/2)$ implies that HS contributions
to $\alpha_2^u(K\pi)$ are mainly imaginary and interfere constructively with the
imaginary part of the vertex corrections.  At the best-fit point, all
observables have zero pull values except for a $0.8\sigma$ pull in $C(B^- \to
\bar{K}^0 \pi^-)$ and $R_n^B$ which had been discarded from the fit.

As can be seen in \reffig{BAT_SM_XH_BKpi}, the contours of constant
$\xi_2^H(K\pi)$ exhibit a different dependence on $\rho_H^{K\pi}$ as compared to
$\xi_3^A$ for $\rho_A^{K\pi}$. Already in the conventional approach
($|\rho_H^{K\pi}| = 1.0$), $\xi_2^H(K\pi) = 1$ is admitted in estimates of
theoretical uncertainties, which is a remnant artefact of the parametrization
$X_H \sim (1 + \rho_H^{K\pi})$. In the fit this contour line lies within the
95\% CR for $1.8 \lesssim |\rho_H^{K\pi}| \lesssim 2.8$ and for the smallest
$|\rho_H^{K\pi}| = 1.8$ at $\phi_H^{K\pi} = 4.6$, the pull of $\Delta C$
decreases, to $-1.0 \sigma$, compared to $2.8\sigma$ in the SM.  Concerning the
WA corrections shown in \reffig{BAT_SM_XAwithXH_BKpi}, $|\rho_A^{K\pi}|$ is
shifted towards lower values compared to \reffig{BAT_SM_Kpi_b}, which allows
also for smaller $\xi_3^A(K\pi)$.  The fit shows that these lower values of
$|\rho_A^{K\pi}|$ are correlated with large values of $|\rho_H^{K\pi}|$.

Assuming that HS corrections are in fact responsible for the observed
discrepancy in $\Delta C$, similar effects should be observed for related
decays, as for example in CP asymmetries $B_d \to K^0\pi^0$ and analogously $B_s
\to K^0\pi^0$. In the latter decay, such effects should be enhanced due to a
different hierarchy of CKM elements $|\lambda_u^{(s)} / \lambda_c^{(s)}| \ll
|\lambda_u^{(d)} / \lambda_c^{(d)}|$. The predictions of both CP asymmetries are
shown in \reffig{BAT_SM_pred_CPA_BdK0pi0} and \reffig{BAT_SM_pred_CPA_BsK0pi0},
respectively, with color coding as in \reffig{BAT_SM_XH_BKpi} and
\reffig{BAT_SM_XAwithXH_BKpi}. Once measured, respectively measured with higher
precision, both will allow to test the assumption of large HS contributions to
$B_{d,s}\to K\pi$ decays
\begin{align}
\begin{aligned}
  C(B_d \to K^0\pi^0)^{{\rm fit} \, \rho_H} \hspace{0.28cm} & = + 0.14^{+0.06}_{-0.05} \,,
\\
  C(B_d \to K^0\pi^0)^{{\rm scan}\, \rho_H} & \in [-0.04,\, 0.04] \,,
\end{aligned}
\intertext{and}
\begin{aligned}
  C(B_s \to K^0\pi^0)^{{\rm fit}\, \rho_H} \hspace{0.28cm} & = - 0.70^{+0.20}_{-0.19} \,,
\\
  C(B_s \to K^0\pi^0)^{{\rm scan}\, \rho_H} & \in [-0.17,\, 0.48] \,.
\end{aligned}
\end{align}
The predictions labeled ``fit $\rho_H$'' and ``scan $\rho_H$'' are shown in
purple and brown respectively, whereas the QCDF prediction for the conventional
approach (with scanned $\rho_A$) are labeled ``QCDF'' in
\reffig{BAT_SM_pred_CPA_BdK0pi0} and \reffig{BAT_SM_pred_CPA_BsK0pi0}. At the
current stage, the measurement of $C(B_d \to K^0\pi^0)$ prefers smaller HS
contributions although the uncertainty is still too large to draw a definite
conclusion.

A similar analysis of enhanced HS contributions \cite{Cheng:2009eg} has found a
best-fit point at $\rho_H^{K\pi} = 4.9 \exp(4.9\, i)$. Bearing in mind that
different numerical input, e.g. $\lambda_B = 0.35$ GeV, has been used, their
result lies in the ballpark of our 68\% CR. The very recent work
\cite{Chang:2014rla} also deals with fits of WA and HS parameters $\rho_{A,H}$ 
in $B\to PP$ decays $(PP = \pi\pi,\, K\pi,\, KK)$ in the SM in the framework of QCDF.
In our study one $\rho_A^{M_1 M_2}$ is considered for each of the three decay systems 
separately. Instead, in \cite{Chang:2014rla} one $\rho_A$ for building  block $A_3^f$
(see \refeq{WAamp:WC:Aif}), $\rho_A^f$, and one for building blocks $A_1^i \approx
A_2^i$, $\rho_A^i$, are used simultaneously, neglecting thus $SU(3)$-breaking
corrections for all three $b\to d$ and $b\to s$ decay systems. Also in this case
one finds that $\rho_A^f$ is rather strongly constrained with two solutions similar
to the ones shown in \reffig{BAT_SM_Kpi}. Concerning $\rho_H$, similar regions
are found as in our \reffig{BAT_SM_XH_BKpi} in scenario~III of \cite{Chang:2014rla}.

%
%
\subsection{Results for $B \to K \rho,\, K^*\pi,\, K^*\rho$}

\begin{figure*}
  \begin{subfigure}[t]{0.32\textwidth}
    \centering
    \includegraphics[width=\textwidth]{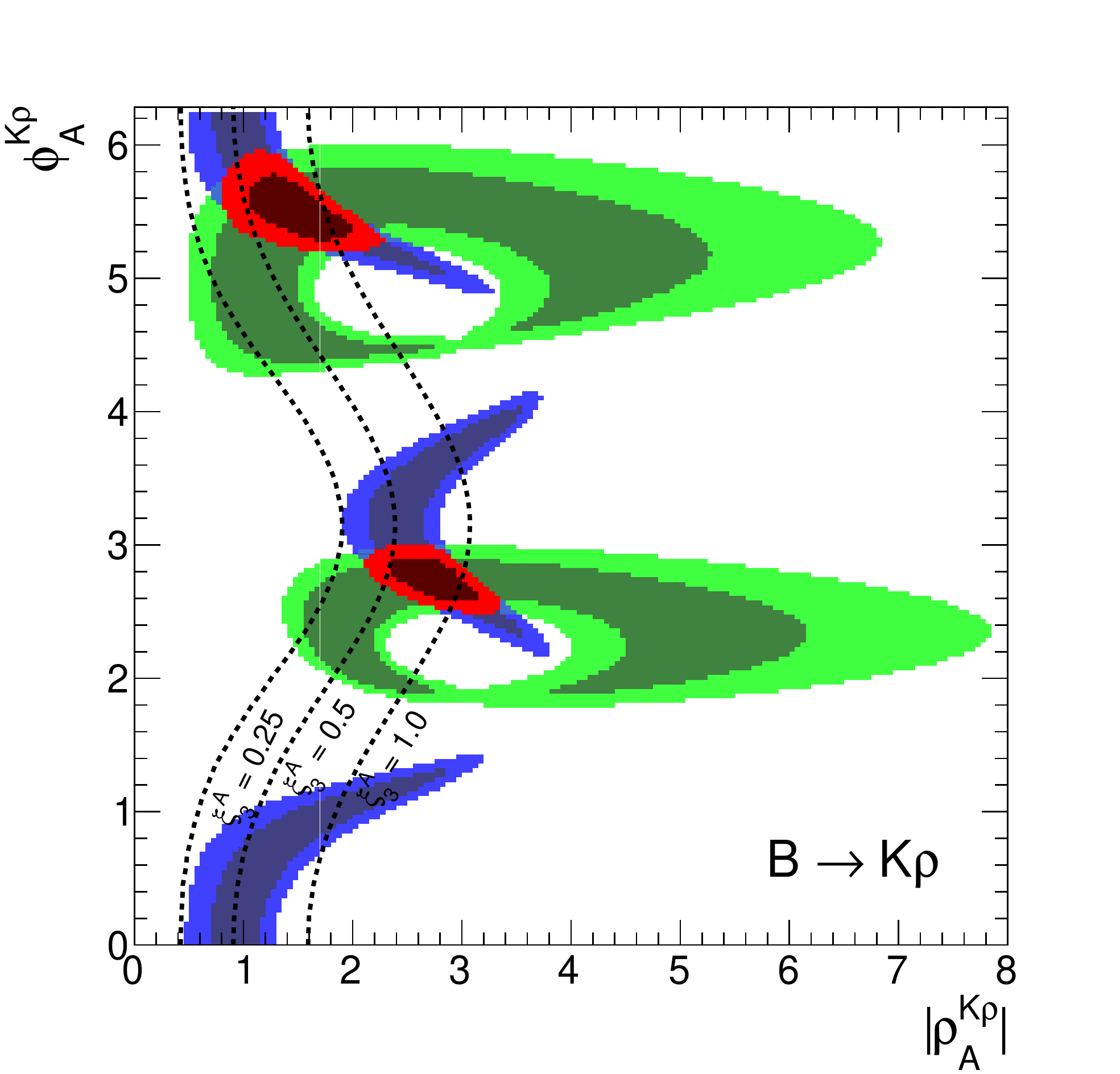}
    \caption{}
    \label{fig:BAT_SM_Krho_a}
  \end{subfigure}
  \begin{subfigure}[t]{0.32\textwidth}
    \centering
    \includegraphics[width=\textwidth]{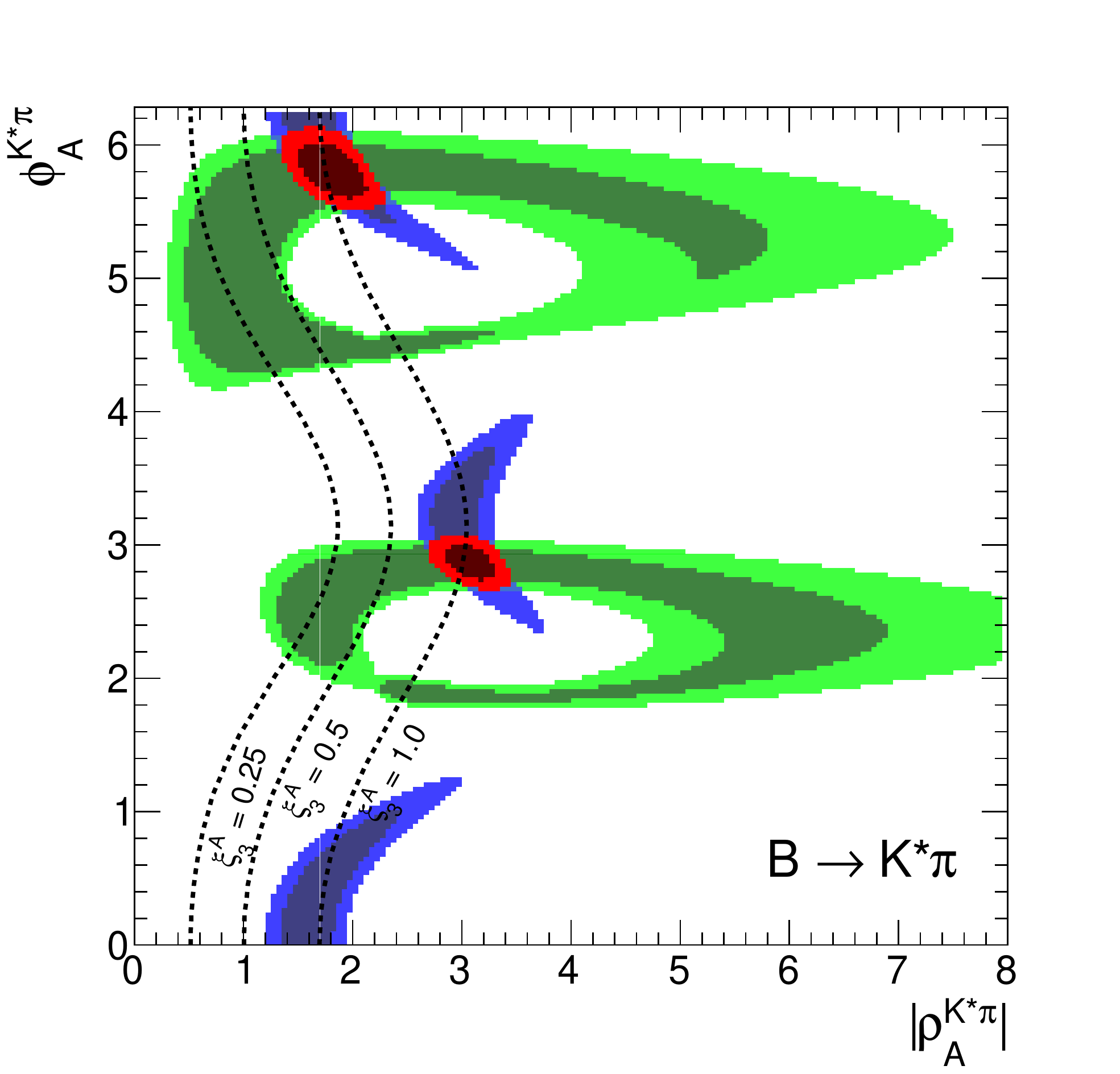}
    \caption{}
    \label{fig:BAT_SM_Kstpi_a}
  \end{subfigure}
  \begin{subfigure}[t]{0.32\textwidth}
    \centering
    \includegraphics[width=\textwidth]{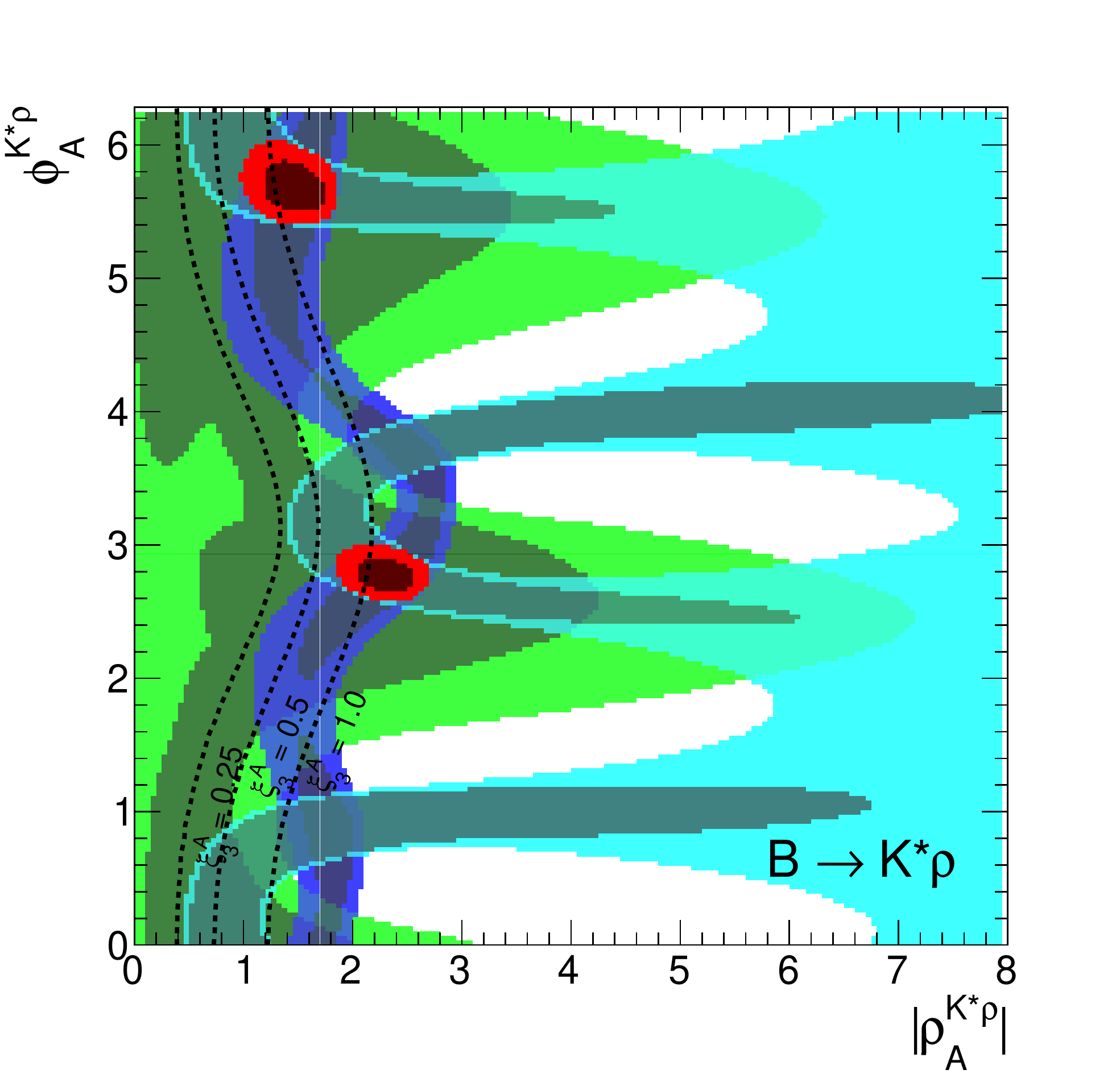}
    \caption{}
    \label{fig:BAT_SM_Kstrho_a}
  \end{subfigure}
\\  
  \begin{subfigure}[t]{0.32\textwidth}
    \centering
    \includegraphics[width=\textwidth]{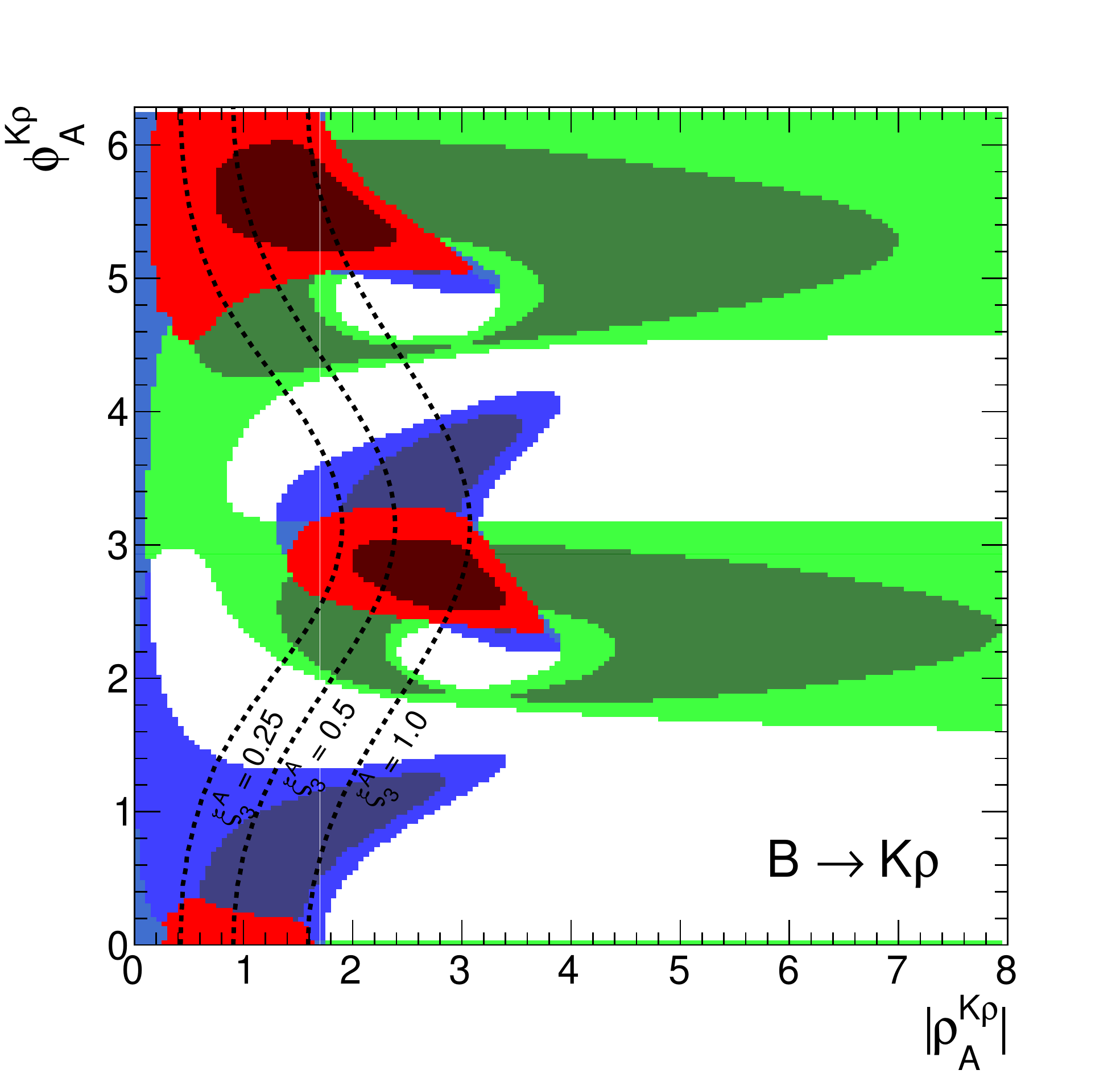}
    \caption{}
    \label{fig:BAT_SM_Krho_b}
  \end{subfigure}
  \begin{subfigure}[t]{0.32\textwidth}
    \centering
    \includegraphics[width=\textwidth]{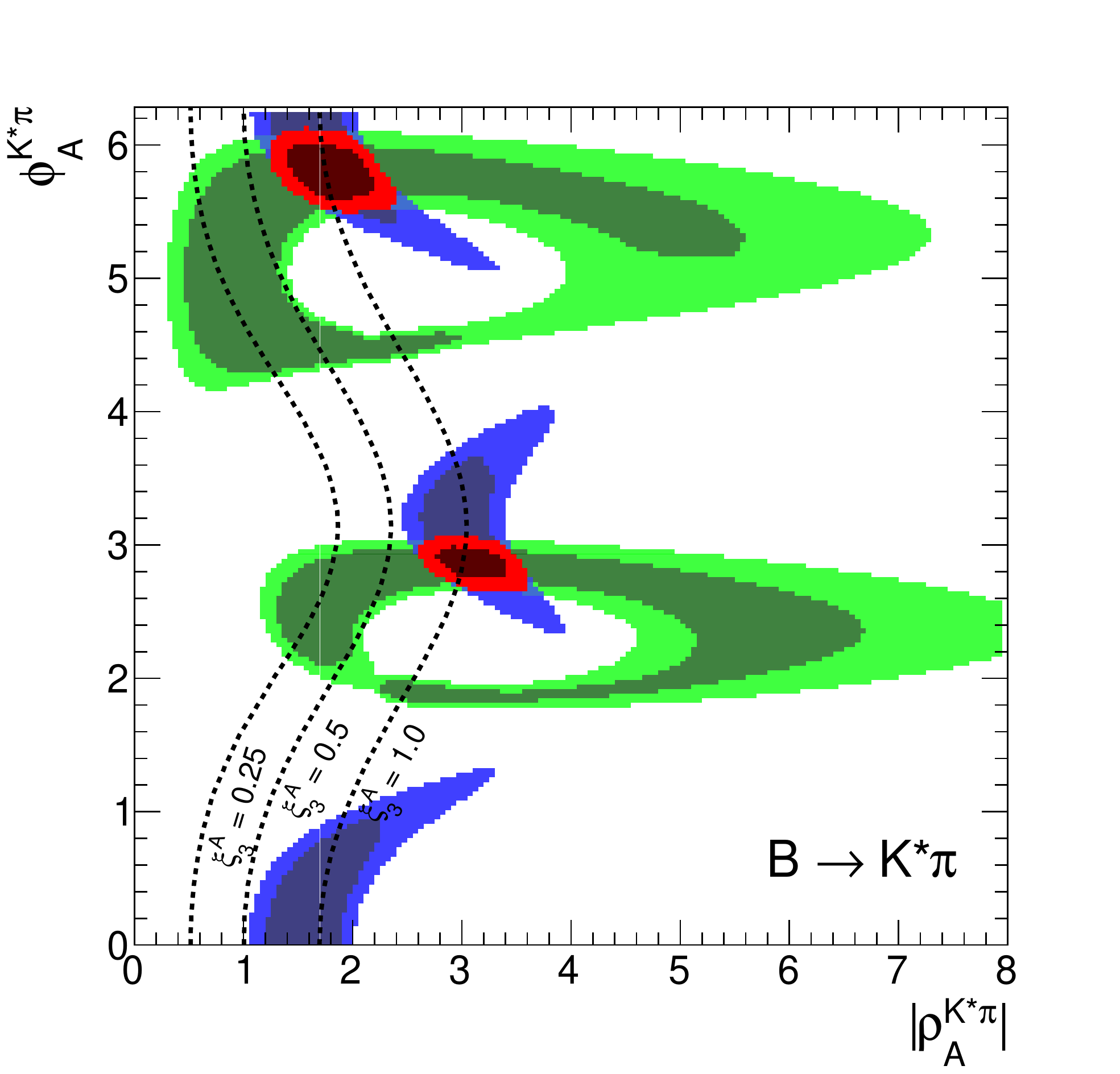}
    \caption{}
    \label{fig:BAT_SM_Kstpi_b}
  \end{subfigure}  
  \begin{subfigure}[t]{0.32\textwidth}
    \centering
    \includegraphics[width=\textwidth]{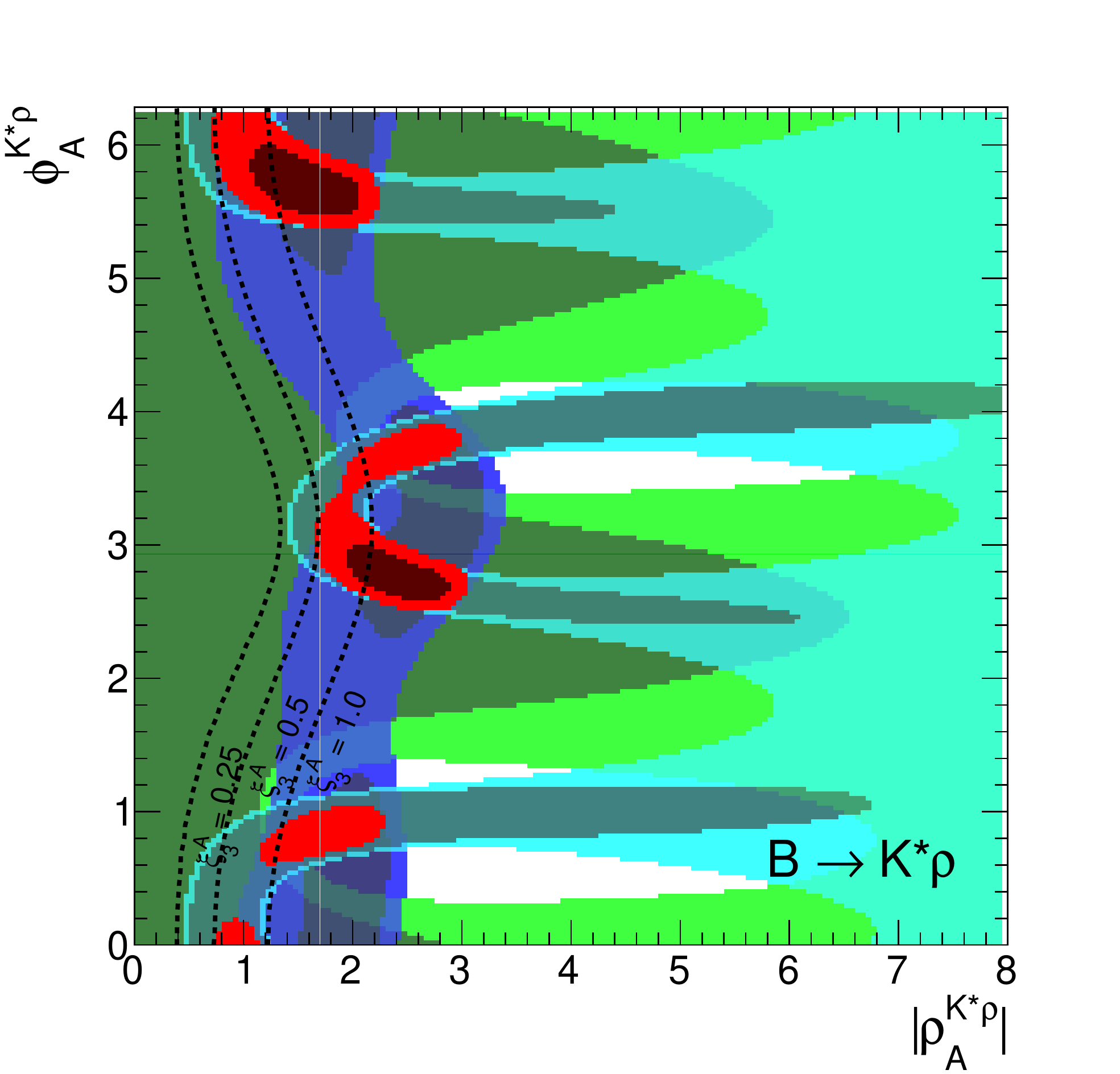}
    \caption{}
    \label{fig:BAT_SM_Kstrho_b}
  \end{subfigure}  

  \caption{ The 68\% (dark) and 95\% (bright) CRs of $\rho_A^{M_1 M_2}$ from a
    fit of observables in Set I (upper) and Set II (lower) of $B \to K \rho$
    (left), $B \to K^* \pi$ (middle), and $B \to K^* \rho$ (right).  Allowed
    regions are shown for $\Br$ and $R_{n,c}$ (blue), $C$ and $\Delta C$
    (green), $f_L$ (cyan) and their combination (red).  The dashed lines
    correspond to constant $\xi^A_3 = (0.25,\, 0.5,\, 1.0)$ from left to right.
  }
\label{fig:BAT_SM_Krhoetc}
\end{figure*}

In this section we discuss decay systems obtained from the replacement of a
pseudoscalar in $B\to K\pi$ by its vector meson equivalent $\pi \leftrightarrow
\rho$ and $K\leftrightarrow K^*$. Indeed, QCDF implies some qualitative
differences when changing the spin of the final state particles, but since the
parametrization of the decay amplitudes of all four decay systems is equal, one
might expect the discussed features of the $B\to K\pi$ system to appear also in
$B\to K^*\pi$ ($PV$), $B\to K \rho$ ($VP$) \footnote{The classification of
  decays into $M_1M_2 = PV$ and $VP$ refers to the amplitude $\alpha_4^c(M_1
  M_2)$, which indeed exclusively occurs in that combination in all decay
  amplitudes of both decay systems. Nevertheless, some other $\alpha_i$ with
  $i\neq 4$ also contain contributions in which the pseudoscalar and vector
  mesons are interchanged.} and $B\to K^*\rho$ ($VV$). Currently the
experimental measurements are not as precise as for $B\to K\pi$, and no striking
tensions are found as can be seen from the $p$ values and pulls of observables
in \reftab{SM-pulls}.

The allowed regions of $\rho_A^{M_1 M_2}$ are shown in \reffig{BAT_SM_Krhoetc}
for the observable Set~I (upper panels) and Set~II (lower panels). As before,
the 68\% and 95\% CRs allowed by fits from only $\Br/R_{c,n}$ or only $C/\Delta
C$ and in addition for $M_1 M_2 = VV$ also only $f_L$ are color coded as blue,
green and cyan, whereas the combined regions are depicted in red. As in the case
of $B\to K\pi$, the combined constraints on $\rho_A^{M_1 M_2}$ from Set~I and
Set~II observables are compatible with each other, but more stringent from
Set~I, especially for $B\to K\rho$ and $B\to K^* \rho$. Remarkably, the data of
all four decay systems $M_1M_2 = K\pi,\, K^*\pi,\, K\rho,\, K^*\rho$ prefers the
same regions of $\phi_A^{M_1 M_2} \sim \pi,\, 2\pi$, excluding large destructive
interference of $\alpha_4^c(M_1 M_2)$ and $\beta_3^c(M_1 M_2)$.  There is
overlap at the 68\% probability level for all three systems for the solution
$\phi_A \sim 2\pi $ and at 95\% probability for $\phi_A \sim \pi$.  This is also
supported by the data of $f_L$ in $B\to K^*\rho$, where the measurements of CP
asymmetries are not very precise yet and otherwise no stringent constrains on
$\rho_A^{K^*\rho}$ could have been obtained from branching fraction measurements
alone.

The relative amount of power corrections to the leading contribution for $PV$,
$VP$ and $VV$ final states is collected in \reftab{SM-xi-values} and indicated
in \reffig{BAT_SM_Krhoetc} by contour lines of constant $\xi_3^A(M_1 M_2) =
0.25,\, 0.5,\, 1.0$. It is typically larger by a factor of $2-3$ compared to the
$PP$ final state in $B\to K\pi$, which is a qualitative feature of QCDF.  The
leading QCD-penguin flavor amplitude is a linear combination of the vector
amplitude, $a_4$, and the chirally enhanced scalar QCD-penguin amplitude, $a_6$,
\cite{Beneke:2003zv}
\begin{align}
  \alpha_4 (M_1 M_2)& = a_4(M_1 M_2) \pm r_\chi^{M_2} a_6(M_1 M_2)
\end{align}
where the ``$+$'' sign applies to $M_1 M_2 = PP,\, PV$ and the ``$-$'' sign to
$M_1 M_2 = VP,\, VV$ final states. The two contributions interfere destructively
in the case $M_1 M_2 = VP$ leading to smaller QCD-penguin amplitudes than for
$M_1M_2 = PP$. Further, the tree level contribution to $a_6(M_1 M_2)$ vanishes
for $M_2 = V$, again reducing $\alpha_4$ in $M_1M_2 = PV,\, VV$ compared to
$M_1M_2 = PP$ giving implicitly rise to larger ratios $\xi_3^A(M_1 M_2)$.
Values as low as $\xi_3^A(K\rho) = 0.50$, $\xi_3^A(K^*\pi) = 0.82$ and
$\xi_3^A(K^*\rho) = 0.93$ can be reached within the 68 \% CRs of Set~I
observables, whereas even smaller values are allowed from Set~II, see
\reftab{SM-xi-values}. Concerning decays with $K^*$ in the final state, the
largish values of $\xi_3^A$ are required mainly by measurements of branching
fractions, whereas CP asymmetries and polarization fractions $f_L$ would allow
for smaller values of $\xi_3^A$, see \reffig{BAT_SM_Kstpi_a} and
\reffig{BAT_SM_Kstrho_a}.

The largest pull values arise for CP asymmetries $C(B^- \to \bar{K}^{*0} \pi^-)$
with $+1.0\sigma$ and $C(B^- \to \bar{K}^0 \rho^-)$ with $+0.7\sigma$. As in the
case of $C(B^- \to \bar{K}^0 \pi^-)$, these CP asymmetries almost vanish in QCDF
and it is difficult to increase the predictions beyond $1\%$, even in the
presence of large power corrections.

The advantage of observable Set~II strongly depends on cancellation of theory
uncertainties, as for example the form factors in the ratios of branching
fractions. Especially in cases where WA contributions are large compared to the
leading amplitude, i.e., large $\xi_3^A$, the reduction of uncertainties is less
effective and there is no unambiguous preference for the use of either Set~I nor
Set~II. Furthermore, the outcome of fits of Set~I and Set~II might differ
depending strongly on the experimental measurements. Apart from that we are not
aware of a specific reason for the qualitative differences between fits of Set~I
and Set~II for the $B \to K\pi,\, K^* \pi$ systems compared to $B \to K\rho,\,
K^*\rho$ systems. As can be seen from \reftab{SM-pulls}, pull values from Set~II
are in general slightly larger than from Set~I.

%
%
\subsection{Other decays and comments on $B\to K\omega,\, K^*\phi$}

\begin{figure*}
  \begin{subfigure}[t]{0.32\textwidth}
    \centering
    \includegraphics[width=\textwidth]{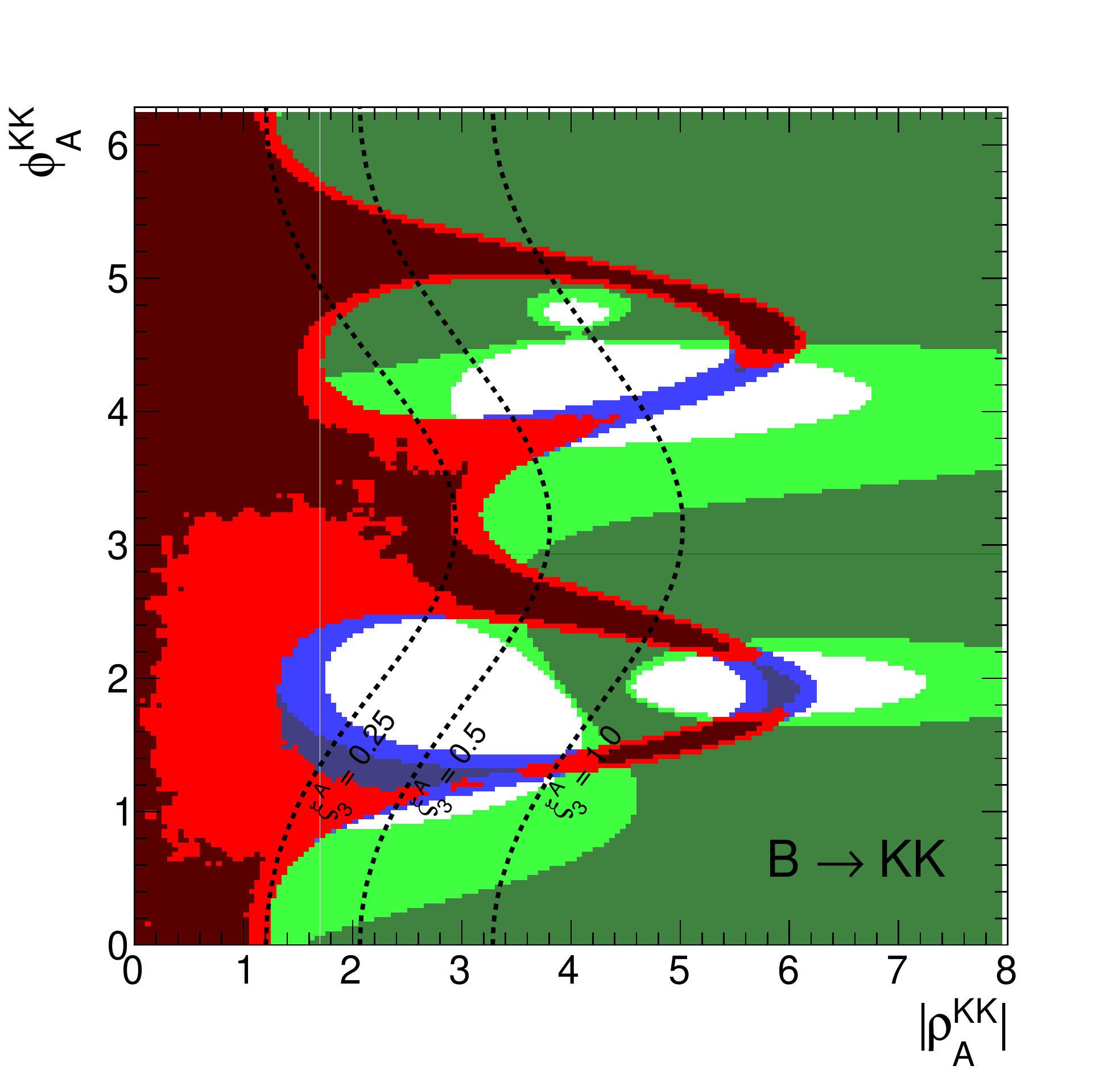}
    \caption{}
    \label{fig:BAT_SM_KKPengDom}
  \end{subfigure}
  \begin{subfigure}[t]{0.32\textwidth}
    \centering
    \includegraphics[width=\textwidth]{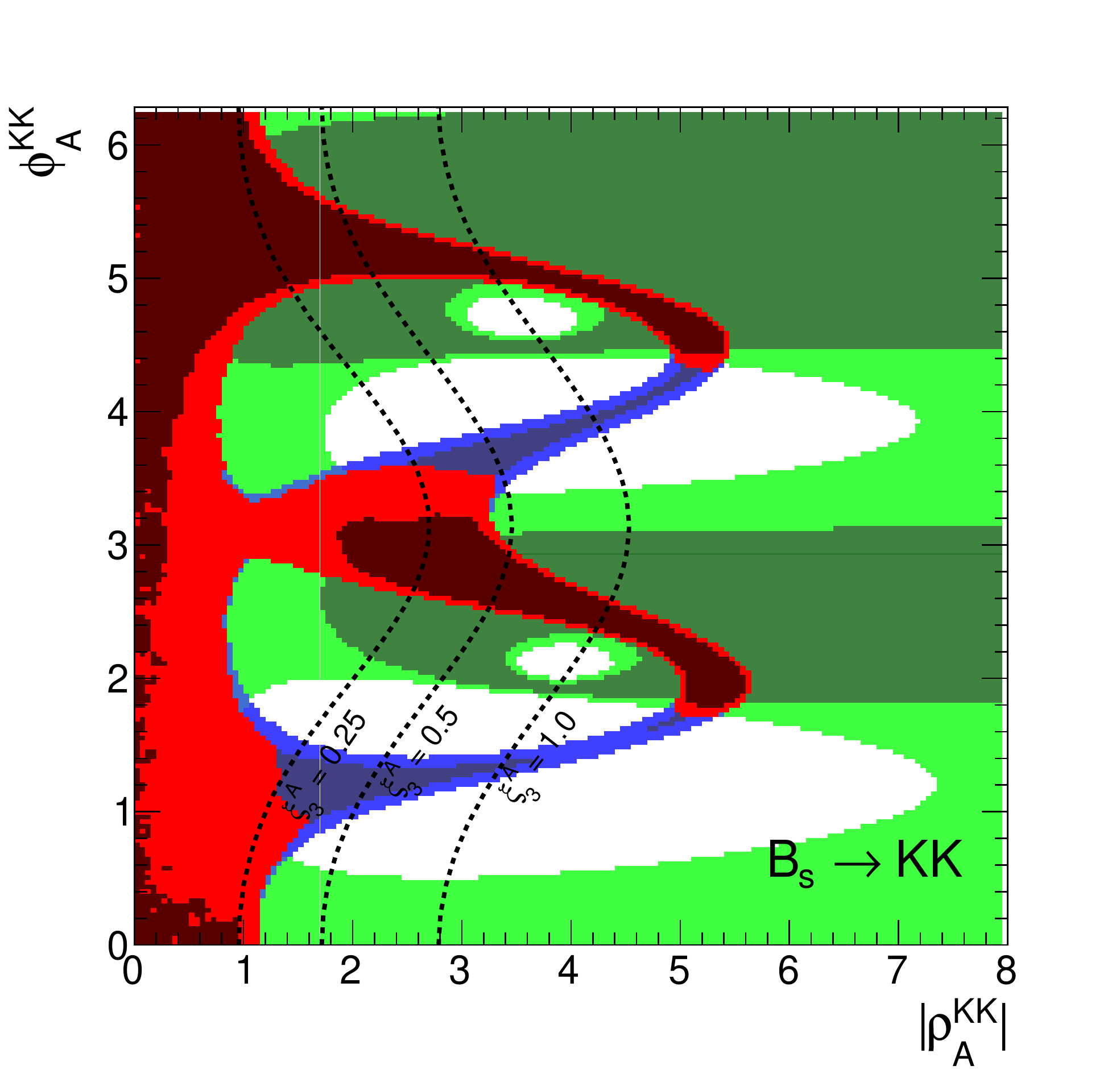}
    \caption{}
    \label{fig:BAT_SM_BsKK}
  \end{subfigure}
  \begin{subfigure}[t]{0.32\textwidth}
    \centering
    \includegraphics[width=\textwidth]{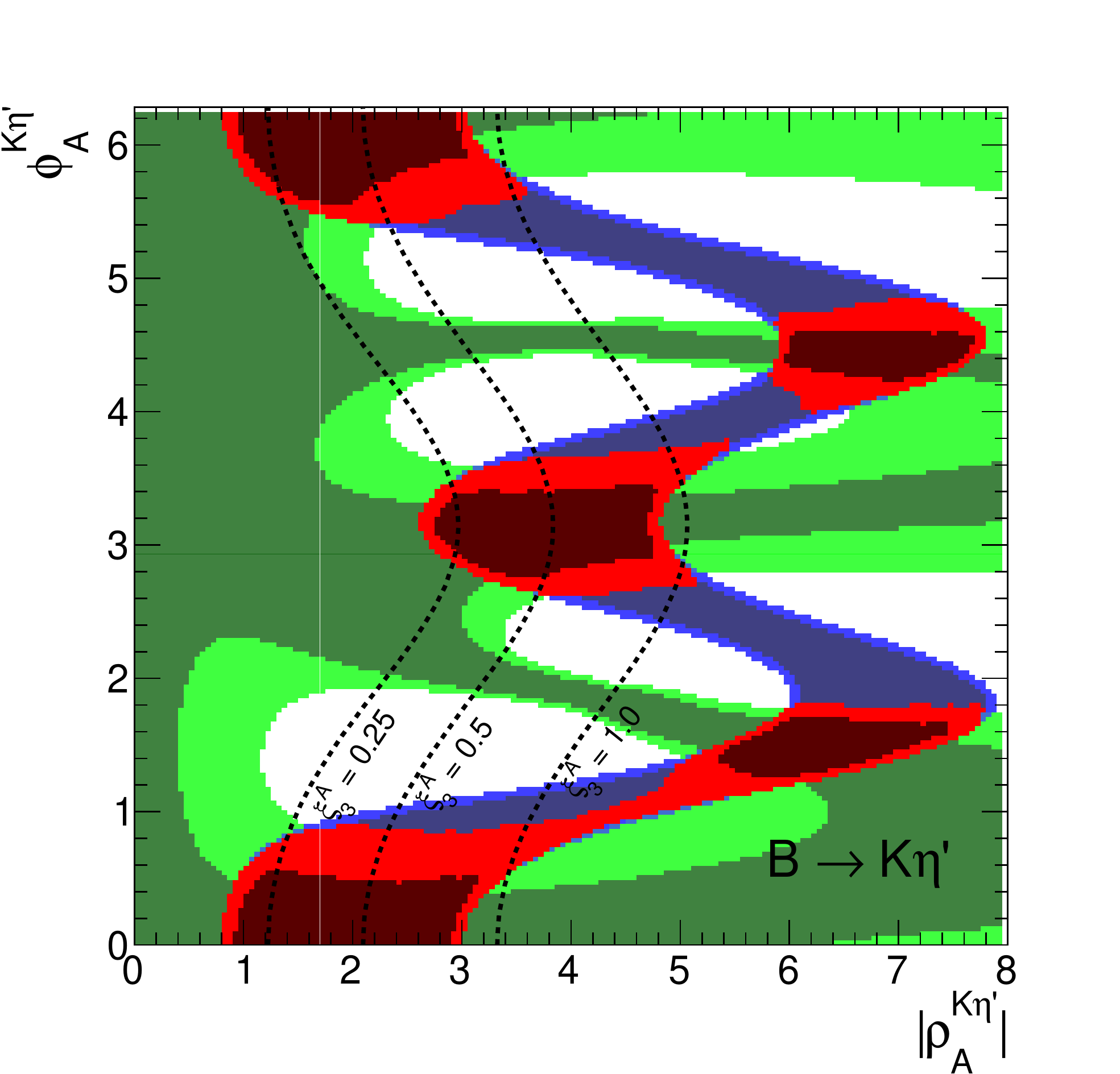}
    \caption{}
    \label{fig:BAT_SM_Ketap}
  \end{subfigure}

  \caption{ The 68\% (dark) and 95\% (bright) CRs of $\rho_A^{M_1 M_2}$ from a
    fit of observables in $B \to PP$: (a) $B\to KK$ (penguin dominated), (b)
    $B_s \to KK$ and (c) $B\to K \eta'$.  Allowed regions are shown for $\Br$
    (blue), $C$ (green) and their combination (red).  The dashed lines
    correspond to constant $\xi^A_3 = (0.25,\, 0.5, \, 1.0)$.  }
\label{fig:BAT_SM_PPetc}
\end{figure*}

\begin{figure*}
  \begin{subfigure}[t]{0.32\textwidth}
    \centering
    \includegraphics[width=\textwidth]{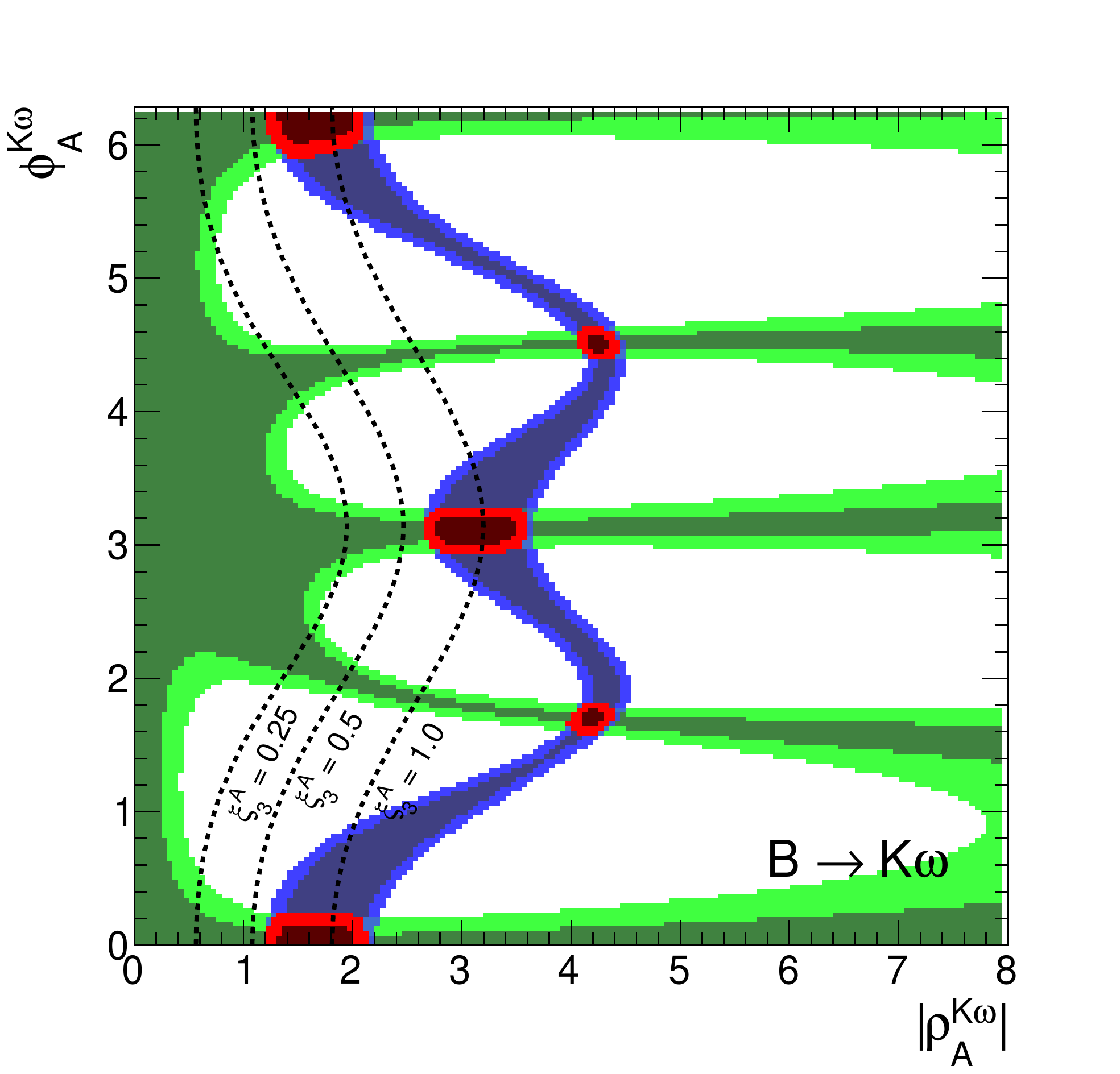}
    \caption{}
    \label{fig:BAT_SM_Komega}
  \end{subfigure}
  \begin{subfigure}[t]{0.32\textwidth}
    \centering
    \includegraphics[width=\textwidth]{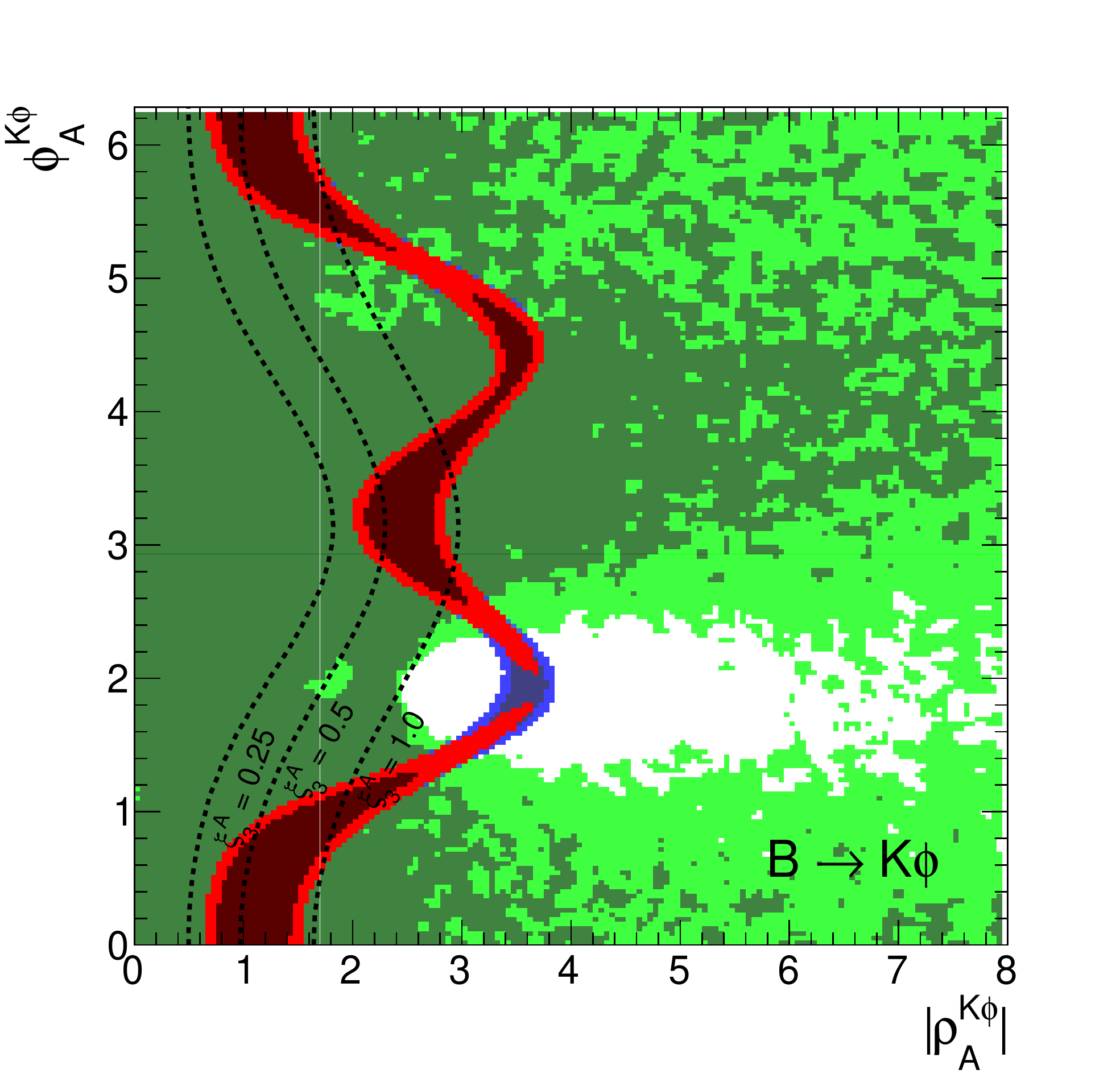}
    \caption{}
    \label{fig:BAT_SM_Kphi}
  \end{subfigure}  
  \begin{subfigure}[t]{0.32\textwidth}
    \centering
    \includegraphics[width=\textwidth]{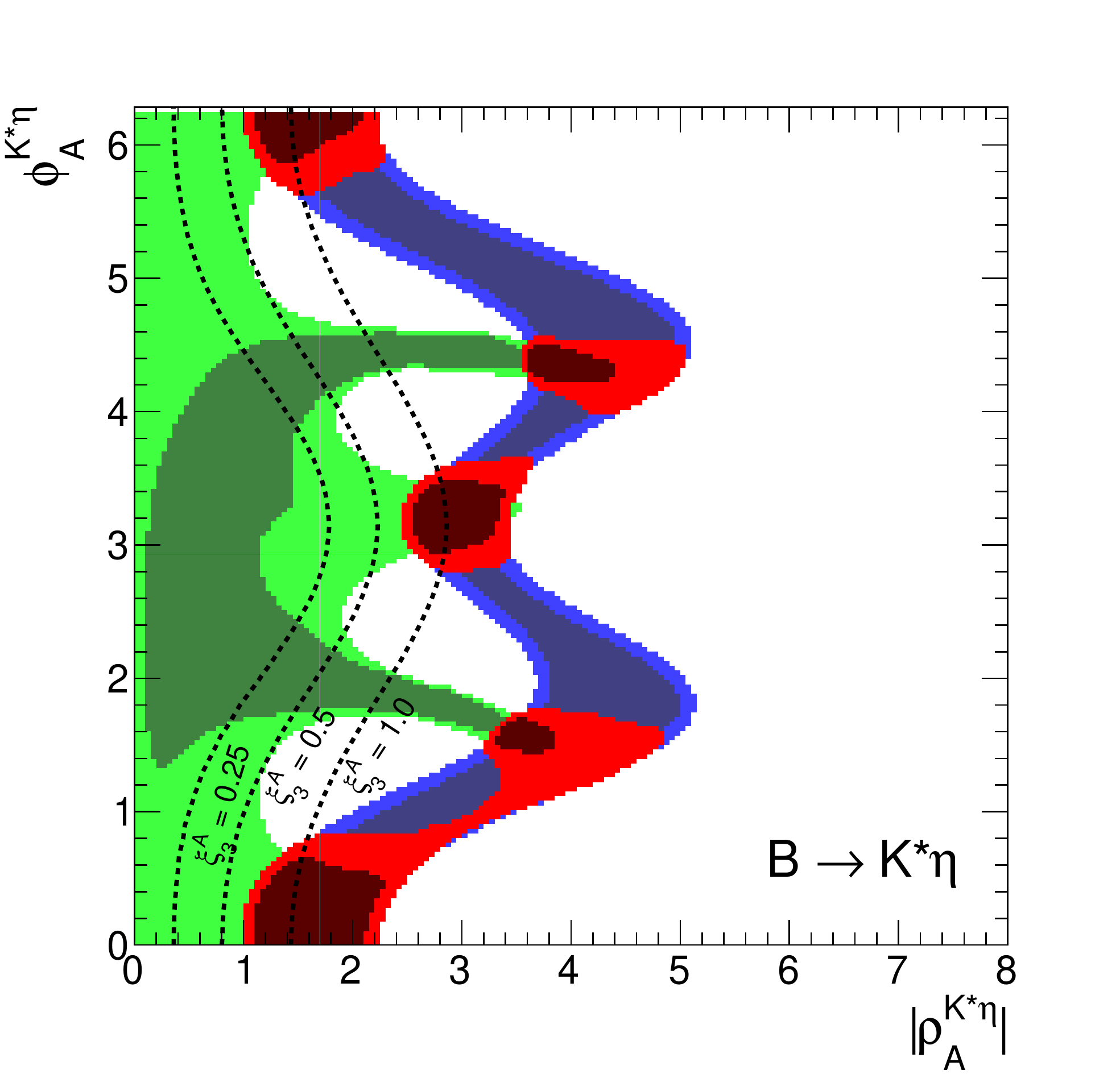}
    \caption{}
    \label{fig:BAT_SM_Kstareta}
  \end{subfigure}

  \caption{ The 68\% (dark) and 95\% (bright) CRs of $\rho_A^{M_1 M_2}$ from a
    fit of observables in $B \to PV$: (a) $B\to K\omega$, (b) $B \to K\phi$ and
    (c) $B\to K^* \eta$.  Allowed regions are shown for $\Br$ (blue), $C$
    (green) and their combination (red).  The dashed lines correspond to
    constant $\xi^A_3 = (0.25,\, 0.5,\, 1.0)$.  }
\label{fig:BAT_SM_PVetc}
\end{figure*}

\begin{figure*}
  \begin{subfigure}[t]{0.32\textwidth}
    \centering
    \includegraphics[width=\textwidth]{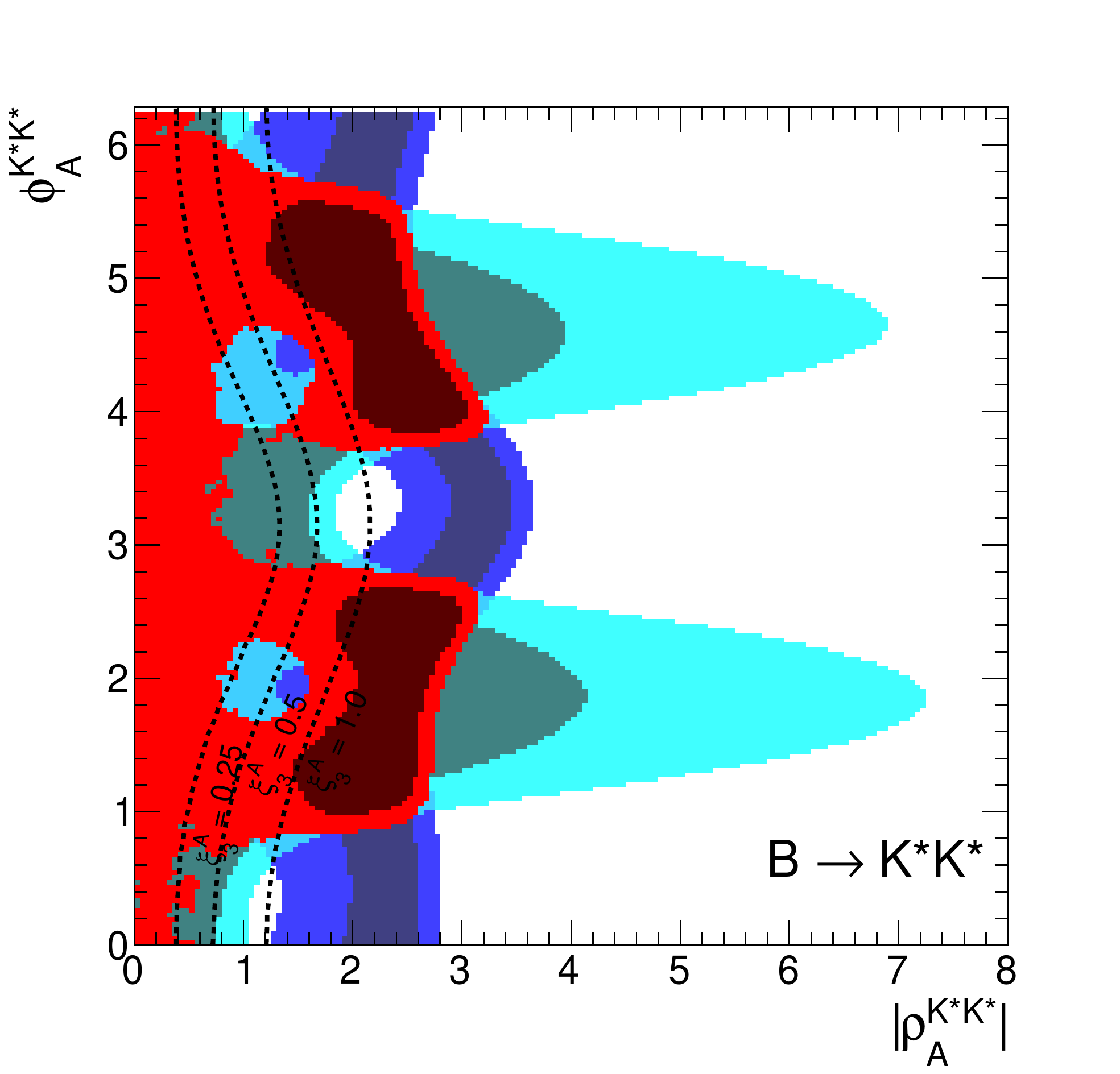}
    \caption{}
    \label{fig:BAT_SM_KstarKstar}
  \end{subfigure}
  \begin{subfigure}[t]{0.32\textwidth}
    \centering
    \includegraphics[width=\textwidth]{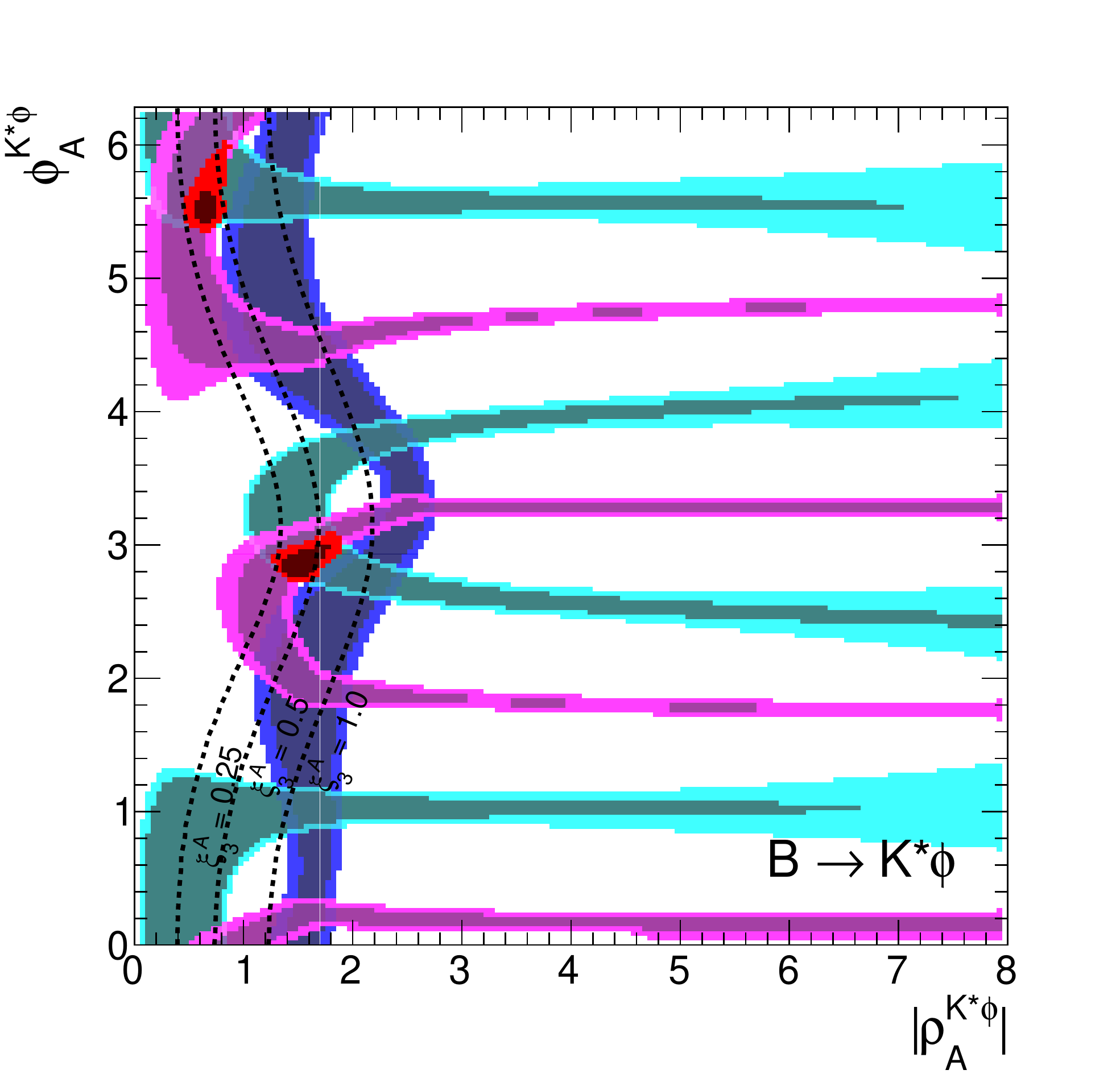}
    \caption{}
    \label{fig:BAT_SM_Kstarphi}
  \end{subfigure}
  \begin{subfigure}[t]{0.32\textwidth}
    \centering
    \includegraphics[width=\textwidth]{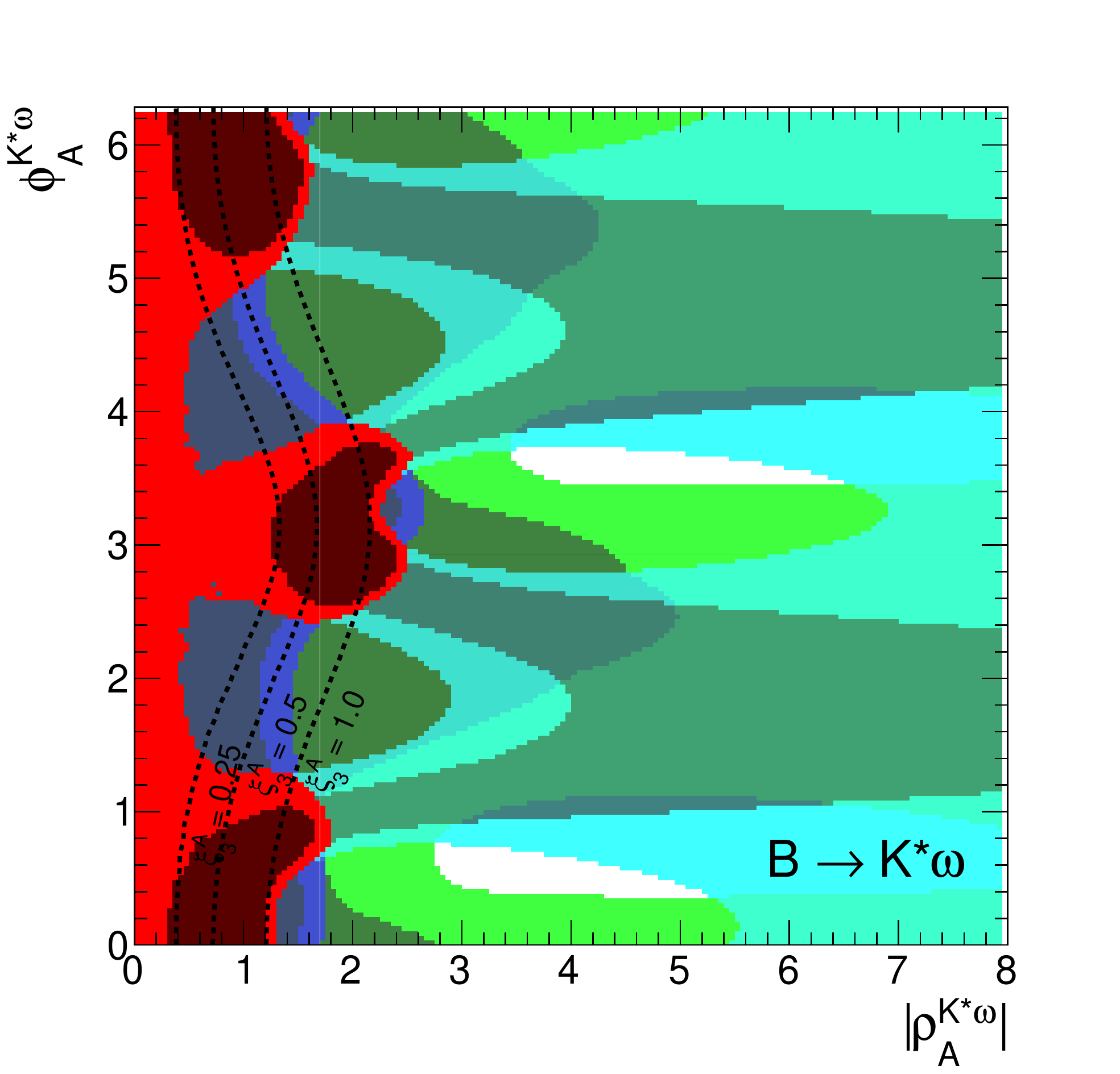}
    \caption{}
    \label{fig:BAT_SM_Kstaromega}
  \end{subfigure}
\\  
  \begin{subfigure}[t]{0.32\textwidth}
    \centering
    \includegraphics[width=\textwidth]{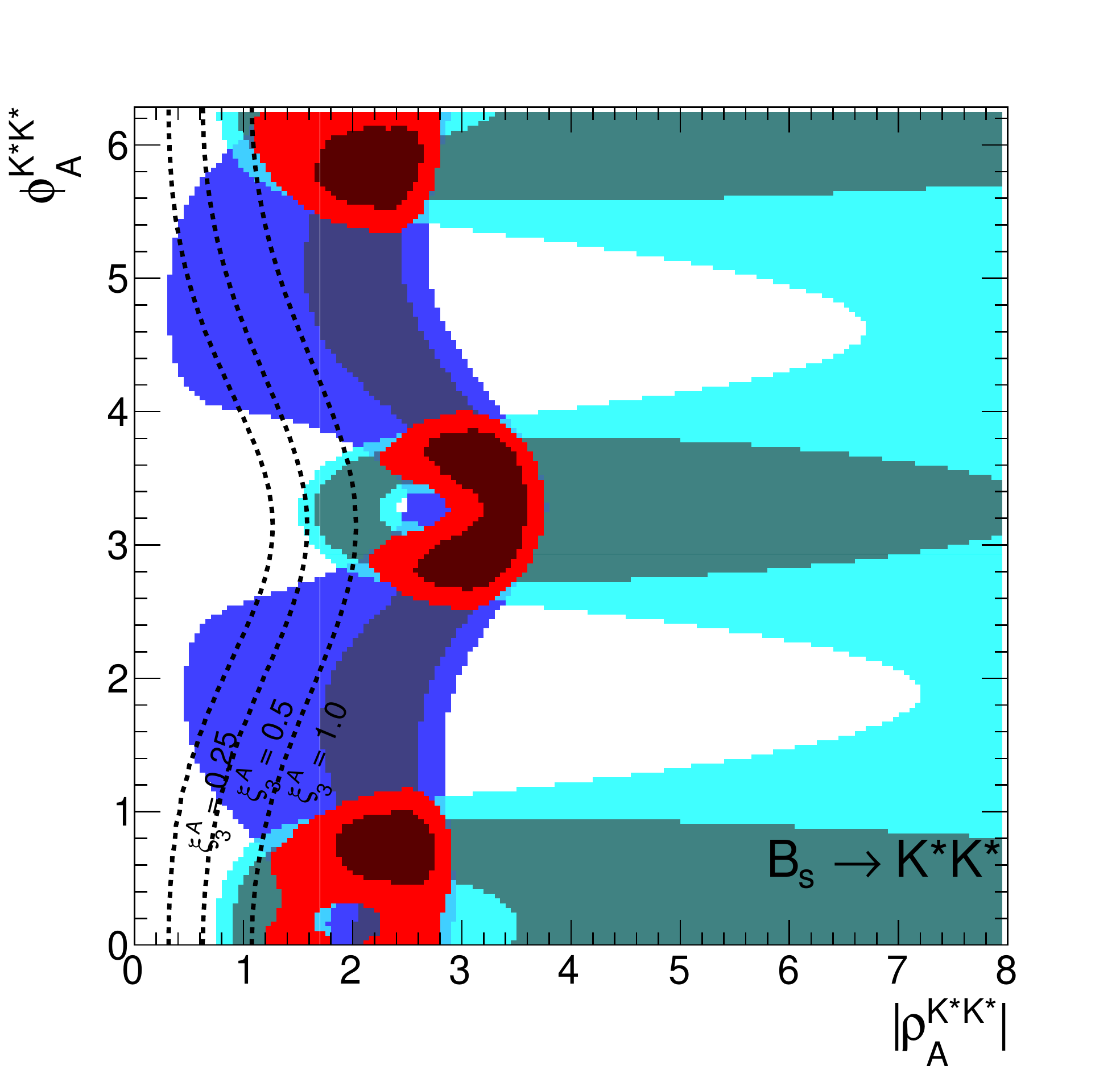}
    \caption{}
    \label{fig:BAT_SM_Bs_KstarKstar}
  \end{subfigure}
  \begin{subfigure}[t]{0.32\textwidth}
    \centering
    \includegraphics[width=\textwidth]{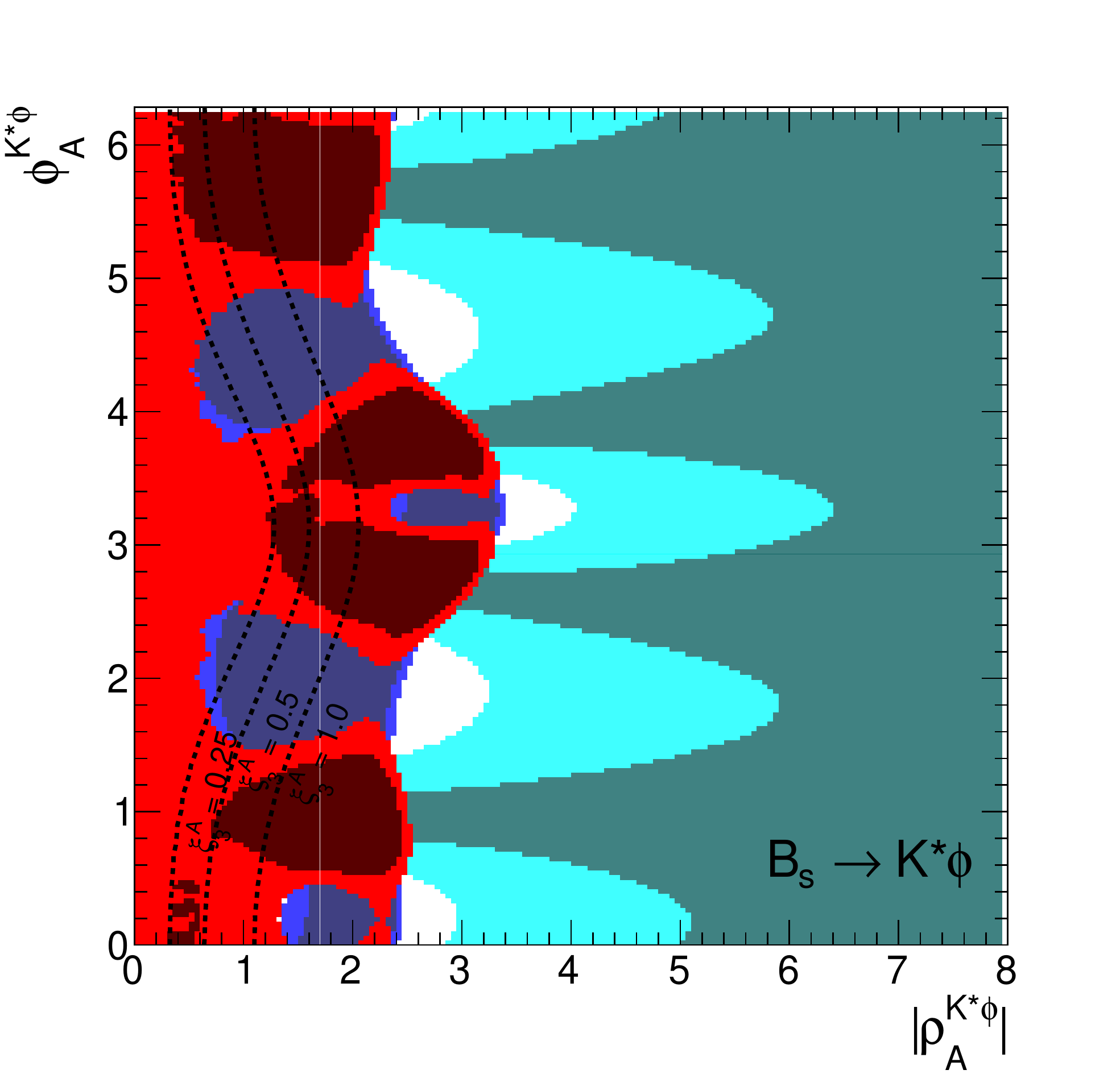}
    \caption{}
    \label{fig:BAT_SM_Bs_Kstarphi}
  \end{subfigure}  
  \begin{subfigure}[t]{0.32\textwidth}
    \centering
    \includegraphics[width=\textwidth]{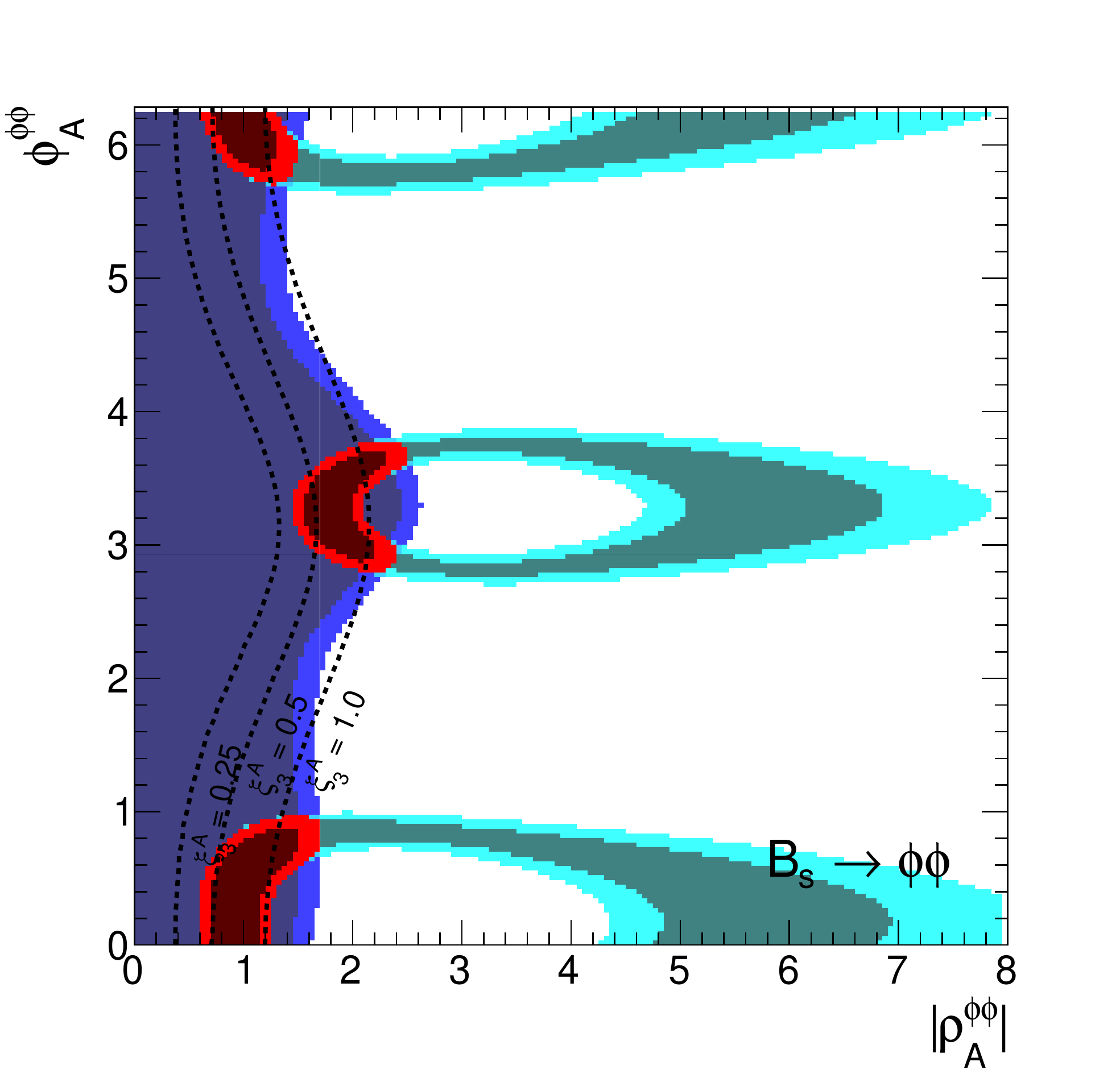}
    \caption{}
    \label{fig:BAT_SM_Bs_phiphi}
  \end{subfigure}  

  \caption{ The 68\% (dark) and 95\% (bright) CRs of $\rho_A^{M_1 M_2}$ from a
    fit of observables in $B \to VV$: (a) $B\to K^*K^*$, (b) $B \to K^*\phi$,
    (c) $B\to K^* \omega$ and (d) $B_s\to K^*K^*$, (e) $B_s \to K^*\phi$, (f)
    $B_s\to \phi\phi$.  Allowed regions are shown for $\Br$ (blue), $C_h$
    (green), $f_h$ (cyan), $\phi_h$ (purple)and their combination (red).  The
    dashed lines correspond to constant $\xi^A_3 = (0.25,\, 0.5,\, 1.0)$ from
    left to right.  }
\label{fig:BAT_SM_VVetc}
\end{figure*}

We tested our assumption of universal WA also with data listed in
\reftab{exp:input:obs:PP}, \reftab{exp:input:obs:PV} and
\reftab{exp:input:obs:VV} for other QCD-penguin dominated decay modes mediated
by $b\to (d, s)$ transitions. For these decays, the analysis is restricted to
observable Set~I, where in most cases the experimental accuracy is poorer than
for previously studied $B\to K\pi,\, K\rho,\, K^*\pi,\, K^*\rho$ systems. The
ranges for the ratios $\xi_3^A(M_1 M_2)$ that are required by data are listed in
\reftab{SM-xi-values}, which have been commented previously.  For all systems,
again preferred regions appear for $\phi_A^{M_a M_b} \sim \pi,\, 2\pi$, and in
some cases also $\phi_A^{M_a M_b} \sim \pi/2,\, (3\pi/2)$ is still allowed.

The allowed regions for $B\to PP$ systems $B\to KK,\, K\eta'$ and $B_s \to KK$
are shown in \reffig{BAT_SM_PPetc} and for $B_s \to K^-\pi^+$ in
\reffig{BAT_SM_BsKpi}.  We do not show $B\to K \eta$ for which the data is even
less constraining.  Except for $B_s \to K^-\pi^+$, the measurements of CP
asymmetries are very poor and provide only little additional constraints to the
ones of branching fractions. The preferred regions of WA contributions for
$B_{d,s} \to KK$ look very alike supporting the assumption of universal WA for
$B_d$ and $B_s$ decays into same final states, entertained in
\refsec{WA:same:final}. In comparison, for $B_{d,s} \to K\pi$
(\reffig{BAT_SM_Kpi_a} and \reffig{BAT_SM_BsKpi}) this might not seem the case,
however here one should compare the result of the fit to $B_d \to K^-\pi^+$ only
rather than the combination of all $B\to K\pi$ decays shown in
\reffig{BAT_SM_Kpi_a} and \reffig{BAT_SM_Kpi_b}.

In \reffig{BAT_SM_PVetc} the allowed regions for $B\to VP$ systems $B\to
K\omega,\, K\phi,\, K^*\eta$ are shown, whereas $B\to K^*\eta'$ has been omitted
due to the poor constraints from the respective data. The measurements of
branching fractions provide in all three cases already appreciable
constraints. Concerning $B\to K\omega$, no tensions are observed. In case of
$C(\bar{B}^0\to \bar{K}^0 \omega^0)$, we included the HFAG average of the two
incompatible measurements of Belle: $C = 0.36\pm 0.19 \pm 0.05$
\cite{Chobanova:2013ddr} and BaBar: $C = -0.52^{+0.22}_{-0.20} \pm 0.03$
\cite{Aubert:2008ad}, which differ by $2.9\sigma$. The HFAG value $C = -0.04 \pm
0.14$ \cite{Amhis:2012bh} indeed coincides with the theory prediction at the
best-fit point ($C = -0.02 \pm 0.08$).  One might hope that improved
measurements at Belle II will settle this problem.  As $\Delta C(K\pi)$, this CP
asymmetry is sensitive to the analogous color-suppressed tree amplitude
$\alpha_2^u(K\omega)$ and might provide further tests of large HS contributions,
which would be clearly visible.  As in fact the largest uncertainty in the
theory prediction is due to $\rho_H^{K\omega}$.  It must be noted that although
the best-fit point at $\rho_A^{K\omega} = 4.2 \exp(i\, 1.7)$ corresponds to a
large $\xi_3^A(K\omega) = 2.7$, other solutions at 68\% probability with
$\xi_3^A(K\omega) \lesssim 1$ provide equally vanishing pulls of observables.

Finally, the allowed regions for $B\to VV$ systems $B\to K^*K^*,\, K^*\phi,\,
K^*\omega$ and $B_s \to K^*K^*,\, K^*\phi,\, \phi\phi$ are shown in
\reffig{BAT_SM_VVetc}.  For $MM = K^*K^*$ final states the measurements of
branching fractions require rather large WA contributions, contrary to the other
considered $VV$ final states. In all cases, the polarization fractions provide
orthogonal constraints, which prefer $\phi_A^{M_a M_b} \sim \pi,\, 2\pi$, except
for $B \to K^*K^*$. For the moment measurements of CP asymmetries are only
available for $B\to K^*\omega$ and the very recent LHCb measurements for $B\to
K^*\phi$ \cite{Aaij:2014tpa}. They are compatible with zero and do not provide
constraints yet since the theory predicts also rather small values.

Concerning $B\to K^*\phi$, we include in addition also available measurements of
relative amplitude phases $\phi_{\perp,\parallel}$ (purple). The combined
allowed region from all observables does not overlap with regions from only
branching fractions nor only amplitude phases at 68\% probability, giving rise
to large pull values of the branching fraction $\Br(B^0 \to K^{*0} \phi)$:
$1.7\sigma$ from BaBar \cite{Aubert:2008zza} and $2.6\sigma$ from Belle
\cite{Prim:2013nmy}; for $C_L(B^-\to K^{*-}\phi)$ of $-1.5\sigma$ from HFAG
\cite{Amhis:2012bh}; for $C_\perp(\bar{B}^0\to \bar{K}^{*0}\phi)$ of $1.2\sigma$
from Belle \cite{Prim:2013nmy}, but not for BaBar ($0.2\sigma$) and LHCb
($-0.6\sigma$); and for $\phi_\perp(\bar{B}^0\to \bar{K}^{*0}\phi)$ of
$1.1\sigma$ from LHCb \cite{Aaij:2014tpa}, but not for BaBar and Belle (both
$0.0\sigma$). However, the $p$ value of 0.95 of the fit is very high as we
include many other measurements that are described consistently in the fit.

Due to a hierarchy of the helicity amplitudes in QCDF $A_L : A^- : A^+ = 1 :
1/m_b : 1/m_b^2$ \cite{Beneke:2006hg} for the SM operator basis \refeq{curcur}
the following relation should hold
\begin{align}
  \phi_\perp & = \phi_\parallel \,.
\end{align}
The experimental situation supports this within current errors. Since the
hierarchy of helicity amplitudes does not hold in the presence of
chirality-flipped operators beyond the SM, the measurement provides strong
constraints on such scenarios. Further, QCDF predicts only small differences for
neutral and charged decay modes such that one expects similar predictions for
observables in both modes, even in the presence of NP contributions.

%
%
\subsection{WA dominated $B \to K^+ K^-$ and $B_s \to \pi^+ \pi^-$}
\label{sec:SM:WA-dominated}

\begin{figure*}
  \begin{subfigure}[t]{0.32\textwidth}
    \centering
    \includegraphics[width=\textwidth]{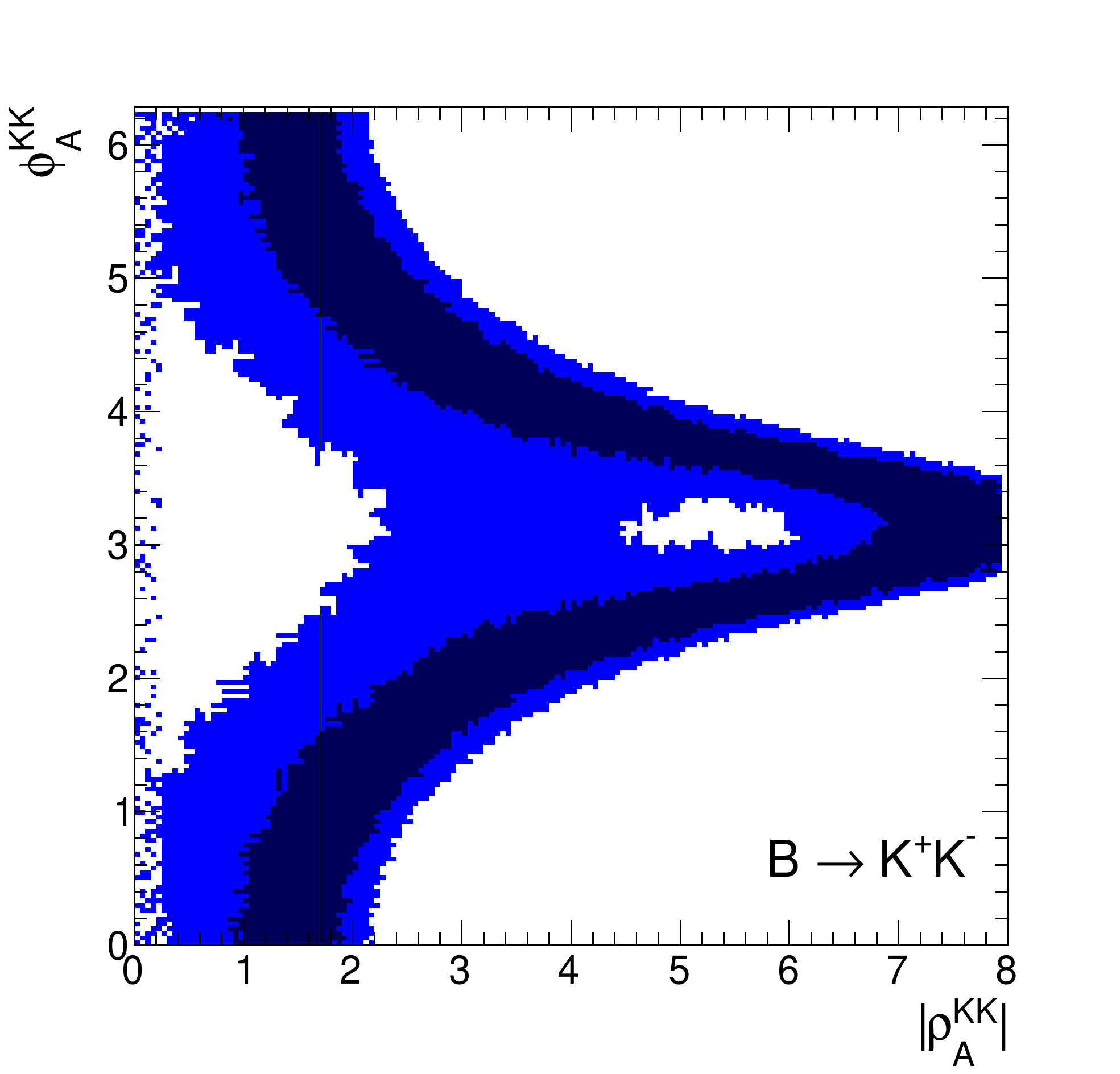}
    \caption{}
    \label{fig:BAT_SM_WA_KK_a}
  \end{subfigure}
  \begin{subfigure}[t]{0.32\textwidth}
    \centering
    \includegraphics[width=\textwidth]{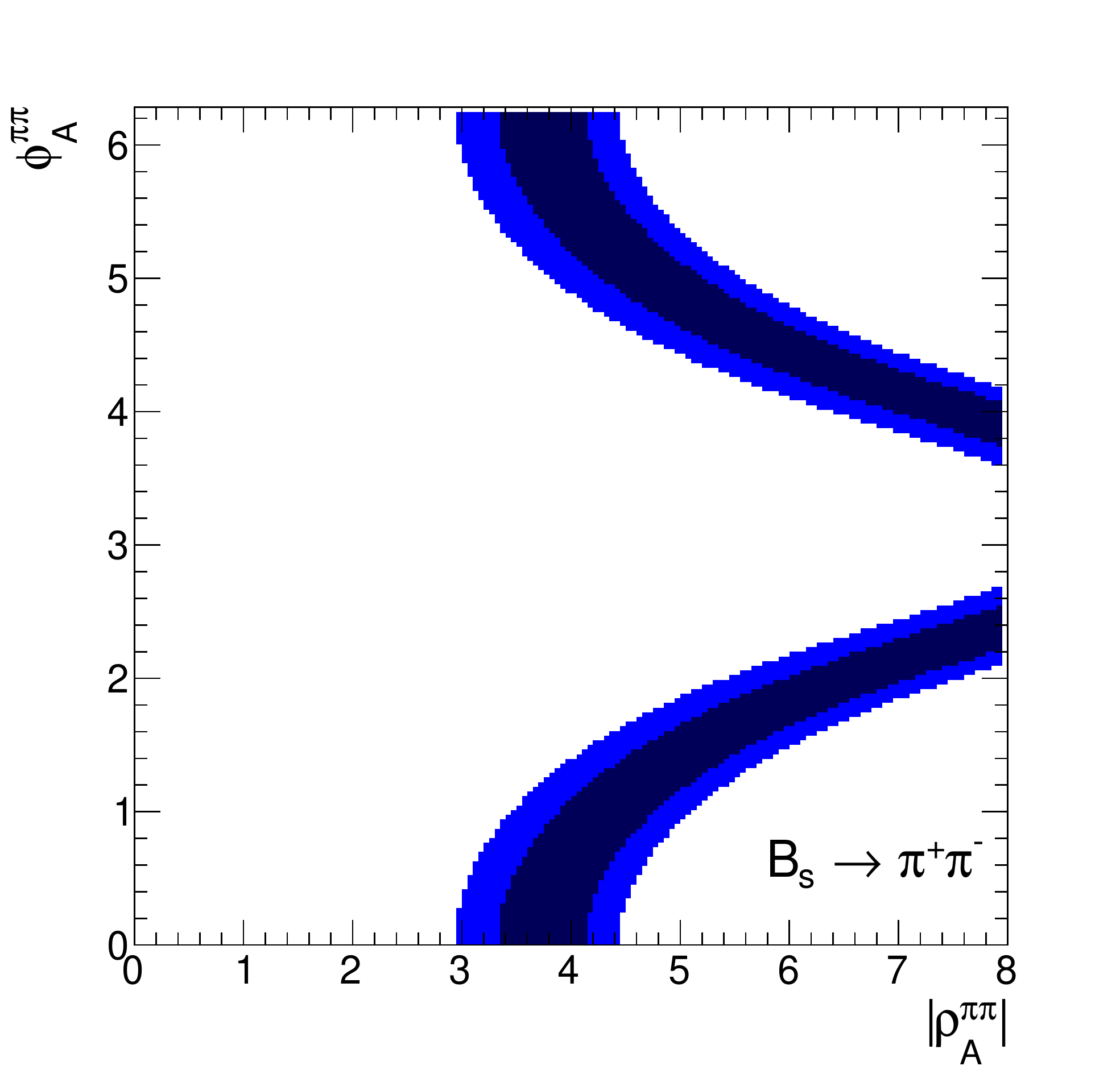}
    \caption{}
    \label{fig:BAT_SM_WA_pipi_b}
  \end{subfigure}
  \caption{ The 68\% (dark) and 95\% (bright) CRs for $\rho_A^{M_1 M_2}$
    obtained from the branching fraction of WA dominated decays (a) $B \to
    K^+K^-$ and (b) $B_s \to \pi^+\pi^-$.  }
\label{fig:BAT_SM_WA-dominated}
\end{figure*}

So far we discussed decays that are dominated by QCD-penguin topologies. They
share the feature that leading WA contributions $\beta_3^c(M_1 M_2)$ are
dominated by the building block $A_3^f$ (see \refeq{WAamp:WC:Aif}), which
originates from gluon emission off the quark current in the final
state. Furthermore, we grouped the decays that are related by $(u\leftrightarrow
d)$-quark exchange, and assumed for each group one universal WA parameter
$\rho_A$.

Now we are interested in decay modes that are governed solely by WA topologies.
The only measured systems are so far $B\to K^+ K^-$ and $B_s\to \pi^+ \pi^-$.
Their amplitudes are given by
\begin{equation}
  \label{eq:WA-decays:ampl}
\begin{aligned}
  {\cal A}(B \to K^+ K^-) & \simeq
  f_{B_d} f_K^2 \sum_p \lambda_p^{(d)} B^p_{K^+ K^-} \,,
\\
  {\cal A}(B_s \to \pi^+ \pi^-) & \simeq
  f_{B_s} \, f_\pi^2 \, \sum_p \lambda_p^{(s)} B^p_{\pi^+\pi^-} \,,
\end{aligned}
\end{equation}
with
\begin{align}
  \label{eq:WA-decays:ampl:II}
  B^p_{M_1 M_2} &
  = \left( \delta_{pu} b_1 + 2 b_4^p + \frac{1}{2} b_{4, \rm EW}^p \right) \,.
\end{align}
Since they are independent of quantities like form factors and the inverse
moment of the $B$-meson DA, which cause usually large uncertainties, the
precision of the determination of $\rho_A^{M_1 M_2}$ from the fit is mainly
dictated by the experimental precision. The involved coefficients $b_i(M_1 M_2)$
depend exclusively on the building blocks $A_{1,2}^i(M_1 M_2)$ (see
\refeq{WAamp:WC:Aif}) where the gluon is emitted off the quark current of the
initial state, and are thus in principle different from $A_3^f$ that dominates
the penguin-dominated decays.  Moreover, $A_1^i \approx A_2^i$ for $MM = PP$
final states when restricting to the asymptotic forms of the light-meson DAs
\cite{Beneke:2003zv}.

The contours of $\rho_A^{K^+K^-}$ and $\rho_A^{\pi^+\pi^-}$ from the branching
fraction measurement are shown in \reffig{BAT_SM_WA_KK_a} and
\reffig{BAT_SM_WA_pipi_b}, respectively. Contrary to the penguin-dominated
decays, the shape of the contour from the branching fractions is different,
reaching large values $|\rho_A^{M_1 M_2}|$ for phases $\phi_A^{M_1 M_2} \sim
\pi$, whereas for $\phi_A^{M_1 M_2} \sim 0$ the absolute value can be
restricted: $|\rho_A^{K^+K^-}| \in [0.9, 1.9]$ and $|\rho_A^{\pi^+\pi^-}|\in
[3.4, 4.1]$ at 68\% probability.  While it is possible to have
$|\rho_A^{K^+K^-}| \lesssim 2$ for small phases, as is the case for the
previously considered penguin-dominated decays, the data requires
$|\rho_A^{\pi^+\pi^-}| \gtrsim 3$ for any value of $\phi_A^{\pi^+ \pi^-}$, and
indeed, the contours of $\rho_A^{K^+K^-}$ and $\rho_A^{\pi^+\pi^-}$ do not
overlap within the 95\% CR. Our results are in agreement with similar fits
\cite{Wang:2013fya}.

Apart from the mismatch of WA contributions for different initial and final
states, there might be another interesting aspect, which can be studied in these
decays. Namely, the amplitudes \refeq{WA-decays:ampl} are proportional to one
overall CP conserving strong phase due to the fact that the single amplitudes
$b_i^p(M_1 M_2)$ (see \refeq{WAamp:WC:Aif}) depend on the same CP-conserving
strong phase of $A_1^i$ due to the aforementioned relation $A_1^i \approx
A_2^i$. Hence \refeq{WA-decays:ampl:II} becomes
\begin{align}
  B^p_{PP} &
  \approx \frac{C_F}{N_c^2} 
  A_1^i \left( \delta_{pu} C_1 + 2 (C_4 + C_6)
    + \frac{C_{10} + C_8}{2} \right) \,.
\end{align}
This is contrary to the requirement of at least one relative strong phase
between the CP conserving and CP violating part of the amplitude in order to
have a non-vanishing CP asymmetry. Since QCDF predicts vanishing CP asymmetries,
up to even further suppressed power corrections, no complementary information
can be gained on the phases $\phi_A$ apart from the one of the branching
fractions. An observation of direct CP violation would put into question the
regularization of endpoint divergences \refeq{def:X_A} introduced in QCDF.

\begin{figure*}
  \begin{subfigure}[t]{0.32\textwidth}
    \centering
    \includegraphics[width=\textwidth]{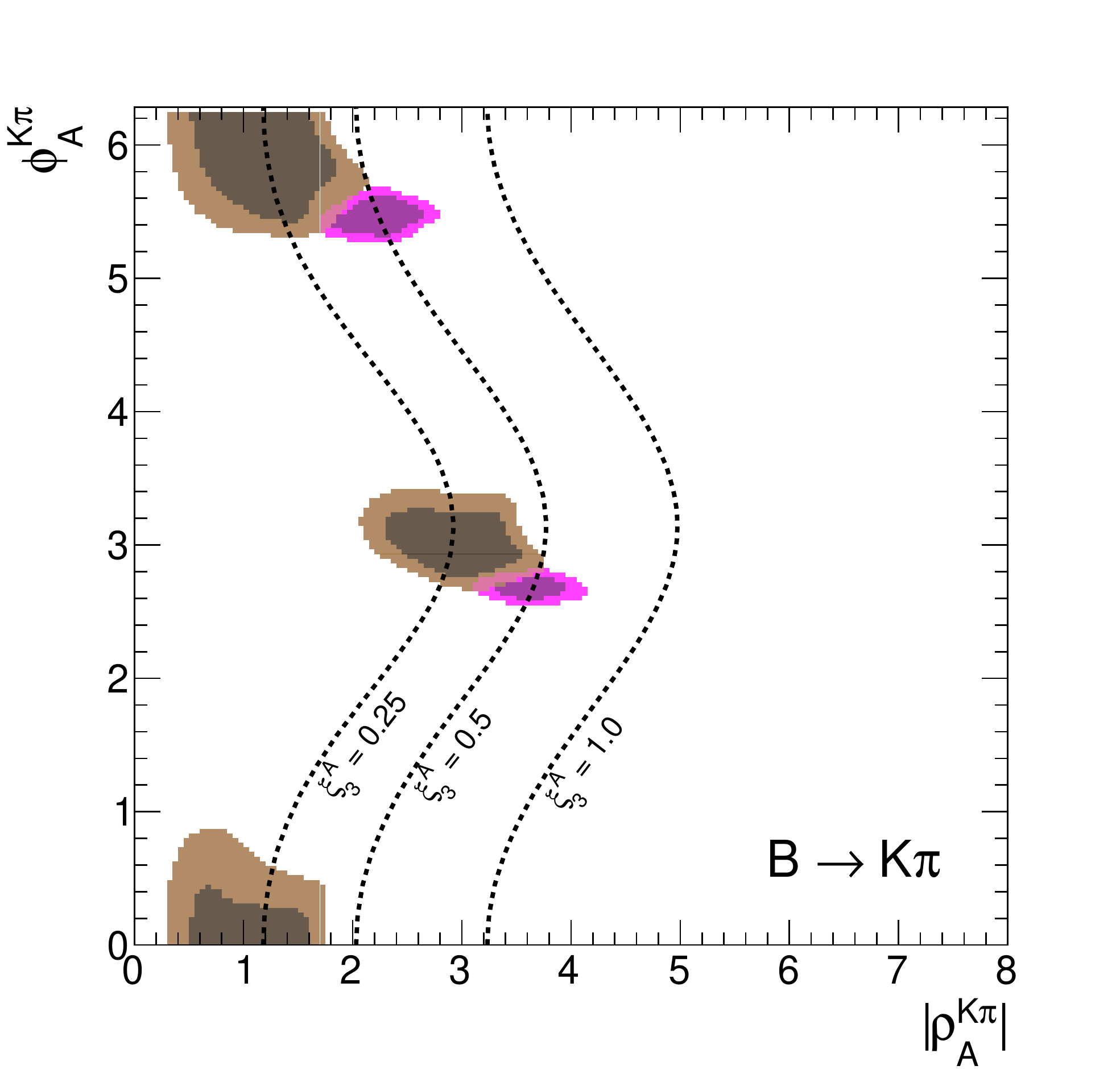}
    \caption{}
    \label{fig:BAT_SM_WA_BKpi_separately}
  \end{subfigure}
  \begin{subfigure}[t]{0.32\textwidth}
    \centering
    \includegraphics[width=\textwidth]{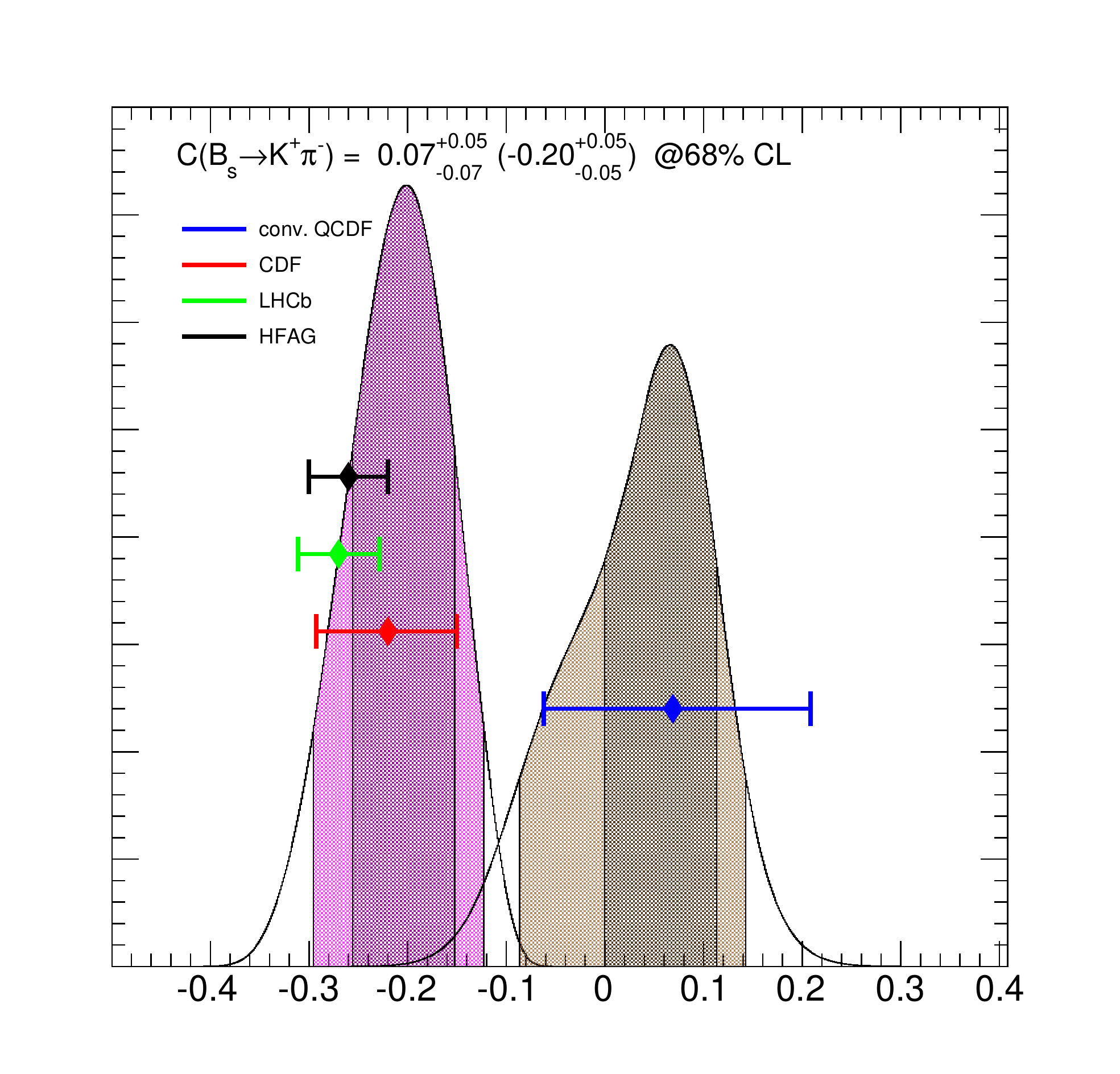}
    \caption{}
    \label{fig:BAT_SM_Cpred_BsKpi}
  \end{subfigure}
\caption{
  The 68\% (dark) and 95\% (bright) CRs for $\rho_A^{K\pi}$ (left), obtained from a
  fit with the reduced observable set for $B^- \to K^-\pi^0$ (brown) and
  $B_d \to K^+\pi^-$ (purple) (see text) assuming the SM. The dashed lines correspond
  to constant $\xi_3^A(K\pi) = (0.25,\, 0.5,\, 1.0)$. The right panel shows the predictions
  for the direct CP asymmetry $C(B_s \to K^+ \pi^-)$ for the two fit regions of
  $\rho_A^{K\pi}$ in the left panel using the same color-coding. Experimental results
  are shown with $1\sigma$ errors and the prediction from QCDF with conventional
  uncertainty estimates is labeled ``QCDF''.
}
\label{fig:BAT_SM_same-finalstates}
\end{figure*}

The measurement of other WA dominated decay modes with $PV$ and $VV$ final
states can help to further scrutinize WA contributions. For example for $M_1 M_2
= PV$, one has $A_1^i \approx - A_2^i$ \cite{Beneke:2003zv} yielding
\begin{align}
  B^p_{PV} &
  \approx \frac{C_F}{N_c^2} A_1^i \left( \delta_{pu} C_1 + 2 (C_4 - C_6)
    + \frac{C_{10} - C_8}{2} \right) \,,
\end{align}
whereas for $MM = VV$ final states $A_1^{i, h} \approx A_2^{i, h}$ ($h = L, +,
-$) \cite{Beneke:2006hg}. In the SM, the Wilson coefficients interfere
destructively for $MM = PV$ and constructively in $MM = (PP,\, VV)$ decay modes.
These decays are in principal sensitive to physics beyond the SM in $O_1$ and
the color-octet operators $O_{4,6,8,10}$.

%
%

\subsection{Universal WA for $B_d$ and $B_s$ decays to same final states}
\label{sec:WA:same:final}

So far, we have assumed one universal parameter for WA contributions of
QCD-penguin dominated decays that are related by $(u \leftrightarrow d$)-quark
exchange, i.e., those groups of decays gathered in \reftab{exp:input:obs:PP},
\reftab{exp:input:obs:PV} and \reftab{exp:input:obs:VV}. For the purpose of this
section, we will study effects which arise from the additional assumption of a
universal WA parameter $\rho_A$ for decays into same final states mediated by
the same quark currents at the weak interaction vertex. This implies in general
relations between $|\Delta S| = 1$ and $|\Delta D| = 1$ decays.

In QCDF this assumption might be justified bearing in mind that WA contributions
in QCD-penguin dominated decay amplitudes are numerically dominated by
topologies in which the gluon is emitted from the quark current that hadronizes
into the final states, namely $A_3^f$ in \refeq{WAamp:WC:Aif}. In this case the
momentum transfer from the initial $B$ meson is solely present at the weak
interaction vertex, rendering the final-state hadronization independent of the
flavor of the initial-state spectator quark. Therefore, one can expect that the
difference between WA amplitudes in $B_d \to M_a M_b$ and $B_s \to M_a M_b$
decays might be of the order $\sim (m_{B_s} - m_{B_d})/m_{B_s} \approx
m_s/m_b$. Similar arguments had been presented for the decays $B_d \to K^+
\pi^-$ and $B_s \to K^+ \pi^-$ in \cite{Lipkin:2005pb}.

Currently experimental information is limited for $B_s$ decays to final states
$M_a M_b = K\pi,\, KK,\, K^*\phi,\, K^*K^*$, whereas for $M_a M_b = \phi\phi$
the corresponding measurements for the $B_d$ is lacking. We do not consider $B_s
\to \pi^+ \pi^-$, which is WA-dominated and was discussed in
\refsec{SM:WA-dominated}, and further the corresponding $B_d \to \pi^+ \pi^-$
decay is tree-dominated. For $B_{d,s} \to KK$, the 68\% CRs overlap nicely as
can be seen from the comparison of \reffig{BAT_SM_KKPengDom} and
\reffig{BAT_SM_BsKK}. In the case of $B_{d,s} \to K^* K^*$, branching-fraction
measurements are compatible, but regions from polarization-fraction measurements
that are favored for $B_d$ decays are excluded for $B_s$ decays as shown in
\reffig{BAT_SM_KstarKstar} and \reffig{BAT_SM_Bs_KstarKstar}. In consequence,
68\% CRs in $B_{d,s} \to K^* K^*$ overlap only marginally.

This leaves us mainly with the final state system $K\pi$ to explore in more
detail the consequences of the assumption of universal WA in decays with same
final states, since for $K^*\phi$ the experimental information for the $B_s$
decay is not yet accurate enough to derive conclusive insights on this
assumption. Especially we would like to test whether the CP asymmetry $C(B_s \to
K^+\pi^-)$, which had been measured recently by CDF \cite{Aaltonen:2014vra} and
LHCb \cite{Aaij:2013iua}, can be predicted correctly from WA contributions
determined in $B \to K\pi$ decays.

As discussed before in \refsec{SM:BKpi}, the fit for $B\to K\pi$ does not allow
for a simultaneous explanation of the two CP asymmetries $C(B^- \to K^-\pi^0)$
and $C(\bar{B}^0 \to K^-\pi^+)$. For this purpose we determine $\rho_A^{K\pi}$
separately from the combination of the branching fraction and CP asymmetry for
each of the two contradicting decays. In addition we used $R_c^K(K\pi)$ to
suppress solutions from the large WA scenario. The best-fit regions of
$\rho_A^{K\pi}$ are shown in \reffig{BAT_SM_WA_BKpi_separately} where the
contour from $\bar{B}^0 \to K^-\pi^+$ coincides nicely with the one in
\reffig{BAT_SM_Kpi_b}, where all constraints had been combined, due to the
higher statistical weight of $C(\bar{B}^0 \to K^-\pi^+)$.  We note that
$|\rho_A^{K\pi}| > 1$ does originate from the precise measurement of
$C(\bar{B}^0 \to K^-\pi^+)$, contrary to $C(B^- \to K^-\pi^0)$ that allows also
smaller values of $|\rho_A^{K\pi}|$ as can be seen in
\reffig{BAT_SM_WA_BKpi_separately}.

Based on our assumption, we predict from both fits the CP asymmetry $C(B_s \to
K^+\pi^-)$, see \refapp{predictions} for details. As shown in
\reffig{BAT_SM_Cpred_BsKpi}, the measurements agree with the prediction from the
$(K^-\pi^+)$-fit whereas it fails at more than $4\sigma$ for the
$(K^-\pi^0)$-fit.  In this case data supports the assumption that WA might be
universal for decays with the same final states. It will be interesting to test
these assumption further against improved measurements in the future.  On the
other hand this result shows that giving up the universality of the WA parameter
for final states related by $(u\leftrightarrow d)$ exchange, but still insisting
on a universal parameter for same final states would also resolve the ``$\Delta
\ACP$ puzzle''.

%
%
%

\section{New Physics scenarios 
\label{sec:NP-fit-results}}

In the framework of the SM, our analysis in the previous \refsec{SM-fit-results}
has shown that the data of all investigated systems can be described with one
universal WA parameter per system of decays that are related by
($u\leftrightarrow d$) quark exchange, apart from stronger tensions in $B\to
K\pi$ and in $B\to K^*\phi$. This section is devoted to the attempt to constrain
new-physics parameters in fits of the data simultaneously with the determination
of one universal WA parameter per system using data from $B\to K\pi,\, K\rho,\,
K^*\pi,\, K^*\rho$, and $K^*\phi$, i.e., in total five WA parameters
$\rho_A^{M_a M_b}$. In the presence of additional degrees of freedom of the NP
parameters, one can expect that tensions present in the SM fit will be relaxed
and the size of power corrections ($\xi_3^A$) can be decreased further.

We choose a model-independent approach, assuming NP contributions to Wilson
coefficients of operators present in the SM operator basis \refeq{heff-b-one}
and their chirality-flipped counterparts obtained by
$(1-\gamma_5)\leftrightarrow (1 + \gamma_5)$ interchange. The $B\to M_1 M_2$
matrix elements of the chirality-flipped operators can be obtained from the
non-flipped ones via parity transformations \cite{Kagan:2004ia}
\begin{align}
\label{eq:HME:chi-flip}
  \langle M_1 M_2 | O_i'| B \rangle & = - \eta_{M_1 M_2} \langle M_1 M_2 | O_i'|
  B \rangle
\end{align}
with $\eta_{M_1 M_2} = + 1$ for $M_1 M_2 = PP,\, VV$ final states and $\eta_{M_1
  M_2} = -1$ for $M_1 M_2 = PV,\, VP$ final states. In this case $b_i'(M_1 M_2)
= b_i(M_1 M_2)[C_i \to C_i']$, see \refeq{WAamp:WC:Aif}, and analogous relations
hold for $a_i'(M_1 M_2)$. In the case of positive/negative polarized final
states, form factors and decay amplitudes have to be replaced by their
helicity-flipped counterpart e.g., $F_\pm \leftrightarrow F_\mp$ and $A_\pm(M_1
M_2) \leftrightarrow A_\mp(M_1 M_2)$.

In \refsec{NP:QEDpenguins} we explore new physics contributions to the Wilson
coefficients of color-singlet QED-penguin Wilson coefficients $C_{7,9}$ and
their chirality-flipped counterparts $C_{7,9}'$.  They are well-known solutions
of the ``$\Delta \ACP$ puzzle'' in $B\to K\pi$ \cite{Buras:2003dj, Buras:2004ub}
and here we further investigate the compatibility of such NP contributions with
data of the four other aforementioned decay systems. As a second model-independent
scenario we consider NP contributions in the Wilson coefficients of the tree-level
$b\to s\,\bar{u}u$ operators in \refsec{NP:bsuu}. In the SM, they are doubly 
Cabibbo-suppressed $\sim \lambda_u^{(s)}/\lambda_c^{(s)}$ in all CP-averaged
observables in $b\to s$ transitions, but give leading contributions to CP
asymmetries. The investigation of further scenarios that involve also
complementary constraints from exclusive $b\to s\,(\gamma,\, \bar\ell\ell)$
decays are given in \cite{Vickers:2014else}.

%
%

\subsection{NP in QED penguins}
\label{sec:NP:QEDpenguins}

The QED-penguin operators $O_{7,\ldots, 10}$, see \refeq{heff-b-one}, and their
chirality-flipped counterparts $O'_{7,\ldots, 10}$ are isospin-violating.
Compared to the SM, NP contributions can relax the encountered tensions in
$\Delta C(K\pi)$ and $R_n^B(K\pi)$ and here we combine $B\to K\pi$ data with
additional measurements from the aforementioned decay systems.  We will focus on
the color-singlet operators $i = 7,7',9,9'$ since the matching contributions to
Wilson coefficients of the color-octet operators $i = 8,8',10,10'$ are
suppressed by the strong coupling $\alpha_s$.  Moreover, in the SM the chirality
structure yields very small $C_7$ and large $C_9$, which must not be the case
for NP scenarios.  Depending on the final state, the two linear combinations
$\overline{C}_i \equiv (C_i + C'_{i})$ and $\Delta C_i \equiv (C_i- C'_{i})$ can
be tested in $MM = PV$ and $MM = PP,\, V_L V_L$, respectively.

We introduce NP contributions to the Wilson coefficients at the matching scale
$\mu_0 = M_W$ that we set to the mass of the $W$-boson and for practical
purposes we rescale them with the SM value $C_9^{\rm SM}(\mu_0) = -1.01
\alpha_e$
\begin{align}
  C_i(\mu_0) & = C_i^{\rm SM}(\mu_0) + |C_9^{\rm SM}(\mu_0)|\, {\cal C}_i
\end{align}
for $i = 7,7',9,9'$. We consider several sub-scenarios
\begin{itemize}
\item single operator dominance \\
  Sc$-i$ : ${\cal C}_i \neq 0$ and ${\cal C}_{j \neq i} = 0$ for $i = 7,7',9,9'$
\item parity (anti-)symmetric scenario \\
  Sc$-77'$ : ${\cal C}_{7,7'} \neq 0$ and ${\cal C}_{9,9'} = 0$ \\
  Sc$-99'$ : ${\cal C}_{9,9'} \neq 0$ and ${\cal C}_{7,7'} = 0$   
\item (axial-)vector coupling scenario \\
  Sc$-79\,\,\,$ : ${\cal C}_{7,9}\,\,\, \neq 0$ and ${\cal C}_{7',9'} = 0$  \\
  Sc$-7'9'$ : ${\cal C}_{7',9'} \neq 0$ and ${\cal C}_{7,9}\,\,\, = 0$
\item generic scenario \\
  Sc$-77'99'$ : ${\cal C}_i \neq 0$
\end{itemize}
with complex-valued ${\cal C}_i$. Although we introduce a NP parameterization at
the matching scale, RG evolution will not lead to mixing of QED penguin
operators into QCD and tree-level operators $i = 1,\ldots,6$ at the order
considered here. Thus NP contributions will not modify the leading amplitude
$\hat\alpha_4^c$, but only $\alpha_{3(4),{\rm EW}}^p$ and the WA amplitudes
$\beta_{3(4),{\rm EW}}^p$. Consequently, branching fractions will become
modified only slightly, whereas CP asymmetries can deviate substantially from
their SM predictions for nonzero CP violating phases.

As long as NP contributions do not become very large compared to
$\hat\alpha_4^c$ one might still employ the expansion in small mode-dependent
ratios
\begin{align}
  r_i & = r_{i,{\rm SM}} + \sum_j r_{i,j}\,\overline{\cal C}_j\,,
\end{align}
see \refeq{BKpi:BR}, in which the NP contributions $r_i$ depend linearly on the
complex NP parameters ${\cal C}_j \equiv |{\cal C}_j|e^{i \delta_j}$.  In
particular \cite{Hofer:2010ee}
\begin{equation}
  \label{eq:CPA:BKpi:exp}
\begin{aligned}
  C(B^- \to \bar{K}^0 \pi^-) & 
  \simeq 
         \sum_{j = 7,9}\! \mbox{Im} \left( 
            \frac{2}{3} r_{{\rm EW},j}^{\rm C} 
          - \frac{4}{3} r_{{\rm EW},j}^{\rm A} \right) \mbox{Im}\, \overline{\cal C}_j\, ,
\\[0.2cm]
  \Delta C - \Delta C^{\rm SM} & 
  \simeq 
          \sum_{j = 7,9}\! \mbox{Im} \left(
          -2r_{{\rm EW},j}^{} 
          -2r_{{\rm EW},j}^{\rm A} \right) \mbox{Im}\, \overline{\cal C}_j \,.
\end{aligned}
\end{equation}
Numerically one has approximately
\begin{equation}
  \label{eq:Numerics:ratios:QED}
  \begin{array}{l|cc}
    \mbox{Im} (r_{i,j}) \times 10^2 & j = 7 & j = 9 
  \\
  \hline
    i = \mbox{EW}          
  & \hphantom{+}2.0^{+0.7}_{-0.8}  & -2.0^{+0.8}_{-0.7} 
  \\
    i = \mbox{EW,C}       
  & -1.7^{+0.6}_{-0.5} & \hphantom{+}0.3^{+1.2}_{-1.2}  
  \\[0.2cm]
    i = \mbox{EW,A}\; (\rho_A^{\mbox{fit}})  
  & \hphantom{+}6.5^{+0.5}_{-0.5} & -0.06 \pm 0.04 
  \\
    i = \mbox{EW,A}\; (\rho_A^{\mbox{scan}})  
  & \hphantom{+}0.8^{+5.4}_{-7.4} & \hphantom{+}0.1^{+0.9}_{-0.9} 
  \end{array}
\end{equation}
in which we used the best-fit point of $\rho_A^{K\pi}$ that was obtained from
the SM fit with Set~II in the case of $i \in (\mbox{EW};\, \mbox{EW,C})$ and no
variation of $\phi_A^{K\pi}$ is included in the determination of theory
uncertainties. The case of $r_{\mathrm{EW}, j}^{\rm A}$ is more involved due to
the explicit dependence on the WA parameter and we provide two points: $i)$ for
$\rho_A^{K\pi}$ obtained from the SM fit as above, denoted as
$\rho_A^{\mbox{fit}}$ in \refeq{Numerics:ratios:QED} and $ii)$ for
$\phi_A^{K\pi} = 0$, denoted as $\rho_A^{\mbox{scan}}$ in
\refeq{Numerics:ratios:QED}, as usually chosen in conventional QCDF as central
value including the variation of $\phi_A^{K\pi}$ into the error
estimation. Several observations can be made:
\begin{enumerate}
\item Given that $\mbox{Im}\, \overline{\cal C}_{7,9} \sim {\cal O}(1)$, the
  numerical coefficients imply that the total amount of CP violation from
  $r_{i,j}$ of $i \in (\mbox{EW};\, \mbox{EW,C})$ does not exceed $3.5\%$,
  whereas $r_{\rm{EW},9}^{\rm C}$ is numerically negligible.
\item An accidental cancellation can be observed in $(r_{\rm{EW}, 7}^{} +
  r_{\rm{EW},7}^{\rm C})$ as well as in $(r_{\rm{EW}, 7} + r_{\rm{EW}, 9})$ if
  $\mbox{Im}\, \overline{\cal C}_{7} \approx \mbox{Im}\, \overline{\cal C}_{9}$.
\item The amount of CP violation from $\mbox{Im}\, \overline{\cal C}_9$ to
  $r_{\rm EW}^{\rm A}$ can be neglected in both cases $i)$ and $ii)$, whereas
  the contribution of $\mbox{Im}\, \overline{\cal C}_7$ can indeed become large.
\item Since the measurement of $C(B^- \to \bar{K}^0 \pi^-) = (1.5 \pm 1.9)\%$ is
  rather accurate, it forbids too large CP-violating contributions from
  $\mbox{Im}\, \overline{\cal C}_7$ if $\rho_A$ is fitted.
\end{enumerate}

\begin{figure}
  \centering
  \includegraphics[width=0.32\textwidth]{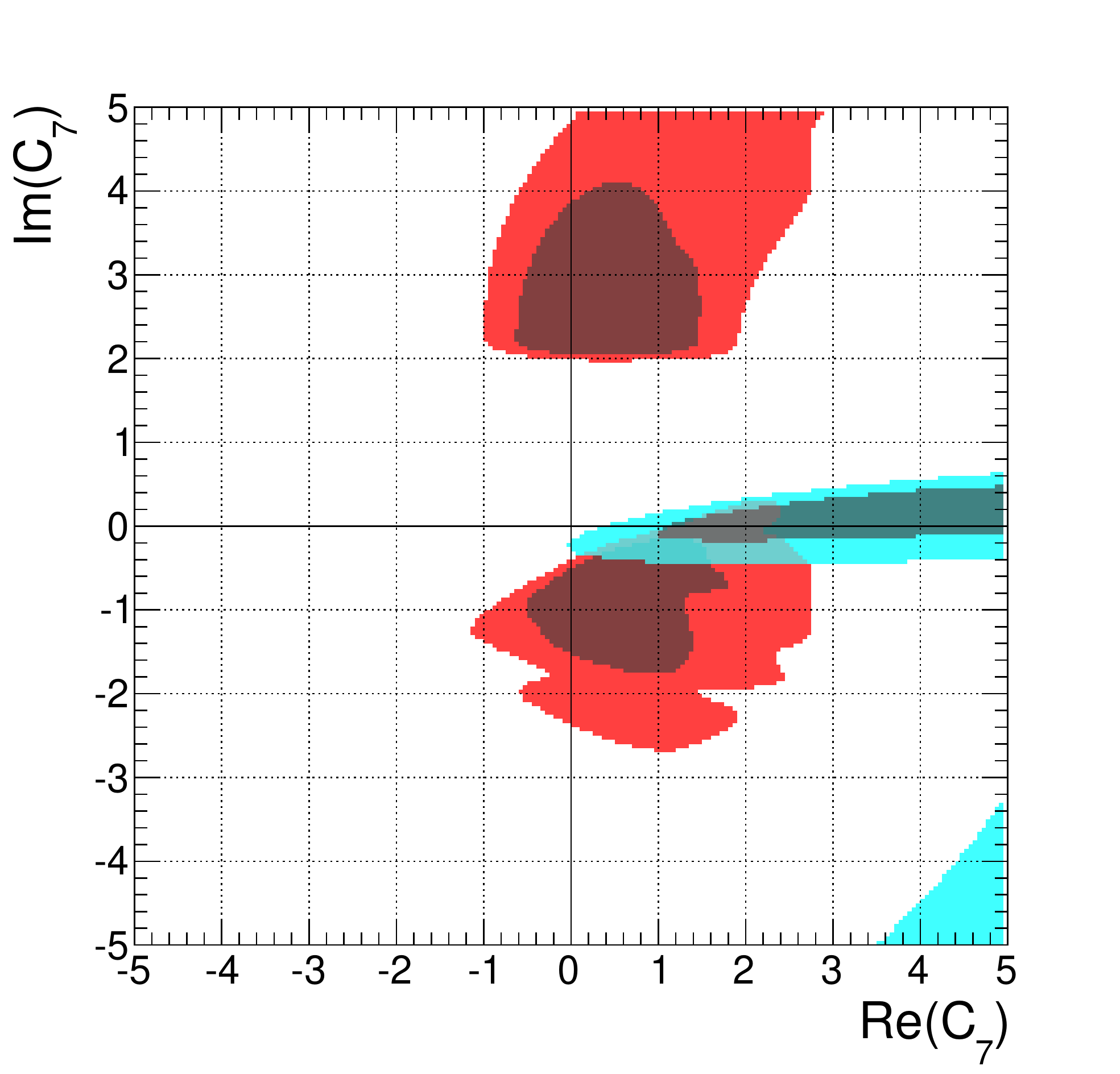}
  \caption{ The 68\% (dark) and 95\% (bright) CRs for ${\cal C}_7$ in scenario
    Sc$-7$, obtained from a fit of observable Set II of the $B \to K\pi$ system
    when treating $\rho_A^{K\pi}$ either as a fit parameter (cyan) or as a
    nuisance parameter (red).  }
\label{fig:fit-versus-scan}
\end{figure}

\begin{figure*}
  \begin{subfigure}[t]{0.32\textwidth}
    \centering
    \includegraphics[width=\textwidth]{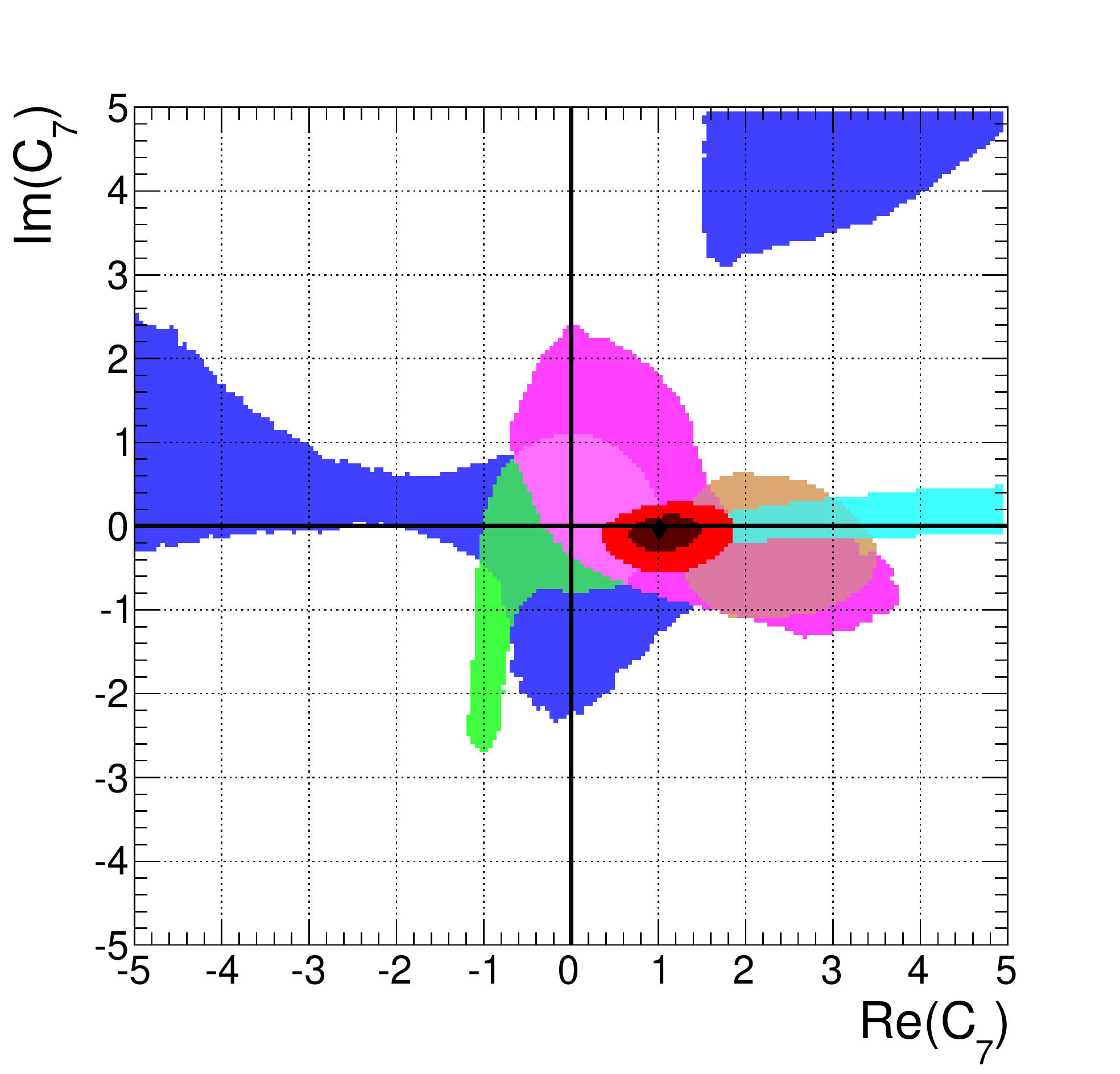}
    \caption{}
    \label{fig:C7}
  \end{subfigure}
  \begin{subfigure}[t]{0.32\textwidth}
    \centering
    \includegraphics[width=\textwidth]{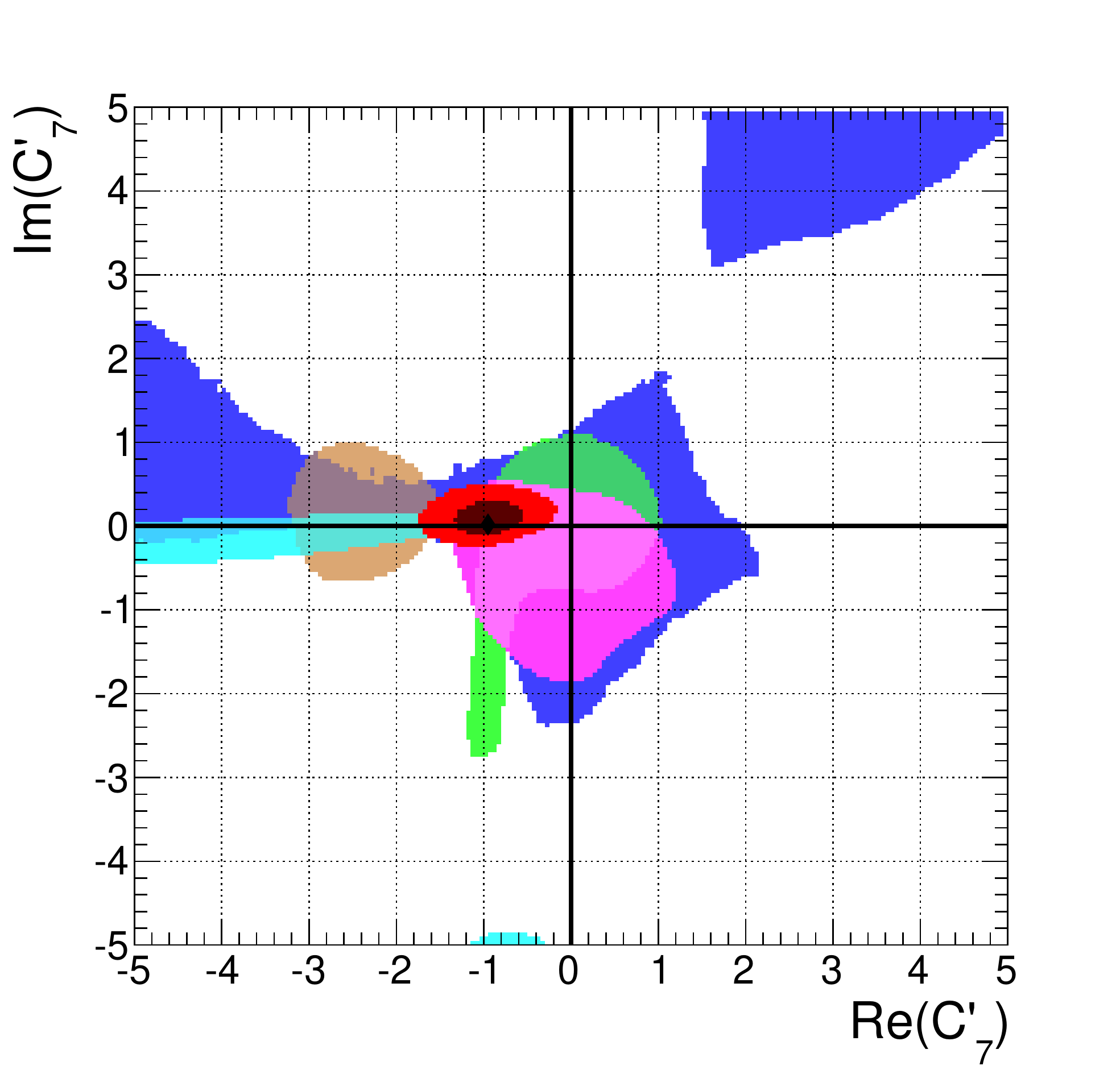}
    \caption{}
    \label{fig:C7p}
  \end{subfigure} \\
  \begin{subfigure}[t]{0.32\textwidth}
    \centering
    \includegraphics[width=\textwidth]{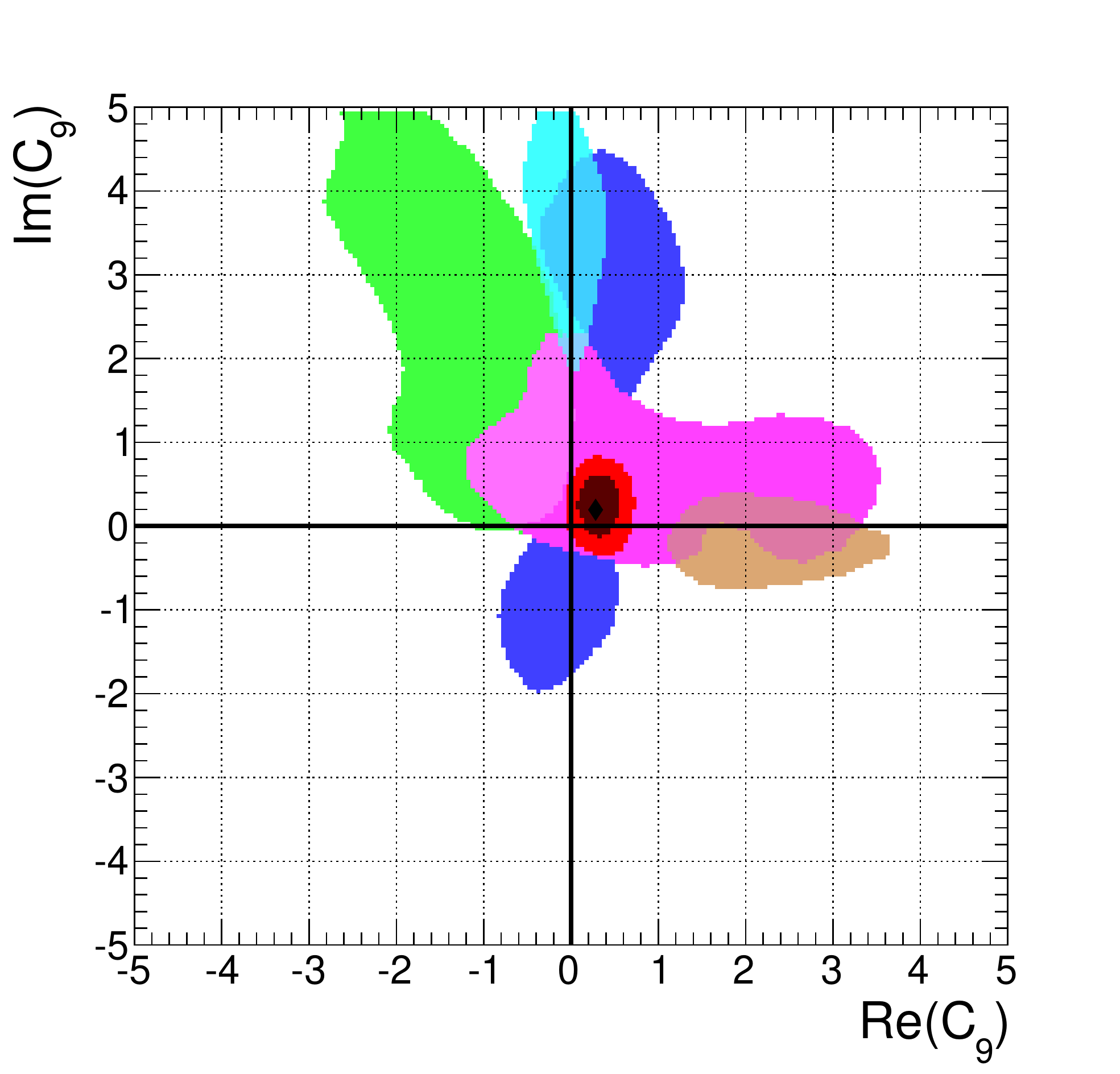}
    \caption{}
    \label{fig:C9}
  \end{subfigure}
  \begin{subfigure}[t]{0.32\textwidth}
    \centering
    \includegraphics[width=\textwidth]{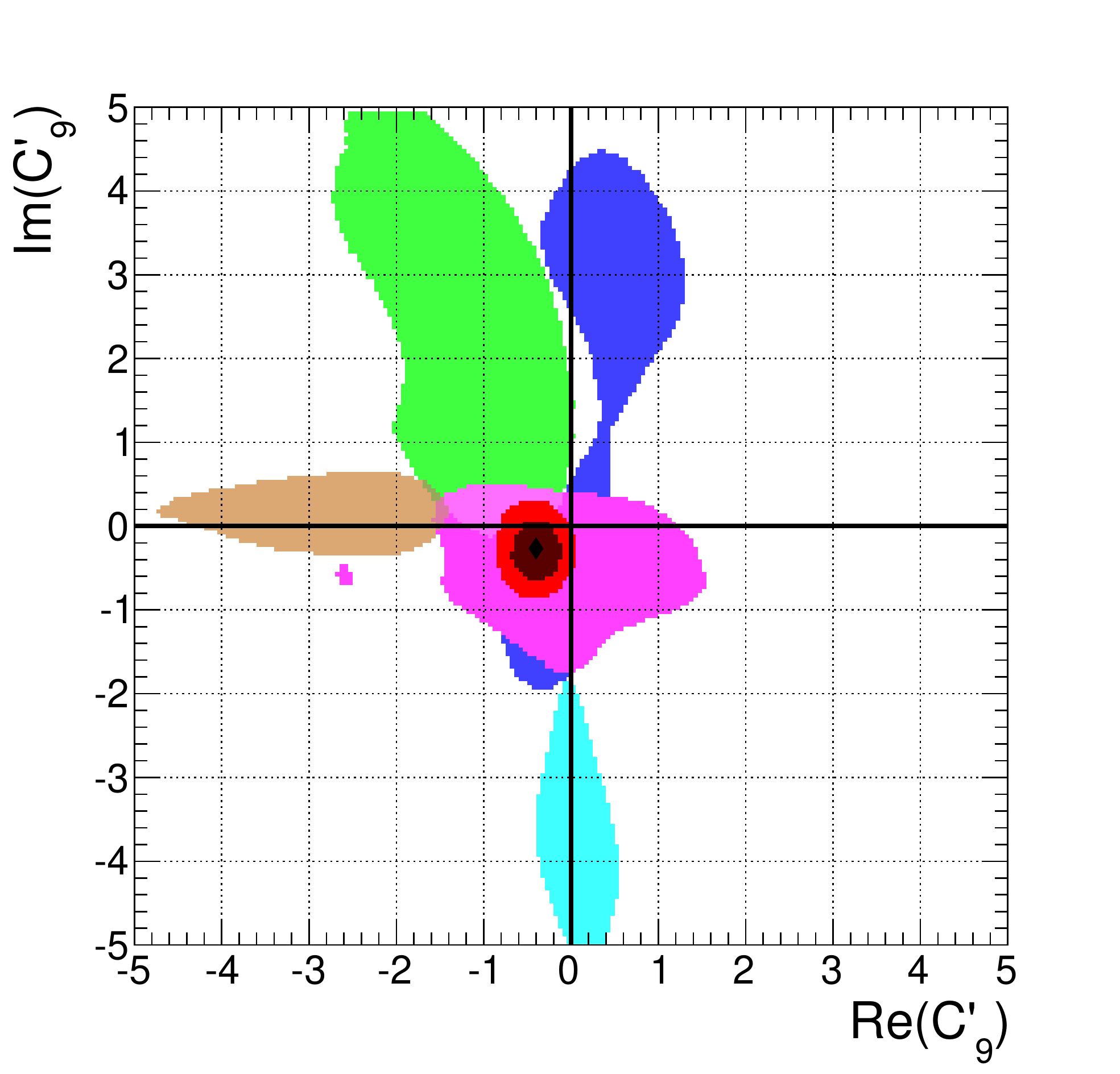}
    \caption{}
    \label{fig:C9p}
  \end{subfigure}
  \caption{ $68\%$ CR for the complex Wilson coefficients
    $\mathcal{C}^{(\prime)}_{7,9}$ in the scenarios Sc$-7,7',9,9'$.  Constraints
    are obtained from the decay systems $B \to K\pi$ (cyan), $B \to K\rho$
    (blue), $B \to K^*\pi$ (green), $B \to K^*\rho$ (purple), and $B \to
    K^*\phi$ (brown). The combined contour (red) is shown for a probability of
    $68\%$ and $95\%$.  The $\blacklozenge$ corresponds to the best-fit point of
    the combined fit.}
   \label{fig:C7-only_C9-only}
\end{figure*}

\begin{table*}
\begin{center}
\renewcommand{\arraystretch}{1.4}
   \begin{tabular}{l||c|ccccccc|c}
\hline\hline
 & {$\;\;\,\mbox{Re}({\cal C}_i^{(\prime)})\!\!\;\; \mbox{Im} ({\cal C}_i^{(\prime)})$}
 & $\Delta C(K\pi)$
 & $R_n^B(K\pi)$
 & $C_L(K^{*-} \phi)$ 
 & $\mathcal{B}(\bar{K}^{*0} \phi)$ 
 & $f_L(K^{*-} \rho^0)$
 & $R_n^B(K^{*}\pi)$
 & $R_n^B(K^{*}\rho)$
 & $\Delta \chi^2(\rm SM) $
\\
\hline
SM
 &   
 & $ \bf{-2.8 \sigma}$
 & $ \bf{-1.9 \sigma}$
 & $ -1.5 \sigma$
 & $ \bf{1.7/2.6 \sigma}$
 & $  0.9  \sigma$
 & $  0.6 \sigma$
 & $  0.6 \sigma$
 & 
\\[0.1em]
\cline{2-10}
Sc$-7$
 & $\hphantom{-}1.01,\, -0.04$
 & $ -0.7 \sigma$
 & $ -0.8 \sigma$
 & $ \bf{-1.6 \sigma}$
 & $  0.3/1.2 \sigma$
 & $  1.0 \sigma$
 & $  0.0 \sigma$
 & $  1.1 \sigma$
 & $ 18.7 $
\\
\cline{2-10}
Sc$-7'$
 & $-0.95,\,\hphantom{-}0.02$
 & $ -0.8 \sigma$
 & $ -0.8 \sigma$
 & $ -\bf{1.6} \sigma$  
 & $  1.2/\bf{2.1} \sigma$
 & $  1.2 \sigma$
 & $  0.0 \sigma$
 & $  0.9 \sigma$
 & $ 12.8 $
\\[0.1em]
\cline{2-10}
Sc$-9$
 & $\hphantom{-}0.28,\, \hphantom{-}0.19 $
 & $ \bf{-2.7 \sigma}$
 & $  \hphantom{-}0.0 \sigma$
 & $ -1.3 \sigma$
 & $  \bf{1.7/2.6 \sigma}$
 & $  1.0\sigma$
 & $  1.1 \sigma$
 & $  0.7 \sigma$
 & $ 1.5 $
\\
\cline{2-10}
Sc$-9'$
 & $-0.40,\, -0.27$
 & $ \bf{-2.6 \sigma}$
 & $  \hphantom{-}0.0 \sigma$
 & $ -1.2 \sigma$  
 & $  \bf{1.6/2.5 \sigma}$
 & $  1.1 \sigma$
 & $  0.0 \sigma$
 & $  0.6 \sigma$
 & $ 3.6 $
\\[0.1em]
\cline{2-10}
\multirow{2}{*}{Sc$-77'$}
 & $\hphantom{-}1.94,\,\hphantom{-}0.12$
 & \multirow{2}{*}{$\hphantom{-}0.0 \sigma$}
 & \multirow{2}{*}{$\hphantom{-}0.0 \sigma$}
 & \multirow{2}{*}{$-1.5 \sigma$}
 & \multirow{2}{*}{$0.0/0.0 \sigma$}
 & \multirow{2}{*}{$ 0.8\sigma$}
 & \multirow{2}{*}{$0.3 \sigma$}
 & \multirow{2}{*}{$\bf{1.6 \sigma}$}
 & \multirow{2}{*}{$ 23.9 $}
\\
 & $-1.65,\,\hphantom{-}0.03$
 & 
 & 
 & 
 & 
 & 
 & 
 & 
 & 
\\[0.1em]
\cline{2-10}
\multirow{2}{*}{Sc$-99'$}
 & $-0.05,\,\hphantom{-}2.15$
 & \multirow{2}{*}{$ -\bf{2.2} \sigma$}
 & \multirow{2}{*}{$  \hphantom{-}0.0 \sigma$}
 & \multirow{2}{*}{$ -0.6 \sigma$}
 & \multirow{2}{*}{$  \bf{1.6/2.5 \sigma}$}
 & \multirow{2}{*}{$  0.9 \sigma$ } 
 & \multirow{2}{*}{$  \bf{1.6 \sigma}$}
 & \multirow{2}{*}{$  0.7 \sigma$}
 & \multirow{2}{*}{$ 9.7 $}
\\
 & $-0.42,\,\hphantom{-}1.64$
 & 
 & 
 & 
 & 
 & 
 &   
 & 
 & 
\\[0.1em]
\cline{2-10}
\multirow{2}{*}{Sc$-79$}
 & $\hphantom{-}1.02,\,-0.02$
 & \multirow{2}{*}{$ -0.9 \sigma$}
 & \multirow{2}{*}{$ -0.6 \sigma$}
 & \multirow{2}{*}{$ \bf{-1.6} \sigma$}
 & \multirow{2}{*}{$  0.3/1.2 \sigma$ } 
 & \multirow{2}{*}{$  0.9 \sigma$}
 & \multirow{2}{*}{$  0.0 \sigma$}
 & \multirow{2}{*}{$  1.2 \sigma$}
 & \multirow{2}{*}{$ 19.0 $}
\\
 & $\hphantom{-}0.06,\, \hphantom{-}0.13$
 & 
 & 
 & 
 & 
 &   
 & 
 & 
 & 
\\[0.1em]
\cline{2-10}
\multirow{2}{*}{Sc$-7'9'$}
 & $-1.75,\,-0.02 $
 & \multirow{2}{*}{$ -0.4 \sigma$}
 & \multirow{2}{*}{$ \hphantom{-} 0.3 \sigma$}
 & \multirow{2}{*}{$ \bf{-1.6} \sigma$}
 & \multirow{2}{*}{$  0.0/0.3 \sigma$} 
 & \multirow{2}{*}{$  \bf{2.3} \sigma$}
 & \multirow{2}{*}{$  1.0 \sigma$}
 & \multirow{2}{*}{$ \bf{1.6} \sigma$}
 & \multirow{2}{*}{$ 18.0 $}
\\
 & $-0.93,\,\hphantom{-}0.28$
 & 
 & 
 & 
 & 
 & 
 & 
 & 
 & 
\\[0.1em]
\cline{2-10}
\multirow{4}{*}{Sc$-77'99'$} 
 & $\hphantom{-}1.61,\, \hphantom{-}0.24$
 & \multirow{4}{*}{$ -0.1 \sigma$}
 & \multirow{4}{*}{$ \hphantom{-} 0.0 \sigma$}
 & \multirow{4}{*}{$ -1.2 \sigma$}
 & \multirow{4}{*}{$  0.0/0.0 \sigma$} 
 & \multirow{4}{*}{$  0.6 \sigma$}
 & \multirow{4}{*}{$  0.7 \sigma$}
 & \multirow{4}{*}{$  1.5 \sigma$}
 & $ \multirow{4}{*}{31.2} $
\\
 & $-0.87,\,\hphantom{-}0.11$
 &  &  &  &    &  &  &  & 
\\
 & $\hphantom{-}0.31,\,\hphantom{-}1.65$
 &  &  &  &    &  &  &  & 
\\
 & $-0.60,\,\hphantom{-}1.59$
 &  &  &  &    &  &  &  & 
\\
\hline\hline
\end{tabular}
\caption{ Compilation of best-fit points and pull values with $|\delta| \ge 1.6$
  for the model-independent fits of scenarios with NP in QED-penguin operators.
  $C(\bar{K}^{*0} \phi)$ and $\mathcal{B}(\bar{K}^{*0} \phi)$ are for
  experimental values \cite{Prim:2013nmy}.  }
\label{tab:Chadr:pulls} 
\end{center}
\end{table*}

We start the discussion of our results with the confrontation of our procedure
of fitting simultaneously NP and WA parameters, with the conventional QCDF
approach, where only NP parameters are fitted and WA parameters are treated as
nuisance parameters. As an example, \reffig{fit-versus-scan} provides the
allowed regions of $\mbox{Re}\,{\cal C}_7$ versus $\mbox{Im}\,{\cal C}_7$ in the
scenario Sc$-7$ from the observable Set II of the $B\to K\pi$ system. We
emphasize again that both fits underly very different assumptions, in fact
treating $\rho_A^{K\pi}$ as a nuisance parameter implies that it can be
different for each decay as well as each observable, whereas fitting it imposes
one universal parameter for all observables in the $B\to K\pi$ system. It can be
seen that both approaches yield rather different results that overlap only for a
very small part of the considered parameter space.  The contour from
conventional QCDF (red) allows $\mbox{Im}\,{\cal C}_7$ to be rather large and
even its sign is not dictated by the data. Contrary to that the corresponding
contour that we obtain from a simultaneous fit of NP and WA parameters (cyan)
becomes strongly constrained and fixes ${\cal C}_7$ to be almost purely real.
The different outcomes due to the two treatments of $\rho_A$ originate from
$r_{\rm{EW},7}^{\rm A}$, which is in both cases the leading NP contribution to
$\Delta C$ and $C(B^- \to \bar{K}^0 \pi^-)$ in \refeq{CPA:BKpi:exp}. However,
for case $ii)$, $r_{\rm{EW},7}^{\rm A}$ is assigned with an approximately
vanishing central value and huge symmetric uncertainties, whereas for $i)$ the
central value of $r_{\rm{EW},7}^{\rm A}$ is large and uncertainties are small.
The former implies that both CP asymmetries in \refeq{CPA:BKpi:exp} can be
explained simultaneously due to large uncertainties, which depend linearly on
$\mbox{Im}\,\overline{\cal C}_7$ and enter the determination of the individual
observables uncorrelated. The latter case however implies that a significant
modification of one of the two CP asymmetries inevitably induces a similar large
contribution to the other. Since $C(B^- \to \bar{K}^0 \pi^-)$ is measured rather
accurately and consistent with its value at the best-fit point of the SM fit,
large contributions to $\mbox{Im}\,\overline{\cal C}_7$ are consequently
forbidden (see $4)$. This shows that the bounds on a NP parameter space strongly
depend on the treatment of $\rho_A$.

Bounds on the complex-valued Wilson coefficients ${\cal C}^{(\prime)}_i$ from
fits in scenarios of single operator dominance are shown in
\reffig{C7-only_C9-only} for each of the decay systems $B \to K\pi,\, K\rho,\,
K^*\pi,\, K^*\rho,\, K^*\phi$ at 68\% and their combination at 68\% and 95\%
probability. Due to the different dependence of the spin of the final states on
chirality-flipped operators, see \refeq{HME:chi-flip} and comments below, the
contours for $B \to PP,\, V_L V_L$ systems are mirrored at the origin, whereas
for $B \to PV$ systems they remain invariant, when considering scenarios that
are related by ${\cal C}_i \leftrightarrow {\cal C}'_i$.

As can be expected from the pull values of the SM fit, shown in
\reftab{SM-pulls}, the allowed regions from $B \to K\rho,\, K^*\rho$ contain the
SM, whereas some small pulls in $B \to K^*\pi$ can be reduced with non-SM values
of ${\cal C}_{9,9'}$. Concerning $B \to K\pi$, the data prefers NP contributions
that are almost purely real for Sc$-7,7'$ and imaginary for Sc$-9,9'$, excluding
the SM with a probability of more than 95\%. As already explained above,
experimental data of $C(B^- \to \bar{K}^0 \pi^-)$ forbids large contributions to
$\mbox{Im}\,\overline{\cal C}_7$, implying also small $\Delta C$ in the
approximation of small $r_{i,j}$ as used in \refeq{CPA:BKpi:exp}.  Nevertheless,
in our approach $r_{\mathrm{EW}, 7}^{\rm A}$ can become rather large, see
\refeq{Numerics:ratios:QED}, such that second order interference terms $\propto
r_{\rm T}\, r_{\mathrm{EW}, 7}^{\rm A}\, \mbox{Re}\, \overline{C}_7$, which do
not exactly cancel in $\Delta C$, can provide better agreement with the data.
The improvement of the tension is quantified in \reftab{Chadr:pulls} at the
best-fit point of the combination of all five decay systems. For example,
scenarios Sc$-7,7'$ allow to reduce the pull of $\Delta C$ of $-2.8 \sigma$ in
the SM below $-1 \sigma$, and similarly for $R_n^B(K\pi)$. In scenarios
Sc$-9,9'$ the solution to the ``$\Delta \ACP$ puzzle'' proceeds via $r_{{\rm
    EW},9}$, see \refeq{Numerics:ratios:QED}, requiring large values of
$\mbox{Im}\, \Delta {\cal C}_9$, which are strongly disfavored by measurements
of direct CP asymmetries in $B \to K^* \phi$. In consequence of this strong
tension, Sc$-9,9'$ cannot really improve existing pulls of the SM, except for
$R_n^B(K\pi)$, which results in a very small improvement of $\Delta\chi^2({\rm
  SM})$, shown in \reftab{Chadr:pulls}. Contrary, Sc$-7,7'$ exhibits a large
improvement of $\Delta\chi^2({\rm SM})$ since here the allowed region of the
Wilson coefficient from $B \to K^* \phi$ is compatible with the one from $B \to
K\pi$.

\begin{figure*}
  \begin{subfigure}[t]{0.32\textwidth}
    \centering
    \includegraphics[width=\textwidth]{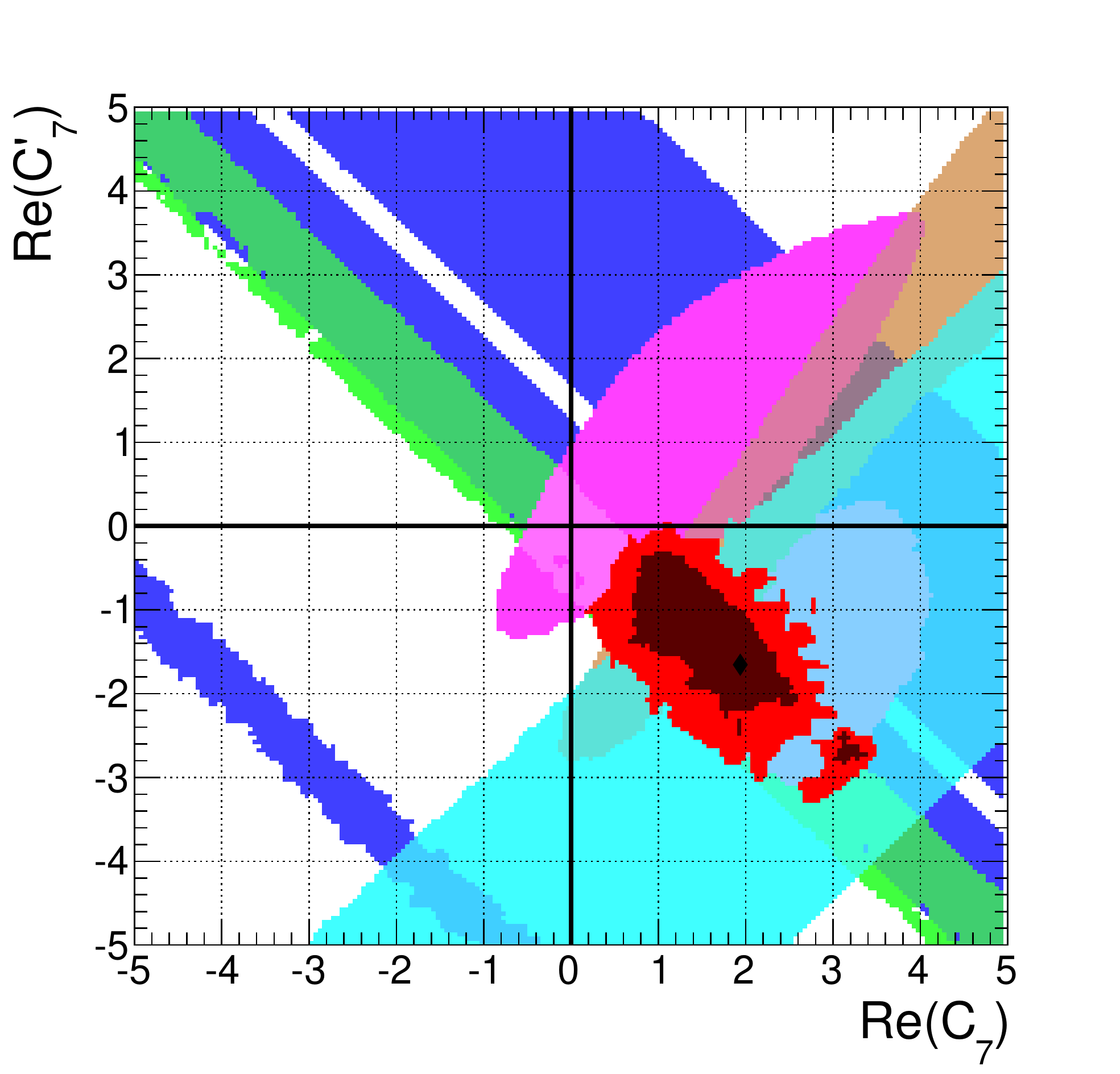}
  \end{subfigure}
  \begin{subfigure}[t]{0.32\textwidth}
    \centering
    \includegraphics[width=\textwidth]{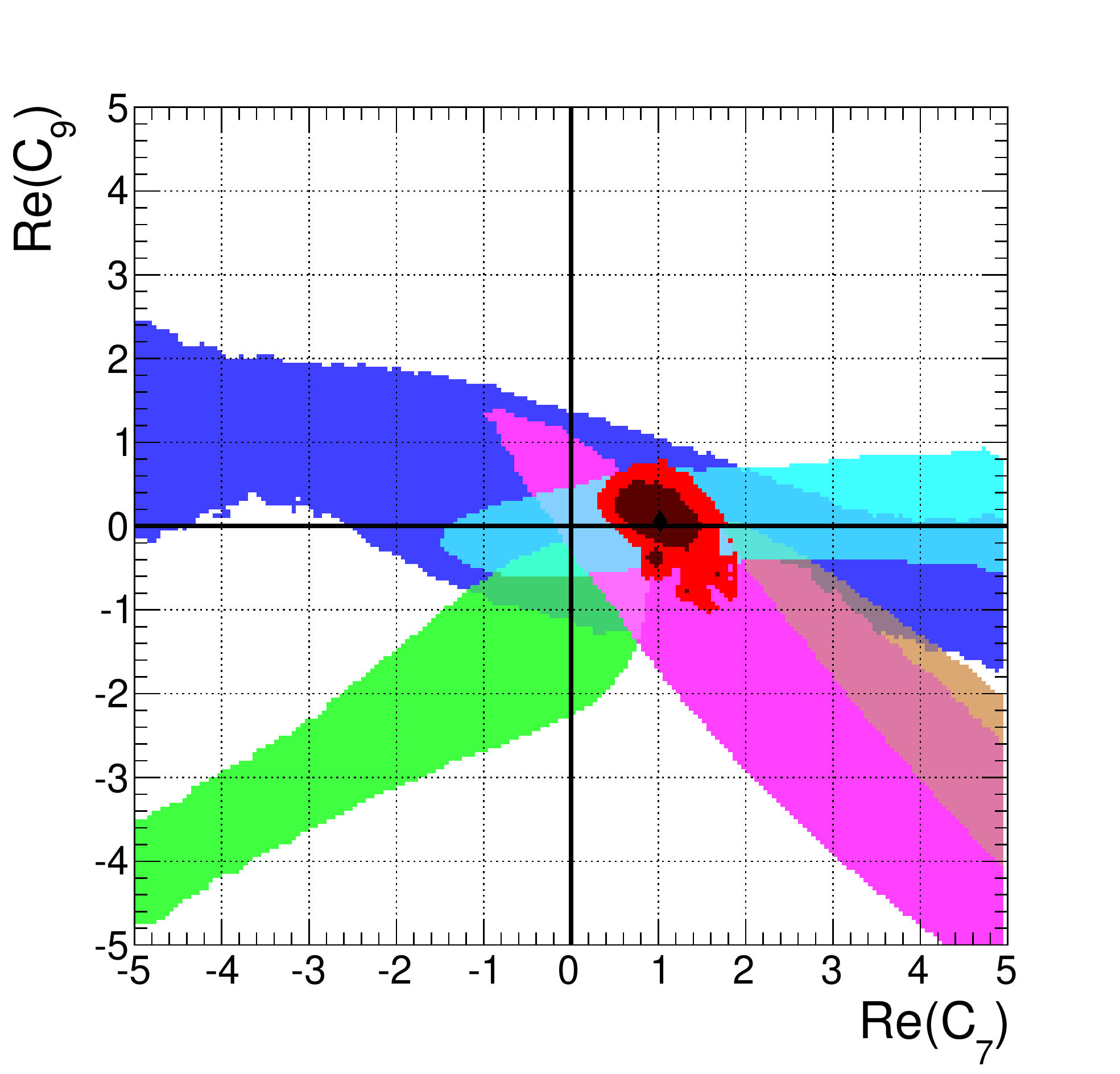}
  \end{subfigure}
  \begin{subfigure}[t]{0.32\textwidth}
    \centering
    \includegraphics[width=\textwidth]{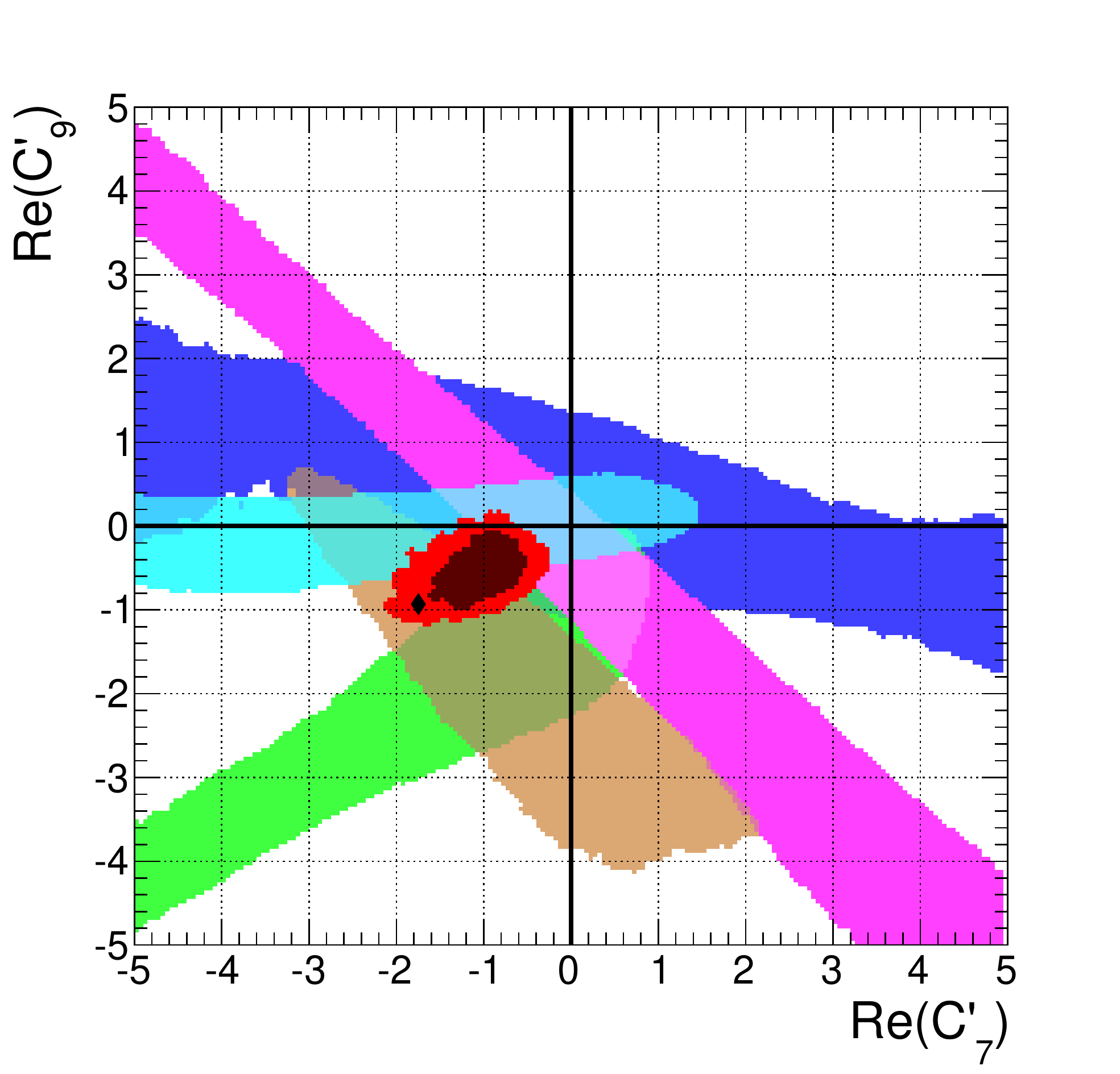}
  \end{subfigure}\\
  \begin{subfigure}[t]{0.32\textwidth}
    \centering
    \includegraphics[width=\textwidth]{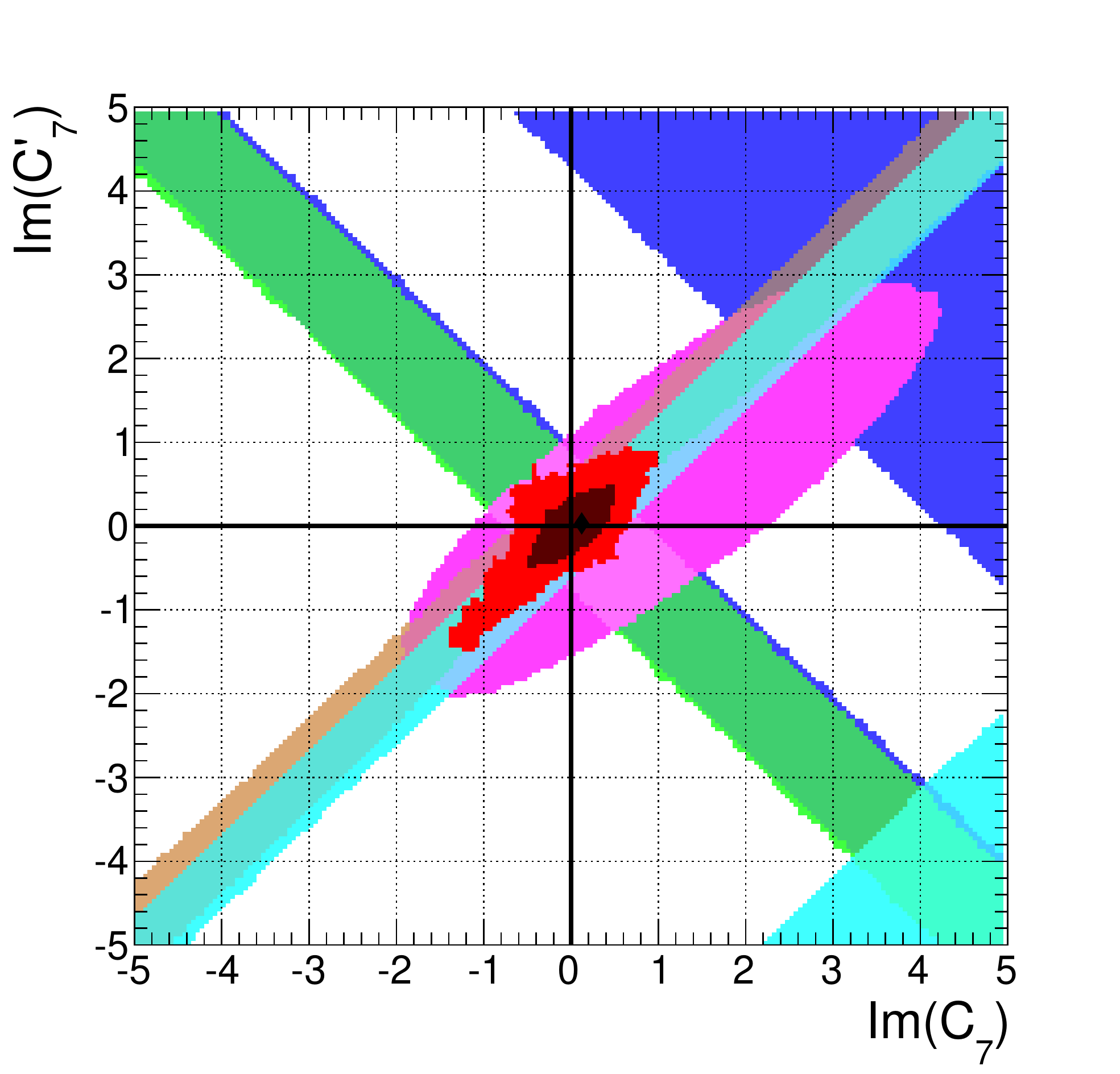}
    \caption{}
    \label{fig:C77p}
  \end{subfigure} 
  \begin{subfigure}[t]{0.32\textwidth}
    \centering
    \includegraphics[width=\textwidth]{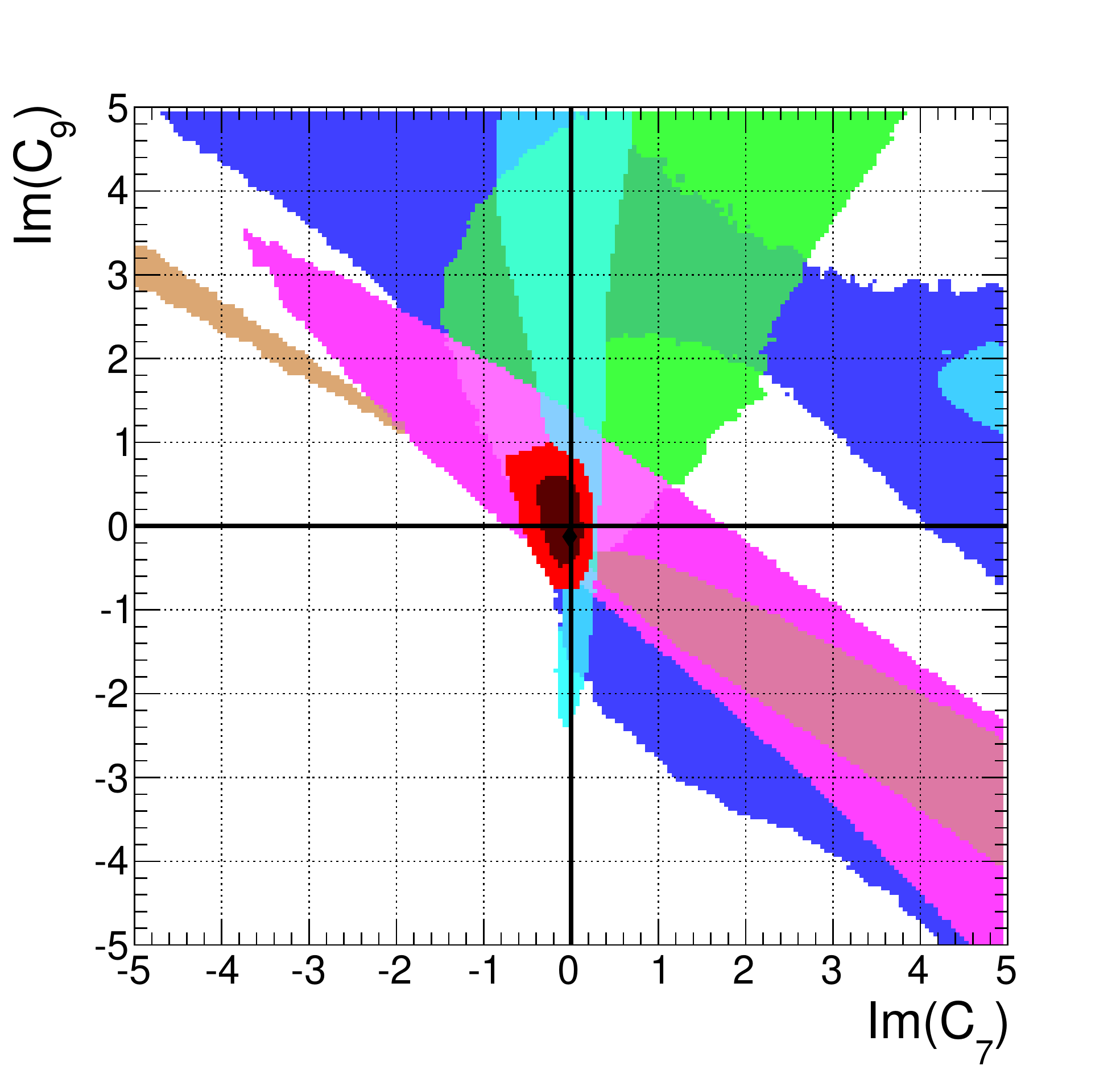}
    \caption{}
    \label{fig:79}
  \end{subfigure} 
  \begin{subfigure}[t]{0.32\textwidth}
    \centering
    \includegraphics[width=\textwidth]{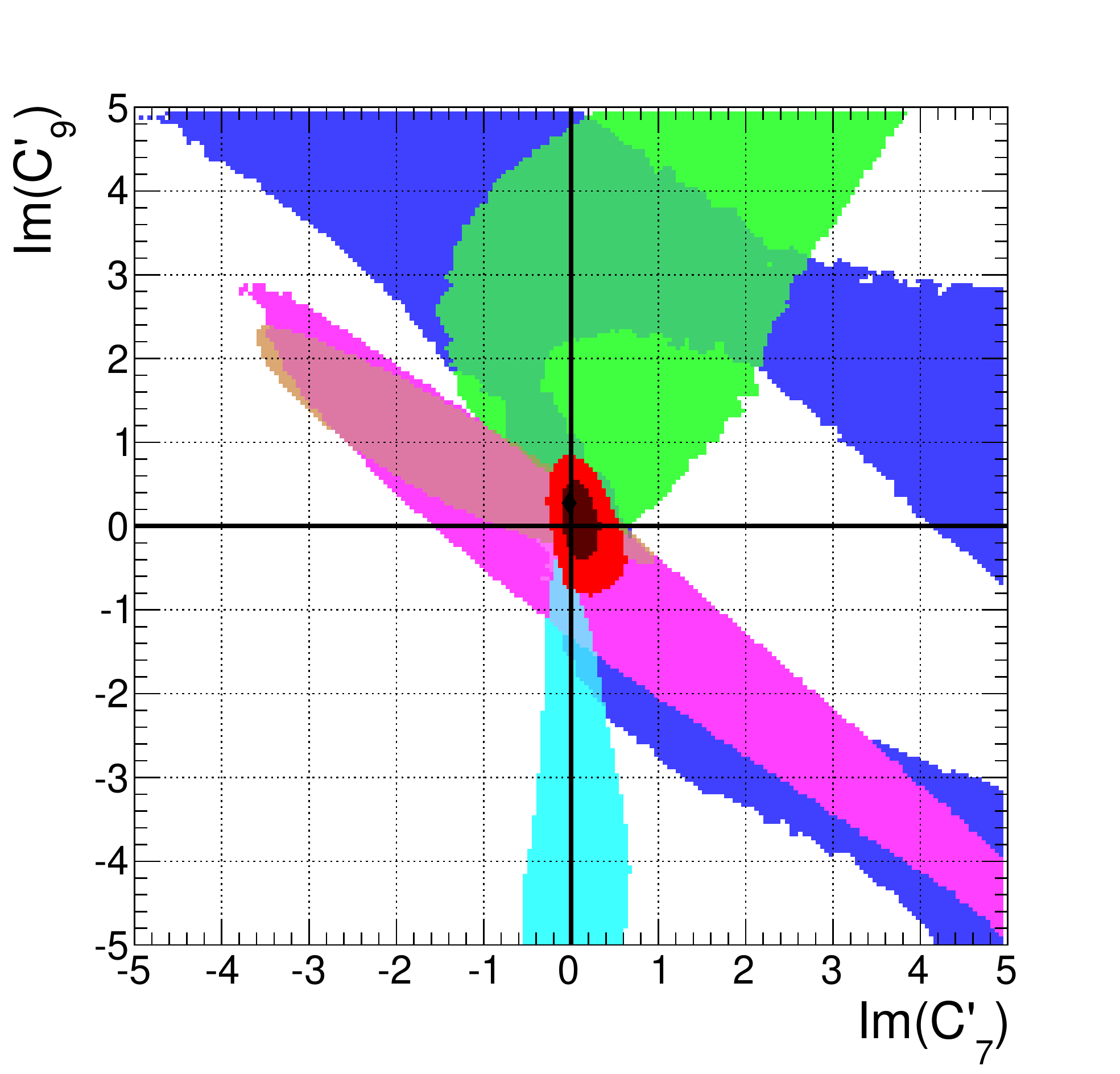}
  \caption{}
    \label{fig:7p9p}
  \end{subfigure}
  \caption{ $68\%$ CR for the complex Wilson Coefficients
    $\mathcal{C}^{(\prime)}_{7,9}$ in the scenarios Sc$-77'$ (left), Sc$-79$
    (middle), and Sc$-7'9'$ (right).  Constraints are obtained from the decay
    systems $B \to K\pi$ (cyan), $B \to K\rho$ (blue), $B \to K^*\pi$ (green),
    $B \to K^*\rho$ (purple), and $B \to K^*\phi$ (brown). The combined contour
    (red) is shown for $68\%$ and $95\%$ CRs.  The $\blacklozenge$ corresponds
    to the best-fit point of the combined fit.}
  \label{fig:C77p_C79_C7p9p}
\end{figure*}
The analysis of scenarios that are dominated by single operators has shown that
NP in QED-penguin operators is suitable to sufficiently address all tensions
present in the SM, though not all in one particular scenario. The benefits of
each single scenario combines in the generalized scenarios, as is evident from
the improvement of $\Delta \chi ({\rm SM})$ in \reftab{Chadr:pulls}. In fact,
the most general considered Sc$-77'99'$ has greatly reduced pull values compared
to the SM and largest $\Delta \chi ({\rm SM})$. Concerning models that allow for
NP in two Wilson coefficients, only Sc$-99'$ cannot resolve tensions in $B \to
K\pi,\, K^*\pi,\, K^*\phi$, showing that NP is required in ${\cal C}_7$,
respectively ${\cal C}'_7$.  In \reffig{C77p_C79_C7p9p}, we show the contours
for $\mbox{Re}\, {\cal C}_i$ versus $\mbox{Re}\, {\cal C}_j$ and $\mbox{Im}\,
{\cal C}_i$ versus $\mbox{Im}\, {\cal C}_j$ of the fits of Sc$-77'$, Sc$-79$,
and Sc$-7'9'$. The features of \reffig{C7-only_C9-only} are present again,
namely large imaginary parts for the Wilson coefficients are excluded, whereas
for ${\cal C}_7^{(\prime)}$ non-SM values for the real parts are allowed,
disfavoring the SM by more than 95\% probability in all three scenarios. On the
other hand large imaginary parts for ${\cal C}_9^{(\prime)}$ can only arise in
Sc$-99'$ and Sc$-77'99'$, since only $\mbox{Im}\, \overline{\cal C}_9$ is bound
to be close to zero by the combination of $B \to K^*\rho,\, K^* \phi$.

Measurements of the mixing-induced CP asymmetries $\Delta S_f$ only exist for
two out of the five considered decay systems: $\bar{B}^0 \to \bar{K}^0 \pi^0$ 
and $\bar{B}^0 \to \bar{K}^0 \rho^0$. Since these are rather imprecisely measured,
we omit $\Delta S_f$ as
constraint from the fit and instead give predictions for each scenario of single
operator dominance together with the SM prediction in \reftab{Chadr:pred}.  In
the case of the SM, we observed that the mixing-induced CP asymmetries are
insensitive to the residual $\rho_A$ parameter space that is allowed from
constraints of branching fractions and direct CP asymmetries. As a consequence,
the SM predictions are dictated by error estimation of the nuisance parameters
and therefore quoted as interval. We have seen from the fits that CP-violating
NP contributions to ${\cal C}^{(\prime)}_7$ are strongly disfavored and to
${\cal C}^{(\prime)}_9$ tightly constrained. Although $\mbox{Im}\, {\cal
  C}^{\prime}_9$ could still become large if ${\cal C}_9$ and ${\cal
  C}^{\prime}_9$ are modified, such scenarios do not significantly increase the
quality of the fit. Hence, mixing-induced CP asymmetries are not strongly
affected in the case of single operator dominance and in most cases the central
values of NP predictions coincide with the SM interval.  Ratios of branching
fractions, respectively branching fractions are more sensitive to, for example,
large real-valued ${\cal C}^{(\prime)}_7$.  In particular, the purely
isospin-breaking branching fractions $B_s \to \phi\pi,\, \phi\rho$ as well as
$R_n^{B{_s}}(KK)$, which predictions are also accumulated in
\reftab{Chadr:pulls}, are sensitive to NP in QED-penguin operators. Indeed, all
four considered scenarios, except for the branching fraction of $B_s \to
\phi\pi$ in Sc$-7$ and Sc$-9'$, predict a further suppression of ${\cal B}(B_s
\to \phi\pi,\,\phi\rho)$, which would unfortunately demand even more
experimental effort to observe these very rare decays. On the contrary, the
prediction of $R_n^{B{_s}}(KK)$ remains unchanged within Sc$-9^{(\prime)}$,
whereas it largely deviates within Sc$-7^{(\prime)}$ compared to the SM.

\begin{table*}
  \begin{center}
\renewcommand{\arraystretch}{1.4}
\resizebox{\textwidth}{!}{
   \begin{tabular}{l||cccccccc|ccc}
\hline\hline
 & $\Delta S(K\pi)$
 & $\Delta S(K\rho)$ 
 & $\Delta S(K^*\pi)$
 & $\Delta S_L(K^*\rho)$ 
 & $\Delta S(K\eta')$ 
 & $\Delta S(K\omega)$
 & $\Delta S(K\phi)$
 & $\Delta S_L(K^*\phi)$
 & $\mathcal{B}(\phi \pi)$ 
 & $\mathcal{B}(\phi \rho)$ 
 & $R_n^{B{_s}}(KK)$
\\
\hline
HFAG
 & $-0.11_{-0.17}^{+0.17} $
 & $-0.14_{-0.21}^{+0.18} $
 & --
 & --   
 & $-0.05_{-0.06}^{+0.06} $
 & $ 0.03_{-0.21}^{+0.21} $
 & $ 0.06_{-0.13}^{+0.11} $
 & --
 & --
 & --
 & --
\\
SM
 & $ [ 0.05,\,  0.13]$
 & $ [-0.19,\, -0.04]$
 & $ [ 0.06,\,  0.17]$
 & $ [-0.15,\,  0.09]$   
 & $ [-0.01,\,  0.04]$
 & $ [ 0.09,\,  0.17]$
 & $ [ 0.01,\,  0.05]$
 & $ [ 0.01,\,  0.04]$
 & $ 0.24_{-0.04}^{+0.07}$  
 & $ 0.68_{-0.10}^{+0.19}$  
 & $ 0.99_{-0.08}^{+0.01}$
\\
\cline{2-12}
Sc$-7$
 & $ 0.13_{-0.13}^{+0.02}$
 & $-0.18_{-0.10}^{+0.11}$
 & $ 0.07_{-0.07}^{+0.09}$
 & $\hphantom{-}0.08_{-0.13}^{+0.07}$
 & $\hphantom{-}0.03_{-0.09}^{+0.04}$
 & $ 0.15_{-0.31}^{+0.04}$
 & $\hphantom{-}0.04_{-0.08}^{+0.05}$
 & $\hphantom{-}0.03_{-0.08}^{+0.06}$
 & $ 0.91_{-0.22}^{+0.28}$
 & $ 0.35_{-0.07}^{+0.14}$
 & $ 0.80_{-0.06}^{+0.04}$
\\
Sc$-7'$
 & $ 0.13_{-0.13}^{+0.02}$
 & $-0.08_{-0.10}^{+0.07}$
 & $ 0.10_{-0.07}^{+0.09}$
 & $\hphantom{-}0.10_{-0.12}^{+0.09}$
 & $\hphantom{-}0.02 _{-0.08}^{+0.06}$
 & $ 0.12_{-0.08}^{+0.06}$
 & $-0.01_{-0.07}^{+0.06}$
 & $\hphantom{-}0.05_{-0.07}^{+0.06}$
 & $ 0.06_{-0.04}^{+0.05}$
 & $ 0.26_{-0.08}^{+0.12}$
 & $ 0.87_{-0.06}^{+0.06}$
\\
Sc$-9$
 & $  0.06_{-0.08}^{+0.06}$
 & $ -0.04_{-0.10}^{+0.09}$
 & $  0.05_{-0.09}^{+0.07}$
 & $-0.01_{-0.18}^{+0.10}$
 & $\hphantom{-}0.00_{-0.07}^{+0.06}$
 & $ 0.09_{-0.07}^{+0.09}$
 & $-0.04_{-0.07}^{+0.07}$
 & $-0.08_{-0.06}^{+0.11}$
 & $ 0.11_{-0.04}^{+0.06}$
 & $ 0.40_{-0.10}^{+0.13}$
 & $ 0.94_{-0.06}^{+0.03}$
\\
Sc$-9'$
 & $ 0.05_{-0.08}^{+0.06}$
 & $-0.20_{-0.11}^{+0.10}$
 & $ 0.17_{-0.07}^{+0.05}$
 & $-0.05_{-0.15}^{+0.13}$
 & $-0.02_{-0.05}^{+0.08}$
 & $ 0.16_{-0.11}^{+0.03}$
 & $\hphantom{-}0.06_{-0.10}^{+0.03}$
 & $-0.03_{-0.12}^{+0.09}$
 & $ 0.49_{-0.12}^{+0.14}$
 & $ 0.32_{-0.10}^{+0.15}$
 & $ 0.93_{-0.04}^{+0.05}$
\\
\hline\hline
\end{tabular}
}
\caption{ Predictions for the mixing-induced CP asymmetry of diverse $B_d$
  decays and for the purely isospin-breaking branching ratios
  $\mathcal{B}(\bar{B}_s \to \phi \pi,\, \phi \rho)$ within the single dominant
  operator scenarios and the SM.}
\label{tab:Chadr:pred} 
\end{center}
\end{table*}

\begin{table*}
  \begin{center}
\renewcommand{\arraystretch}{1.4}
\begin{tabular}{l||cc|cc|cc|cc|cc}
\hline\hline
& \multicolumn{2}{c}{$K\pi$}
& \multicolumn{2}{c}{$K^*\pi$}
& \multicolumn{2}{c}{$K\rho$}
& \multicolumn{2}{c}{$K^*\rho$}
& \multicolumn{2}{c}{$K^*\phi$}
\\
 & $|\rho_A|,\, \phi_A$
 & $\xi_3^A$
 & $|\rho_A|,\, \phi_A$
 & $\xi_3^A$
 & $|\rho_A|,\, \phi_A$
 & $\xi_3^A$
 & $|\rho_A|,\, \phi_A$
 & $\xi_3^A$
 & $|\rho_A|,\, \phi_A$
 & $\xi_3^A$
\\
\hline
SM
 & $ 3.34, 2.71$
 & \multicolumn{1}{c|}{0.39}
 & $ 1.61, 5.84$
 & \multicolumn{1}{c|}{0.89}
 & $ 2.69, 2.68$
 & \multicolumn{1}{c|}{0.78}
 & $ 1.56, 5.66$
 & \multicolumn{1}{c|}{1.33}
 & $ 1.50, 2.82$
 & \multicolumn{1}{c}{0.38}

\\
\cline{2-11}
Sc$-7$
 & $2.14, 5.45$
 & $[0.38,\, 0.60]$
 & $1.80, 5.90$
 & $[0.86,\, 1.39]$
 & $1.88, 5.58$
 & $[0.39,\, 1.64]$
 & $1.41, 5.66 $
 & $[0.70,\, 1.81]$
 & $1.53, 2.85$
 & $[0.29,\, 0.65]$

\\
Sc$-7'$
 & $3.61, 2.68$
 & $[0.34,\, 0.64]$
 & $3.73, 1.84$
 & $[0.72,\, 2.72]$
 & $2.14, 5.36$
 & $[0.54 ,\, 1.46]$
 & $1.29, 5.64$
 & $[0.57,\, 1.75]$
 & $0.71, 5.64$
 & $[0.38,\, 0.59]$
\\
\cline{2-11}
Sc$-9$
 & $1.86, 5.49 $
 & $[0.35,\, 0.60]$
 & $1.63, 5.87$
 & $[0.78,\, 1.49]$
 & $1.52, 5.44$
 & $[0.41,\, 1.38]$
 & $1.54, 5.63$
 & $[0.62,\, 1.97]$
 & $1.53, 2.83$
 & $[0.35,\, 0.52]$
\\
Sc$-9'$
 & $1.85, 5.49$
 & $[0.35,\, 0.62]$
 & $2.99, 2.91$
 & $[0.76,\, 1.55]$
 & $2.71, 2.68$
 & $[0.41,\, 1.49]$
 & $1.53, 5.62$
 & $[0.62,\, 1.91]$
 & $1.55, 2.84$
 & $[0.36,\, 0.53]$
\\
\cline{2-11}
Sc$-77'$
 & $2.45, 5.69$
 & $[0.34,\, 1.18]$
 & $3.03, 2.91$
 & $[0.71,\, 2.98]$
 & $1.51, 5.44$
 & $[0.00,\, 1.61]$
 & $1.71, 6.00$
 & $[0.51,\, 2.37]$
 & $0.95, 6.00$
 & $[0.30,\, 0.88]$
\\
Sc$-99'$
 & $2.39, 5.40$
 & $[0.33,\, 0.71]$
 & $3.34, 2.99$
 & $[0.68,\, 3.24]$
 & $1.44, 0.04$
 & $[0.01,\, 2.78]$
 & $2.31, 2.74$
 & $[0.40,\, 2.28]$
 & $1.54, 2.84$
 & $[0.26,\, 0.71]$
\\
\cline{2-11}
Sc$-79$
 & $3.59, 2.68$
 & $[0.23,\,0.70]$
 & $1.80, 5.90$
 & $[0.78,\, 1.70]$
 & $1.95, 5.60$
 & $[0.24,\, 2.40]$
 & $2.19, 2.79$
 & $[0.31,\, 2.06]$
 & $1.53, 2.86$
 & $[0.17,\, 0.68]$
\\
Sc$-7'9'$
 & $2.17, 5.53$
 & $[0.30,\, 0.66]$
 & $2.89, 2.77$
 & $[0.56,\, 2.74]$
 & $2.19, 5.27$
 & $[0.65,\, 1.84]$
 & $2.43, 2.87$
 & $[0.43,\, 1.92]$
 & $0.99, 6.01$
 & $[0.31,\, 0.83]$
\\
\cline{2-11}
Sc$-77'99'$
 & $2.24, 5.56$
 & $[0.08,\, 1.49]$
 & $1.65, 6.07$
 & $[0.11,\, 3.58]$
 & $1.45, 6.28$
 & $[0.00,\, 2.64]$
 & $1.81, 5.89$
 & $[0.12,\, 2.83]$
 & $0.90, 5.93$
 & $[0.01,\, 1.02]$
\\
\hline\hline
\end{tabular}
\caption{ Compilation of best-fit points for $\rho_A$ and $\xi_3^A$ at $68\%$
  probability. The results are given for the considered decay systems and
  scenarios Sc$-i$. As explained in \refsec{def:sub-leading:ratios}, the
  interval of $\xi_3^{A} ({\rm NP})$ should be compared to $\xi_3^{A}({\rm SM})$
  at the best-fit point of $\rho_A$, listed in the first row.  }
\label{tab:EW-xi4A} 
\end{center}
\end{table*}

Apart from the NP parameters discussed so far, we simultaneously fitted one
universal WA parameter per decay system. The comparison of the best-fit points
of these parameters with the SM fit is summarized in \reftab{EW-xi4A} for each
of the considered scenarios.  These best-fit points lie in the solutions that
were singled out by the SM fit, owing to the fact that NP in QED-penguin
operators does not modify the numerically leading decay amplitude
$\hat{\alpha}_4^c$. We further provide ranges for the ratios $\xi_3^A$ at 68\%
probability that quantify the relative size of subleading WA amplitudes, which
have been determined according to the procedure given in
\refsec{def:sub-leading:ratios}. The presence of NP always allows for smaller
values of $\xi_3^A$ than in the SM fit. In the most general scenario Sc$-77'99'$
the size of power corrections can be lower than 15\% for all considered decay
systems. Especially for $B \to K \rho,\, K^*\rho$ also simpler NP scenarios
already lead to a significant reduction. On the other hand the presence of NP
might allow also for very large values of $\xi_3^A$ in most systems, except for
$B \to K\pi\,(K^*\phi)$, where $\xi_3^A \lesssim 1.5\,(1.2)$.

%
%
\subsection{NP in tree-transitions $b\to s\, \bar{u}u$}
\label{sec:NP:bsuu}

In the case of the SM, isospin-breaking contributions to hadronic $B$ decays
occur either through QED-penguin operators, which were investigated in the
previous section, or through tree-level operators with an up-quark current.  The
latter operators occur in the SM in a color-singlet, $O_1^u$, and -octet,
$O_2^u$, configuration and are the only source of CP violation in the SM for
flavor-violating $b\to s$ transitions of $B$ mesons. Hence, these operators seem
to be suitable to address the tensions of the SM in both $\Delta C(K \pi)$ as
well as $R_n^B(K\pi)$ if they can be enhanced.  We also encountered some
discrepancy in the branching fraction of $B \to K^{*0} \phi$, but these decays
do not directly depend on either of the two tree-level operators, leaving their
explanation, at least in the context of the following discussion, due to
statistical fluctuation or underestimated theory uncertainties. Due to the
strong CKM hierarchy in $b \to s$ transitions, $b \to s\, \bar{u} u$ operators
give only numerically important contributions to CP asymmetries, contrary to $b
\to d\, \bar{u} u$ operators, which are constrained by well-measured branching
fractions and CP asymmetries in tree-dominated decays $B \to \pi\pi,\,
\rho\rho,\, \rho\pi$ \cite{Bobeth:2014rda}.

\begin{figure*}
  \begin{subfigure}[t]{0.32\textwidth}
    \centering
    \includegraphics[width=\textwidth]{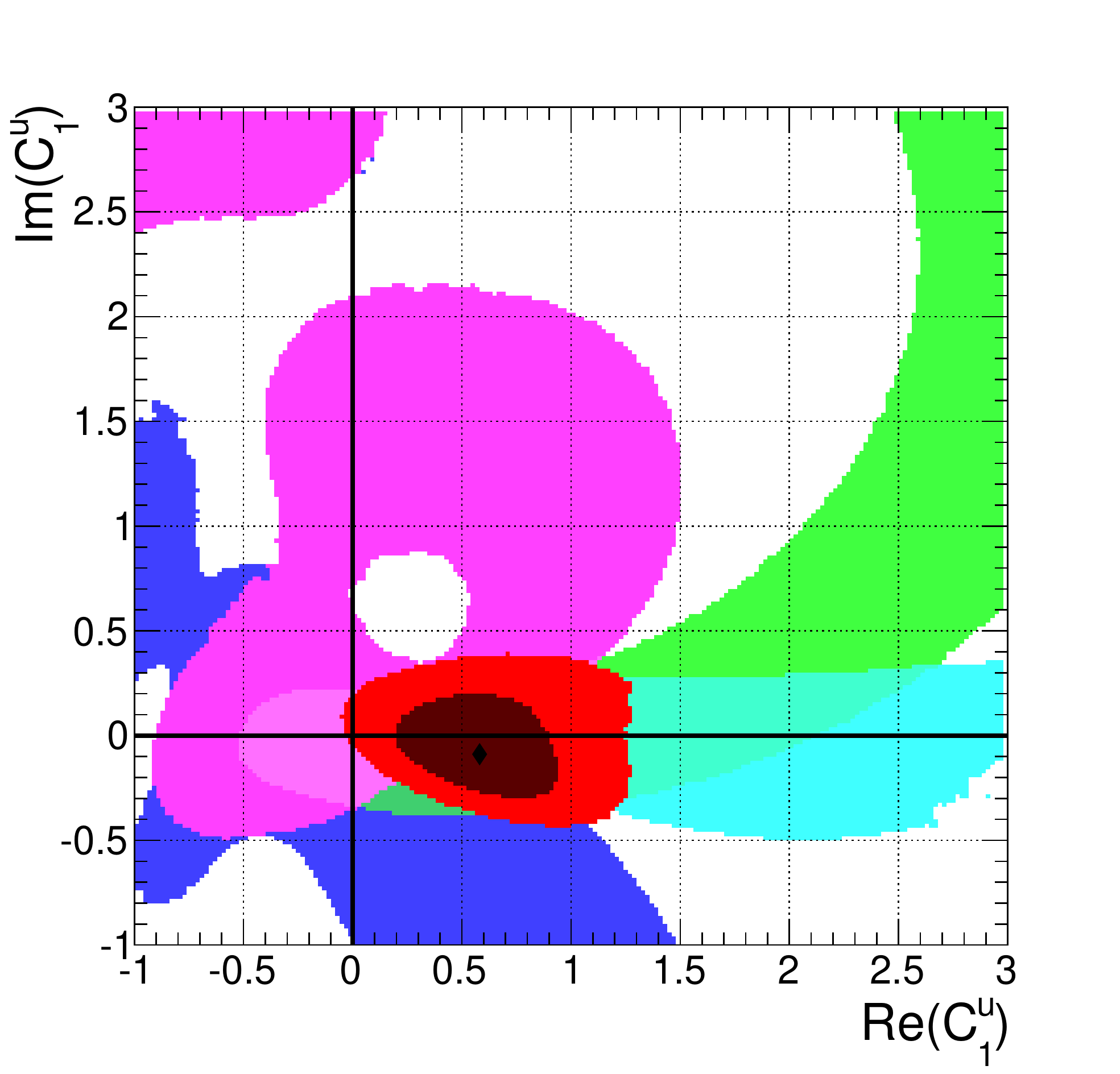}
  \caption{}
    \label{fig:C1-only}
  \end{subfigure}
  \begin{subfigure}[t]{0.32\textwidth}
    \centering
    \includegraphics[width=\textwidth]{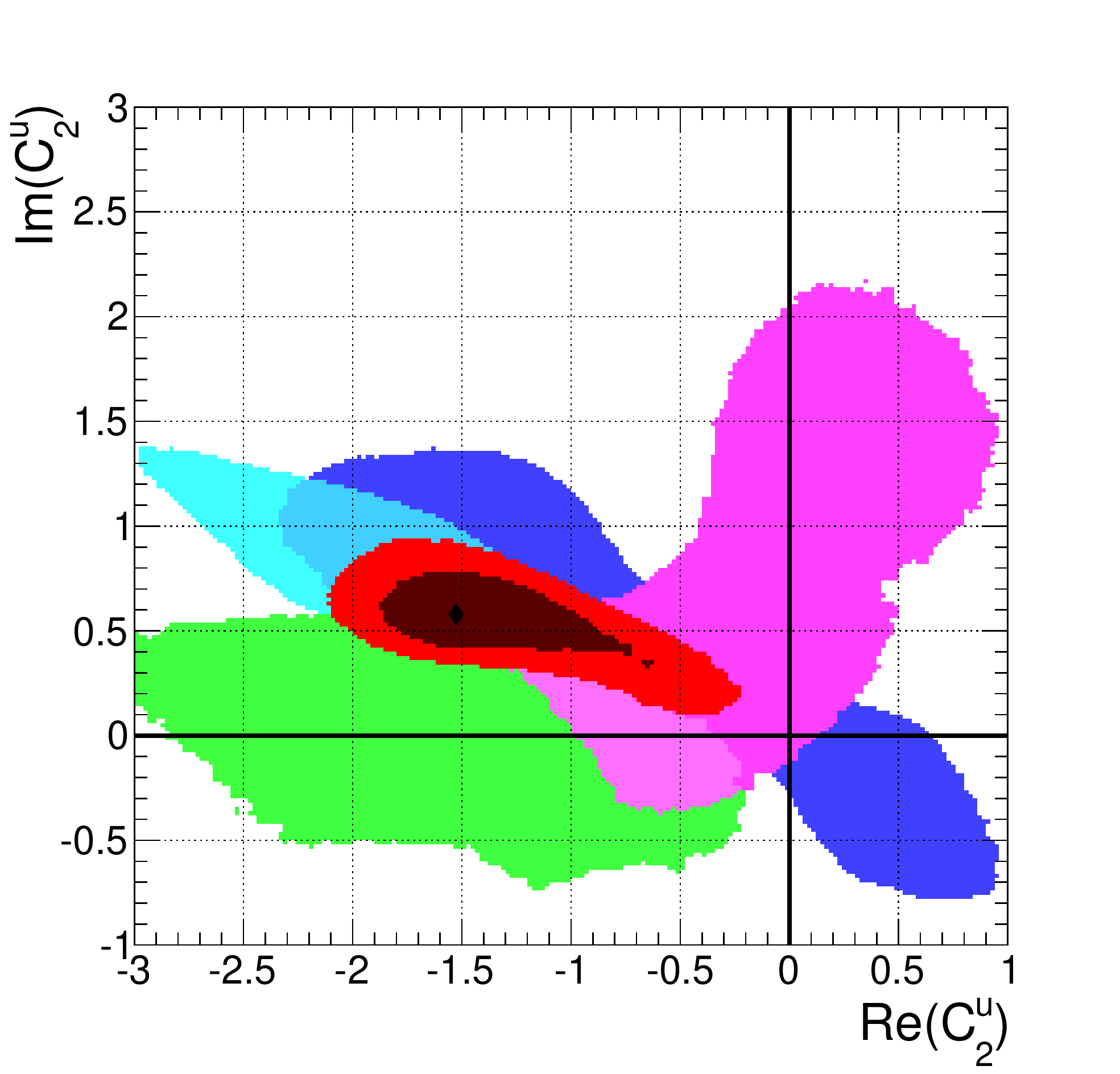}
  \caption{}
    \label{fig:C2-only}
  \end{subfigure}
  \caption{ $68\%$ CR for the complex Wilson Coefficients
    $\mathcal{C}^{u}_{1,2}$ in the scenarios Sc$-1$ (left) and Sc$-2$ (right).
    Constraints are obtained from the decay systems $B \to K\pi$ (cyan), $B \to
    K\rho$ (blue), $B \to K^*\pi$ (green), and $B \to K^*\rho$ (purple). The
    combined contour (red) is shown for $68\%$ and $95\%$ CRs. The
    $\blacklozenge$ corresponds to the best-fit point of the combined fit.}
  \label{fig:C1_only-C2_only}
\end{figure*}

\begin{table*}
  \begin{center}
\renewcommand{\arraystretch}{1.1}
   \begin{tabular}{l||c|ccc|c}
\hline\hline
 & $\mbox{Re}({\cal C}_i^u), \mbox{Im} ({\cal C}_i^u)$
 & $\Delta C(K\pi)$
 & $R_n^B(K\pi)$
 & $R_n^B(K^*\pi)$
 & $\Delta \chi^2(SM)$
\\
\hline
SM
 &   
 & $ \bf{-2.8} \sigma$
 & $ \bf{-1.9} \sigma$
 & $ 0.6\sigma$   
 & $ $
\\
\cline{2-6}
Sc$-1$
 & $ \hphantom{-}0.58,\, -0.09$ 
 & $ -0.9 \sigma $
 & $  \hphantom{-}0.0 \sigma $
 & $  0.2 \sigma $
 & $ 9.2$
\\
\cline{2-6}
Sc$-2$
 & $-1.53,\, \hphantom{-}0.58$ 
 & $ \hphantom{-}0.0  \sigma $
 & $ -0.3 \sigma $
 & $ \bf{1.6}  \sigma $
 & $ 12.4$
\\
\cline{2-6}
\multirow{2}{*}{Sc$-12$}
 & $\hphantom{-}1.47,\,\hphantom{-}0.03 $ 
 & \multirow{2}{*}{$ \hphantom{-}0.0 \sigma $}
 & \multirow{2}{*}{$ -0.5 \sigma $}
 & \multirow{2}{*}{$  1.1 \sigma $}
 & \multirow{2}{*}{$ 15.6$}
\\
 & $-2.25,\, \hphantom{-}0.38$ 
 &  &  &  & 
\\
\hline\hline
    \end{tabular}
    \caption{ Compilation of best-fit points and pull values, with $ |\delta|
      \ge 1.6$, for the model-independent fits of $b\to s\,\bar{u}u$ operators.}
    \label{tab:bsuu:pulls} 
  \end{center}
\end{table*}

We introduce the following NP contribution to the Hamiltonian of
\refeq{heff-b-one}
\begin{equation}
\label{eq:NP:tree-bsuu}
  C^u_{1,2}(\mu_0) = C_{1,2}^{u,{\rm SM}}(\mu_0) + {\cal C}^u_{1,2}\,,
\end{equation}
where we choose $\mu_0 = M_W$ as before.  Although ${\cal C}^u_{1,2}$ mix into
Wilson coefficients of all other SM operators, this contribution is doubly
Cabibbo-suppressed compared to ${\cal C}^c_{1,2}$ and numerically negligible in
all amplitudes, except for $r_{\rm T}^{},\, r_{\rm T}^{\rm C}$. As discussed in
\refeq{CPA:BKpi:exp}, in the SM the latter two are the dominant contributions in
CP asymmetries for decay systems considered below.

In connection with the SM, we already discussed in \refsec{SM:BKpi} the
possibility of large hard scattering solution to the $\ACP(K\pi)$ problem, see
also \cite{Cheng:2009eg}.  Here we show that the assumption of NP in $b \to
s\,\bar{u}u$ operators provide qualitatively different solutions to large hard
scattering. For this purpose we remind of the dependence of CP asymmetries and
ratios of branching fractions \refeq{DefOfRatios} on the tree amplitudes:
\begin{equation}
\begin{aligned}
\label{eq:Obs:expansion}
 C &\propto 2\, \mbox{Im}\, (r^{{\rm (C)}}_{{\rm T}})\,    \sin \gamma + 
            2\, \mbox{Im}\, (r^{{\rm (C)}}_{{\rm T},j})\,  \mbox{Im}\, ({\cal C}_j e^{-i\gamma}), \\
 S &\propto 2\, \mbox{Re}\, (r^{{\rm (C)}}_{{\rm T}})\,    \sin \gamma + 
            2\, \mbox{Re}\, (r^{{\rm (C)}}_{{\rm T},j})\,  \mbox{Im}\, ({\cal C}_j e^{-i\gamma}), \\
 R &\propto 2\, \mbox{Re}\, (r^{{\rm (C)}}_{{\rm T}})\,    \cos \gamma + 
            2\, \mbox{Re}\, (r^{{\rm (C)}}_{{\rm T},j})\,  \mbox{Re}\, ({\cal C}_j e^{-i\gamma}) + \ldots, 
\end{aligned}
\end{equation}
when utilising the expansion in small $r_i$ and the dots stand for contributions
of further $r_i$ that are not affected from NP in the considered scenarios.
Hard scattering enters only the $r_i$, especially $r_{\rm T}^{\rm C}$. Hence,
direct and mixing-induced CP asymmetries become correlated through their common
dependence on $\mbox{Im}\, ({\cal C}_j e^{-i\gamma}) $, whereas they depend
differently on hard scattering. Analogous, qualitative differences exist among
CP asymmetries and the ratios $R$. In consequence, when mixing-induced CP
asymmetries become more precisely measured, it will be possible to distinguish
both scenarios.

\begin{table*}
  \begin{center}
\renewcommand{\arraystretch}{1.4}
\resizebox{\textwidth}{!}{
   \begin{tabular}{l||cccccccc|ccc}
\hline\hline
 & $\Delta S(K\pi)$
 & $\Delta S(K\rho)$ 
 & $\Delta S(K^*\pi)$
 & $\Delta S_L(K^*\rho)$ 
 & $\Delta S(K\eta')$ 
 & $\Delta S(K\omega)$
 & $\Delta S(K\phi)$
 & $\Delta S_L(K^*\phi)$
 & $\mathcal{B}(\phi \pi)$ 
 & $\mathcal{B}(\phi \rho)$ 
 & $R_n^{B_s}(KK)$
\\
\hline
HFAG
 & $-0.11_{-0.17}^{+0.17} $
 & $-0.14_{-0.21}^{+0.18} $
 & --
 & --   
 & $-0.05_{-0.06}^{+0.06} $
 & $ 0.03_{-0.21}^{+0.21} $
 & $ 0.06_{-0.13}^{+0.11} $
 & --
 & --
 & --
 & --
\\
SM
 & $ [ 0.05,\,  0.13]$
 & $ [-0.19,\, -0.04]$
 & $ [ 0.06,\,  0.17]$
 & $ [-0.15,\,  0.09]$   
 & $ [-0.01,\,  0.04]$
 & $ [ 0.09,\,  0.17]$
 & $ [ 0.01,\,  0.05]$
 & $ [ 0.01,\,  0.04]$
 & $ 0.24_{-0.04}^{+0.07}$  
 & $ 0.68_{-0.10}^{+0.19}$  
 & $ 0.99_{-0.08}^{+0.01}$
\\
\cline{2-12}
Sc$-1$
 & $  \hphantom{-}0.13^{+0.07}_{-0.07}$ 
 & $ -0.10^{+0.06}_{-0.22}$ 
 & $  \hphantom{-}0.12^{+0.11}_{-0.06}$ 
 & $ -0.06^{+0.16}_{-0.32}$ 
 & $ \hphantom{-}0.01^{+0.07}_{-0.06}$ 
& $  \hphantom{-}0.20^{+0.06}_{-0.09}$ 
 & $ \hphantom{-}0.09^{+0.02}_{-0.11}$ 
 & $ \hphantom{-}0.09^{+0.02}_{-0.11}$ 
 & $ 0.27^{+0.06}_{-0.07}$ 
 & $ 0.73^{+0.17}_{-0.17}$ 
 & $ 0.92_{-0.09}^{+0.08}$ 
\\
Sc$-2$
 & $ -0.30_{-0.12}^{+0.11}$
 & $ \hphantom{-}0.26_{-0.09}^{+0.01}$
 & $ -0.66_{-0.19}^{+0.20}$
 & $ -1.05_{-0.28}^{+0.30}$
 & $  \hphantom{-}0.06_{-0.05}^{+0.05}$
 & $ -0.55_{-0.22}^{+0.25}$
 & $  \hphantom{-}0.09_{-0.11}^{+0.02}$
 & $  \hphantom{-}0.09_{-0.11}^{+0.01}$
 & $  1.36_{-0.54}^{+0.74}$
 & $  3.94_{-1.40}^{+1.77}$ 
 & $  0.91_{-0.03}^{+0.04}$ 
\\
Sc$-12$
 & $ -0.42_{-0.20}^{+0.15}$ 
 & $ \hphantom{-}0.18_{-0.25}^{+0.09}$ 
 & $ -0.93_{-0.29}^{+0.29}$ 
 & $ -1.59_{-0.11}^{+0.43}$ 
 & $ \hphantom{-}0.07_{-0.05}^{+0.06}$ 
 & $ -0.51_{-0.43}^{+0.21}$ 
 & $ \hphantom{-}0.06_{-0.06}^{+0.05}$ 
 & $ \hphantom{-}0.04_{-0.04}^{+0.06}$ 
 & $ 2.74_{-1.21}^{+1.59}$ 
 & $ 8.78_{-3.67}^{+3.74}$ 
 & $ 0.85_{-0.09}^{+0.11}$ 
\\
\hline\hline
\end{tabular}
}
\caption{ Predictions for the mixing-induced CP asymmetry of diverse $B_d$
  decays and for the purely isospin-breaking branching ratios
  $\mathcal{B}(\bar{B}_s \to \phi \pi,\, \phi \rho)$ within the Sc$-1,2$
  scenarios and the SM.}
   \label{tab:tree-bsuu:pred} 
\end{center}
\end{table*}

\begin{table*}
  \begin{center}
\renewcommand{\arraystretch}{1.1}
   \begin{tabular}{l||cc|cc|cc|cc}
\hline\hline
& \multicolumn{2}{c}{$K\pi$}
& \multicolumn{2}{c}{$K^*\pi$}
& \multicolumn{2}{c}{$K\rho$}
& \multicolumn{2}{c}{$K^*\rho$}
\\
 & $\rho_A$
 & $\xi_3^A$
 & $\rho_A$
 & $\xi_3^A$
 & $\rho_A$
 & $\xi_3^A$
 & $\rho_A$
 & $\xi_3^A$
\\
\hline
SM
 & $ 3.34, 2.71$
 & \multicolumn{1}{c|}{0.39}
 & $ 1.61, 5.84$
 & \multicolumn{1}{c|}{0.89}
 & $ 2.69, 2.68$
 & \multicolumn{1}{c|}{0.78}
 & $ 1.56, 5.66$
 & \multicolumn{1}{c}{1.33}
\\
\cline{2-9}
Sc$-1$
 & $1.54,\, 5.58$
 & $[0.29,\, 0.50]$
 & $1.65,\, 5.87$
 & $[0.73,\, 1.33]$
 & $3.03,\, 2.81$
 & $[0.53,\, 1.32]$
 & $1.39,\, 5.90$
 & $[0.83,\, 1.72]$
\\
Sc$-2$
 & $1.54,\, 5.58$
 & $[0.26,\,0.77]$
 & $1.60,\, 5.92$
 & $[0.75,\,1.33]$
 & $1.11,\, 0.19$
 & $[0.33,\,1.00]$
 & $2.63,\, 3.76$
 & $[0.92,\,2.32]$
\\
\cline{2-9}
Sc$-12$
 & $2.05,\, 5.80$
 & $[0.13,\,0.88]$
 & $1.52,\, 5.90$
 & $[0.54,\,1.47]$
 & $3.76,\, 4.56$
 & $[0.22,\,3.17]$
 & $1.96,\, 3.38$
 & $[0.27,\,2.61]$
\\
\hline\hline
    \end{tabular}
    \caption{ Compilation of best-fit points for $\rho_A$ and $\xi_3^A$ at
      $68\%$ probability. The results are given for the considered decay systems
      and scenarios Sc$-i$. As explained in \refsec{def:sub-leading:ratios}, the
      interval of $\xi_3^{A} ({\rm NP})$ should be compared to $\xi_3^{A}({\rm
        SM})$ at the best-fit point of $\rho_A$, listed in the first row.}
   \label{tab:tree-bsuu:xi3A} 
  \end{center}
\end{table*}

We investigate the effects of the complex-valued Wilson coefficients ${\cal C}_j
= |{\cal C}_j| e^{i\delta_j}$ separately and in combination in the three
scenarios:
\begin{itemize}
\item single operator dominance \\
  Sc$-i$ : ${\cal C}^u_i \neq 0$ and ${\cal C}^u_{j \neq i} = 0$ for $i = 1,2$
\item combined scenario \\
  Sc$-12$ : ${\cal C}^u_{1,2} \neq 0$  \\
\end{itemize}

\reffig{C1_only-C2_only} shows the individual contours for ${\cal C}^u_{1}$
(left) and ${\cal C}^u_{2}$ (right) that were obtained from a fit of each decay
systems $B \to K\pi,\, K\rho,\, K^*\pi,\, K^*\rho$ within the scenarios of a
single operator dominance. In the case of new physics contribution to the
color-singlet operator, the fit prefers a real-valued ${\cal C}^u_{1}$ with a
significant contribution of the order of its SM value.  Due to the
parameterization of the effective weak Hamiltonian in \refeq{heff-b-one} and of
the NP contribution in \refeq{NP:tree-bsuu}, such a solution implies that the CP
violating phase of a particular NP model has to be aligned with the one of the
SM. Hence, the Wilson coefficient is enhanced from $C_1^u(M_W) = 0.98$ in the SM
to $C_1^u(M_W) = |0.98 + (0.58 - i\, 0.09)| \approx 1.56$ at the best-fit point,
tabulated in \reftab{bsuu:pulls}, whereas its weak phase $\gamma$ receives only
marginal corrections from $\delta_1 \approx -8.8^\circ$.  Since all contours
from the individual decay systems nicely overlap with each other, we expect to
resolve the discrepancy that are present for the SM in $B \to K \pi$ without
introducing new tensions in the data of other decay systems. This is confirmed
from the pull values listed in \reftab{bsuu:pulls}. It can also be seen from the
table that the tensions in $\Delta C(K\pi)$ and $R_n^B(K\pi)$ can be well
explained within Sc$-2$ and Sc$-12$ when tolerating a rising tension in
$R_n^B(K^*\pi)$ of $1.6\sigma$, respectively $1.1\sigma$.

The corresponding contours of ${\cal C}_2^u$ are displayed in
\reffig{C2-only}. The combined contour reduces to a common area of the allowed
regions for the decay systems $B \to K\pi,\, K\rho,\, K^*\rho$, whereas the
green contour from $B \to K^*\pi$ is slightly separated from the
combination. The SM value of the color-octet Wilson coefficient, $C_2^{u,
  \rm{SM}}(M_W) = 0.05$, is strongly suppressed compared to its color-singlet
counterpart, but the preferred values that were obtained from our fits shift
$C_2^u(M_W) = |0.05 + (-1.53 + i0.58)| \approx 1.58$ --- competitive to
$C_1^u(M_W)$.  In contrast to Sc$-1$, the weak phase of $C_2^u$ is not aligned
with the SM, but rather receives a significant phase shift of $\delta_2 \approx
159^\circ$.

The pattern that were obtained from the single operator dominance scenarios is
also observed for the combined scenario: $C_{1,2}^u$ becomes further enhanced by
$|0.98 + (1.47 + i0.03)| \approx 2.45$, respectively $|0.05 + (-2.25 + i0.38)|
\approx 2.23$ and $\delta_1 \sim 1^\circ$, whereas $\delta_2$ further tend to
$170^\circ$.

As in the previous analysis of the QED-penguin operators, we quote in
\reftab{tree-bsuu:pred} predictions for several mixing-induced CP asymmetries as
well as for the isospin-sensitive branching fractions of $B_s \to \phi\pi,\,
\phi\rho$ and for $R_n^{B_s}(KK)$.  The impact from an enhanced ${\cal C}_1^u$
on these observables is small and rather challenging to isolate from the SM
background, which is not the case for NP in ${\cal C}_2^u$. Especially the
predictions of the mixing-induced CP asymmetries of the decays $B \to K\pi,\,
K\rho,\, K^*\pi,\, K^*\rho$ and $B \to K\omega$ are visibly different compared
to the SM, making these observables an ideal probe of NP in the color-octet
operator. The same is true for the branching fractions of $B_s \to
\phi\pi,\,\phi\rho$, which we found to be enhanced by a factor of $5$ -- $6$ for
Sc$-2$ and by more than a factor of $10$ in the case of Sc$-12$. Although these
predictions largely deviate from the one of the SM, existing measurements do not
contradict NP in ${\cal C}_2^u$ due to lacking precision.

As before, the NP contributions to the Wilson coefficients have been fitted
simultaneously with WA parameters $\rho_A$ for each decay system in all
considered scenarios. Since NP in $b \to s\,\bar{u}u$ operators do not
contribute directly to the leading decay amplitude $\hat{\alpha}_4^c$ but rather
indirectly through the common dependence on the likelihood function, we expect
moderate changes of WA compared to the results of the SM fit. The best-fit
points of the individual $\rho_A$ as well as the 68\% probability intervals of
$\xi_3^A$ are summarized in \reftab{tree-bsuu:xi3A} for each of the three
scenarios. We observe that almost all best-fit points of $\rho_A$ lie within the
contour regions of the SM fit.  The only exceptions are $\rho_A^{K\rho}$ in
Sc$-12$ and $\rho_A^{K\pi}$ for all considered scenarios. For the latter, the
most likely values of $|\rho_A^{K\pi}|$ in the case of Sc$-1$ and Sc$-2$ are
significantly reduced compared to the SM, whereas $\phi_A^{K\pi}$ tends towards
smaller strong phases in the combined scenario. Due to the additional degrees of
freedom, it is possible that the relative amount of power-suppressed corrections
can be reduced.  In general $\xi_3^A$ is most strongly affected in the combined
scenario, for which we find lower bounds on $\xi_3^A(K\pi) \gtrsim 0.13$,
$\xi_3^A(K\rho) \gtrsim 0.22$, and $\xi_3^A(K^*\rho) \gtrsim 0.27$.  The
potential suppression of $\xi_3^A$ for $B \to K^*\pi$ is less effective and a
relative amount of power-suppressed contribution of at least $0.54$ is required
in any case.  It is worth to notice that the large WA scenario is still
disfavored for $B \to K \pi$, which is in general not true for all other decay
modes.

%
%
%

\section{Conclusion}
\label{sec:conclusion}

In this work we have carried out a phenomenological study of QCD- and
QED-penguin dominated charmless 2-body $B$-meson decays in the framework of QCD
factorization (QCDF).  In particular we investigated whether data supports the
assumption of one universal parameter, $\rho_A$, in weak annihilation (WA)
contributions for decay channels related by $(u\leftrightarrow d)$ quark
exchange in $B_{u,d,s}$ meson decays to $PP$, $VP$ and $VV$ final states, while
the remaining theory uncertainties are incorporated in an uncorrelated
manner. 

We analyse the decay systems of $B_{u,d}$ decays into $PP = K\pi,\,
K\eta^{(')},\, KK$ or $PV = K\rho,\, K\phi,\, K\omega,\, K^*\pi ,\,
K^*\eta^{(')}$ or $VV = K^*\rho,\, K^*\phi,\, K^*\omega,\, K^*K^*$, and further
$B_s$ decays into $PP = \pi\pi,\, KK,\, K\pi$ or $VV = \phi\phi,\, K^*\phi,\,
K^* K^*$ final states and employ the available data (see \reftab{exp:input:obs:PP}, 
\ref{tab:exp:input:obs:PV}, \ref{tab:exp:input:obs:VV})
on branching fractions,
direct CP asymmetries and for $VV$ final states also polarization fractions and
relative phases between polarization amplitudes.

Within the standard model (SM), the data
can be described using one universal WA parameter for each decay system. The only exception is the $B\to
K\pi$ system when using Set II of observables, as specified in \refsec{data},
which includes $\Delta \ACP$ and $R_n^B$, as a manifestation of the ``$\Delta
\ACP$ puzzle'' in our framework.  The only other noticeable pull value of
$2.6\sigma$ $(1.7\sigma)$ arises for the measurement of $\Br(\bar{B}^0\to
\bar{K}^{*0}\phi)$ from Belle (BaBar).  For each system, there are at least two
allowed regions at 68\% CR with the best fit solution residing in one of them
(see \reftab{SM-xi-values}).  These two regions correspond to phases close to
$\pi$ and $2\pi$, outside of regions of large destructive interference of WA
amplitudes with leading amplitudes.  Moreover the ratio of the magnitudes of WA
amplitudes to leading amplitudes, $\xi_3^A$ (see \reftab{SM-xi-values}), is
similar in size in both regions and within the 68\% CR it is possible to have
$\xi_3^A < 1$ (except for $B_s \to K^* K^*$) and for the majority even $\xi_3^A
< 0.5$. QCDF can thus describe the current data without the need of anomalously
large WA contributions.

We emphasize that in our analysis the ``$\Delta \ACP$ puzzle'' is only present
if we assume a universal WA parameter that can be fitted from data.  If we lift
this assumption the anomaly would only reappear if we restrict our analysis to
rather small WA parameters $\rho_A$. Without such a restriction, however the
non-linear dependence of $\xi_3^A$ on $\rho_A$ still permits reasonably small
$\xi_3^A$, which are not larger as currently accepted in the literature.

We studied also ratios of branching fractions and differences of CP asymmetries 
(Set II) for the decay systems $B\to K\pi,\, K^*\pi,\, K\rho,\, K^*\rho$. They are less
sensitive to form-factor and CKM uncertainties or are especially sensitive to numerically
suppressed contributions from tree topologies. The according results listed in
\reftab{SM-pulls} show that currently both sets yield good fits to the data,
except for $B\to K\pi$, where Set~II has a $p$-value of only 4\%. The  data of ratios 
of branching fractions and differences of CP asymmetries 
have been obtained by ourselves from measurements of
observables in Set~I. This neglects correlations and potential
cancellations of systematic uncertainties accessible only in the experimental
analyses. In this regard, future analysis would benefit 
from the direct experimental determination of these composed observables. 

In view of the large pull value of $2.8\sigma$ for $\Delta \ACP$ in
$B\to K\pi$, we performed also a simultaneous fit of the WA and hard-scattering
(HS) phenomenological parameters in the SM. The HS contribution necessary to lower
the pull value of $\Delta \ACP$ to $1.0\sigma$ is not larger
than typically considered in conventional error estimates in the literature ---
$\xi_2^H = 1.0$.
A better description of the data can be achieved with even larger HS contributions. 
A preciser measurement of $C(B_d \to K^0 \pi^0)$ in the future
could be helpful to test a ``large HS''-scenario. Further, larger HS
contributions allow for smaller WA contributions.

We investigate the
feasibility to constrain new-physics (NP) scenarios in view of the aforementioned tensions 
in the SM. Within our framework 
this requires the fit of WA phenomenological parameters
simultaneously with NP parameters from data. In contrast to the conventional
handling of WA contributions within QCDF, we find that the assumption of one
universal parameter per decay system yields stronger constraints on new-physics
parameters for the considered scenarios. We have studied model-independent
scenarios of NP in QED-penguin operators as possible solutions to the ``$\Delta
\ACP$ puzzle'' in $B\to K \pi$ and tensions in $B\to K^*\phi$, taking into
account also data from the systems $B\to K\rho,\, K^*\pi,\, K^*\rho$. As a
second possible solution to the ``$\Delta \ACP$ puzzle'' we investigated NP in
$b\to s\,\bar{u}u$ current-current operators including again data from $B\to
K\rho,\, K^*\pi,\, K^*\rho$.  For each scenario we provide the best fit regions
of the NP contributions to the according Wilson coefficients, reduction of
$\chi^2$ compared to the SM fit, the pull values of observables, and predictions
of mixing-induced CP asymmetries, as well as branching fractions of $B_s \to
\phi\pi,\, \phi\rho$.

In both classes of NP scenarios there is no direct contribution to the
the numerically leading amplitude of QCD-penguin
operators, since we consider only new isospin-violating contributions. 
In consequence, the allowed regions of WA parameters do not differ
qualitatively from those of the SM fit. Yet, the combined fit of NP and WA
allows for smaller $\xi_3^A$ in all scenarios compared to the SM.

It is conceivable that one day factorization theorems will be established even
for WA contributions involving then new nonperturbative quantities. Our studies
suggest that it will be possible to extract these new quantities also from data
in the lack of first principle nonperturbative methods of their calculation. It
will be important to have access to more accurate measurements of the involved
observables which should become available from Belle II and LHCb within the next
decade.

%

\begin{acknowledgments}
  We thank Martin Beneke, Gerhard Buchalla and Yuming Wang for helpful discussions
  and Yuming Wang for comments on the manuscript. We would like to thank Frederik
  Beaujean for his support on BAT \cite{Caldwell:2008fw}. 
  C.B. received support from the ERC Advanced Grant project ``FLAVOUR'' (267104). 
  M.G. acknowledges partial support by the UK Science \& Technology Facilities
  Council (STFC) under grant number ST/G00062X/1.
\end{acknowledgments}

%
%

\appendix

%
%
%

\section{Numerical input}
\label{app:numeric:input}

\renewcommand{\theequation}{A\arabic{equation}}
\setcounter{equation}{0}

Here we collect the numerical input used in our analysis in \reftab{num:input}.
We list two sets of CKM parameters. The first, denoted by ``SM'', is obtained in
a global CKM fit in the framework of the SM \cite{UTFit:post-EPS13} and is used
throughout our SM analyses \refsec{SM-fit-results}. The second, denoted by
``NP'', is obtained from a global fit that includes only tree-level mediated
observables \cite{UTFit:post-EPS13} and which we use throughout the analysis of
scenarios beyond the SM in \refsec{NP-fit-results}. Further, for $B_s$ decays we
use the value of $y_s$ \refeq{exp:value:y_s} as an additional source of error in
\refeq{BR:exp:theo}. Throughout we vary the renormalization scale $\mu_b \in
[m_b/2,\, 2 m_b]$ for the central value of $\mu_b = 4.2$ GeV.  The uncertainty
from endpoint divergences in subleading hard-scattering contributions is
determined by varying $\phi_H \in [0,\, 2\pi]$ for fixed $|\rho_H| = 1$, as
frequently done in the literature.

\begin{table*}
\begin{center}
\renewcommand{\arraystretch}{1.3}
\begin{tabular}{|ccccc|l|}
\hline\hline
\multicolumn{6}{|l|}{\textbf{Electroweak input}}
 \\
 \hline
$G_F [10^{-5}\gev^{-2}]$
& $\Lambda_{\overline {\rm MS}}^{(5)}[\gev]$
& $M_Z [\gev]$
& $\alpha_s^{(5)}(M_Z)$
& $\alpha_e^{(5)}(m_b)$
&
\\
\hline
 $1.16638$
& $0.213$
& $91.1876$
& $0.1184$
& $1/132$
& \cite{Beringer:2012} \cite{Bethke:2009jm}
\\
\hline\hline
\multicolumn{6}{|l|}{\textbf{quark masses [GeV]}}
\\
\hline
$m_t^{\rm pole}$
& $m_b(m_b)$
& $m_c(m_b)$
& $m_s$
& $m_q/m_s$
&
\\
$(173.2 \pm 0.9)$
& $4.2 $
& $(1.3 \pm 0.2)$
& $(0.095 \pm 0.005)$
& $0.0370$
& \cite{Beneke:2003zv} \cite{Beringer:2012}  \cite{Aaltonen:2012ra}
\\
\hline\hline
\multicolumn{6}{|l|}{\textbf{CKM elements}}
\\
\hline
\multicolumn{1}{|c||}{}
& $\lambda $
& $|V_{cb}|$
& $\bar{\rho}$
& $\bar{\eta}$
&
\\
\hline\hline
\multicolumn{1}{|c||}{SM}
& $0.22535 \pm 0.00065$
& $0.04172 \pm 0.00056$
& $0.127 \pm 0.023$
& $0.353 \pm 0.014$
& \multirow{2}{*}{\cite{UTFit:post-EPS13}}
\\
\multicolumn{1}{|c||}{NP}
& $0.2253  \pm 0.0006$
& $0.04061 \pm 0.00097$
& $0.147 \pm 0.045$
& $0.368 \pm 0.048$
&
\\
\hline\hline
\multicolumn{6}{|l|}{\textbf{B-meson input}}
\\
\hline
\multicolumn{1}{|c||}{}
& $B_u$
& $B_d$
& $B_s$
&
&
\\
\hline\hline
\multicolumn{1}{|c||}{$f_B[\mev]$}
& \multicolumn{2}{c}{$190.5 \pm 4.2$}
& $227.7 \pm 4.5$
&
&
\\
\multicolumn{1}{|c||}{$\lambda_B[\mev]$}
& \multicolumn{3}{c}{$200^{+250}_{-0}$}
&
& \cite{Beneke:2006mk}
\\
\multicolumn{1}{|c||}{$\tau_B$[ps$^{-1}$]}
& $1.641$
& $1.519$
& $1.516$
&
& \multirow{2}{*}{\cite{Beringer:2012}}
\\
\multicolumn{1}{|c||}{$M_B[\mev]$}
& $5279.25 $
& $5279.58 $
& $5366.77 $
&
&
\\
\hline\hline
\multicolumn{6}{|l|}{\textbf{Hadronic input -- pseudoscalar mesons}}
\\
\hline
\multicolumn{1}{|c||}{}
& $K$
& $\pi$
& $\eta$
& $\eta'$
&
\\
\hline\hline
\multicolumn{1}{|c||}{$f_P[\mev]$}
& $160$
& $131$
& $(1.07 \pm 0.02) f_{\pi}$
& $(1.34 \pm 0.06) f_{\pi}$
& \cite{Beneke:2002jn}  \cite{Beneke:2003zv}
\\
\hline
\multicolumn{1}{|c||}{$F_0^{B \to P}$}
& $0.33 \pm 0.04^\dagger$
& $0.26 \pm 0.02$
& $0.23 \pm 0.05$
& $0.19 \pm 0.12$
& \cite{Beneke:2002jn} \cite{Ball:2004ye} \cite{Bharucha:2012wy}
\\
\multicolumn{1}{|c||}{$F^{B_s \to P}$}
& $0.30^{+0.04}_{-0.03}$
& -- & --  & --
& \cite{Duplancic:2008tk}
\\
\hline
\multicolumn{1}{|c||}{$\alpha_1(P)$}
& $0.05 \pm 0.02$
& $0.00$
& $0.00$
& $0.00$
& \multirow{2}{*}{\cite{Beneke:2002jn} \cite{Ball:2006wn}}
\\
\multicolumn{1}{|c||}{$\alpha_2(P)$}
& $0.17 \pm 0.10$
& $0.17 \pm 0.10$
& $0.00 \pm 0.3$
& $0.00 \pm 0.3$
&
\\
\hline\hline
\multicolumn{6}{|l|}{\textbf{Hadronic input -- vector mesons}}
\\
\hline
\multicolumn{1}{|c||}{}
& $K^*$
& $\rho$
& $\phi$
& $\omega$
&
\\
\hline\hline
\multicolumn{1}{|c||}{$f_V[\mev]$}
& $218 \pm 4$
& $209 \pm 1$
& $221 \pm 3$
& $187 \pm 3$
& \cite{Beneke:2003zv}
\\
\multicolumn{1}{|c||}{$f_V^\perp[\mev]$}
& $175 \pm 10$
& $156 \pm 9$
& $175 \pm 9$
& $142 \pm 9$
& \cite{Ball:2006eu}
\\
\hline
\multicolumn{1}{|c||}{$A_0^{B \to V}$}
& $0.34 \pm 0.03$
& $0.30 \pm 0.03$
& --
& $0.28 \pm 0.03$
& \multirow{2}{*}{\cite{Ball:2004rg}}
\\
\multicolumn{1}{|c||}{$F_-^{B \to V}$}
&  $0.62 \pm 0.05$
&  $0.58 \pm 0.04$
&  --
& $0.55 \pm 0.04$
&
\\
\multicolumn{1}{|c||}{$F_+^{B \to V}$}
& $0.00 \pm 0.06$
& $0.00 \pm 0.06$
& --
& $0.00 \pm 0.06$
& \cite{Beneke:2006hg}
\\
\hline
\multicolumn{1}{|c||}{$A_0^{B_s \to V}$}
& $0.39 \pm  0.03$
& --
& $0.47 \pm 0.04$
& --
& \multirow{2}{*}{\cite{Ball:2004rg}}
\\
\multicolumn{1}{|c||}{$F_-^{B_s \to V}$}
& $0.59 \pm 0.04$
& --
& $0.72 \pm 0.04$
& --
&
\\
\multicolumn{1}{|c||}{$F_+^{B_s \to V}$}
& $0.00 \pm 0.06$
& --
& $0.00 \pm 0.06$
& --
& \cite{Beneke:2006hg}
\\
\hline
\multicolumn{1}{|c||}{$\alpha_1(V)$}
& $0.02 \pm 0.02$
& $0.00  $
& $0.00  $
& $0.00  $
& \multirow{4}{*}{\cite{Ball:2007zt}}
\\
\multicolumn{1}{|c||}{$\alpha^\perp_1(V)$}
& $0.03 \pm 0.03 $
& $0.00$
& $0.00$
& $0.00$
&
\\
\multicolumn{1}{|c||}{$\alpha_2(V)$}
& $0.08 \pm 0.06$
& $0.10 \pm 0.05$
& $0.13 \pm 0.06$
& $0.10 \pm 0.05$
&
\\
\multicolumn{1}{|c||}{$\alpha^\perp_2(V)$}
& $0.08 \pm 0.06$
& $0.11 \pm 0.05$
& $0.11 \pm 0.05$
& $0.11 \pm 0.05$
&
\\
\hline\hline
\end{tabular}
\caption{Numerical input used for our analysis. Form factors are given at zero
  momentum transfer $q^2 = 0$. Other scale dependent quantities are quoted at
  the scale $\mu = 2\gev$. $^\dagger$For the $B \to K$ form factor we used
  $\alpha_4^K(2.2 \gev) = -0.0089$ \cite{Ball:2004ye} as additional input.}
\label{tab:num:input}
\renewcommand{\arraystretch}{1.0}
\end{center}
\end{table*}

%
%
%

\section{Statistical procedure}
\label{app:statistics}

\renewcommand{\theequation}{B\arabic{equation}}
\setcounter{equation}{0}

This appendix summarizes the statistical methods that are used in order to
obtain probability regions for the parameters of interest, pull values of theory
predictions and corresponding measurements of observables, and $p$-values as a
measure of the goodness of fit.  Further, we describe the determination of
probability distributions of predictions for observables that were not included
in the fit.

%
%
\subsection{Probability regions}
\label{app:prob:region}

\def \bftheta{\boldsymbol\theta}
\def \bfnu{\boldsymbol\nu}

For the purpose of parameter inference we use Bayes theorem to determine the
posterior probability distribution, $P(\bftheta|M,D)$, of the parameters of
interest, $\bftheta = (\theta_1, \theta_2, \ldots)$, given a model $M$ and data
$D$. Parameters of interest in our analysis are {\it i)} the phenomenological
parameters of weak annihilation $\rho_A^{M_1M_2}$ and {\it ii)} parameters of
new physics scenarios. Bayes theorem relates the posterior probability to the
likelihood ${\cal L} (\bftheta) = P(D|M, \bftheta)$, which is the probability of
the data given the model $M$ with parameter values $\bftheta$ and the prior
distributions, $P(M, \bftheta)$, which are the probability of model $M$ with
parameter values $\bftheta$
\begin{align}
  P(\bftheta|M, D) &
  = \frac{P(D|M, \bftheta) \, P(M, \bftheta)}{Z} \,.
\end{align}
Here, the model-dependent normalization factor
\begin{align}
  Z & \equiv \int \!\!
    P(D|M, \bftheta) \, P(M, \bftheta) \, d\bftheta
\end{align}
is known as ``evidence'' or ``marginal likelihood'' that plays an important role
in model comparison within the Bayesian approach. Throughout, the priors of the
$\bftheta$ are chosen as uniform within a certain interval.

It is common to introduce the likelihood function ${\cal L} (\bftheta)$ as the
product of the probabilities $p(O_i = O_i^{\rm th}(\bftheta))$ that each
observable $O_i$ in the data set takes the particular value $O_i^{\rm
  th}(\bftheta)$ predicted at the value of $\bftheta$
\begin{equation}
  \label{eq:def:likelihood}
\begin{aligned}
  {\cal L} (\bftheta) & =
    \prod_{i\, \in\, {\rm data}} p \Big(O_i = O_i^{\rm th}(\bftheta)\Big)
\\ & \quad
    \sim  \exp\left[-\frac{1}{2}Ê\sum_{i \, \in \, {\rm data}}
    \Big(\chi_i(\bftheta)\Big)^2 \right] \,.
\end{aligned}
\end{equation}
The probabilities $p$ are given by the measured probability density functions
pdf$[O_i]$ of each observable $O_i$ and the second part $\sim$ of
\eqref{eq:def:likelihood} indicates the special case of gaussian distributed
pdf's permitting to define a $\chi_i(\bftheta)$.

The expression of ${\cal L} (\bftheta)$ does not yet include the uncertainties
due to nuisance parameters, $\bfnu = (\nu_1, \nu_2, \ldots)$, which enter
theoretical predictions of $B\to M_1 M_2$ decays. In this case, the nuisance
parameters give rise to an interval for the theory prediction $\big[O_i^{\rm th}
- \Delta^-_i,\, O_i^{\rm th} + \Delta^+_i\big]$ with possibly asymmetric
uncertainties $\Delta^\pm_i$ around the central value $O_i^{\rm th}$ that is
obtained for central values of all nuisance parameters.  Here, the theoretical
uncertainty $\Delta^\pm_i$ is determined by adding in quadrature the
uncertainties due to each nuisance parameter $\nu_a$
\begin{align}
  \Delta^\pm_i & =
  \sqrt{ \, \sum_a \left(\Delta^\pm_{i,a}\right)^2} \,,
\end{align}
which arises from the minimal, central and maximal values $\nu_a^{\rm min}$,
$\nu_a^{\rm cen}$ and $\nu_a^{\rm max}$, respectively,
\begin{align}
  \label{eq:theory:uncertainty}
  \Delta^{+(-)}_{i,a} & =
    \big|O_i^{\rm th}(\nu_a^{\rm max(min)})
       - O_i^{\rm th}(\nu_a^{\rm cen})\big| \,,
\end{align}
while keeping all others at their central values. Clearly, this is an
approximation that neglects more complicated interdependences of observables on
several parameters and also possible correlations among different nuisance
parameters. The nuisance parameters are listed in \reftab{num:input}.

In the presence of nuisance parameters, we will adopt the simple procedure to
use the maximal value of the pdf inside the interval of the theory prediction,
hence replacing in~\eqref{eq:def:likelihood}
\begin{equation}
\begin{aligned}
  \mbox{pdf} \Big(O_i = O_i^{\rm th}(\bftheta)\Big) \quad & \to 
\\[0.2cm]Ê \quad
  \max\Big( \mbox{pdf} (O_i) \, \big|  \, O_i \in &
   [O_i^{\rm th} - \Delta^-_i,\, O_i^{\rm th} + \Delta^+_i]
   \Big) \,,
\end{aligned}
\end{equation}
where the dependence of $O_i^{\rm th}$ and $\Delta^\pm_i$ on $\bftheta$ and
$\bfnu$ is not explicitly shown. This procedure is implemented easily for
gaussian distributed pdf's by the modification of the definition of
\begin{widetext}
\begin{align}
  \label{eq:def:chi}
  \chi_i(\bftheta, \bfnu) & 
  = \left\{
  \begin{array}{ccc}
    \displaystyle
    \frac{\big|O_i^{\rm th}(\bftheta,\bfnu) - O_i^{\rm exp}\big|
    - \Delta^+_i(\bftheta,\bfnu)}
         {\sigma^-_i}
  & \mbox{if}
  & O_i^{\rm exp} \geq O_i^{\rm th} + \Delta^+_i
  \\[0.4cm]
    \displaystyle
    \frac{\big|O_i^{\rm th}(\bftheta,\bfnu) - O_i^{\rm exp}\big|
    - \Delta^-_i(\bftheta,\bfnu)}
         {\sigma^+_i}
  & \mbox{if}
  & O_i^{\rm exp} \leq O_i^{\rm th} - \Delta^-_i
  \\[0.4cm]
    0
  & \hskip 0.5cm \mbox{else} \hskip 0.5cm &
  \end{array}
  \right.
\end{align}
\end{widetext}
where $O_i^{\rm exp}$ and $\sigma^\pm_i$ denote the central value and the left
and right standard deviation of the pdf$[O_i]$, respectively. The central value
of the theoretical prediction $O_i^{\rm th}$ is obtained at the particular value
of the parameters of interest $\bftheta$, and the $\bfnu$ are set to their
central values.

Obviously, the modification \eqref{eq:def:chi} is tailored to gaussian pdf's,
which is our interpretation of experimental world averages given by the Particle
Data Group (PDG)~\cite{Beringer:2012} or HFAG \cite{Amhis:2012bh}.  However, the
ratios of gaussian distributed observables -- like the ones defined in
\refeq{DefOfRatios}: $R = \Br_1/\Br_2$ -- follow a gaussian ratio distribution.
In the absence of experimental results of these ratios, one has to resort to the
combination of the two gaussian distributions of numerator and denominator. In
all relevant cases, the $\Br_i$ are gaussian distributed with symmetric errors
(from HFAG) and assuming that their errors are uncorrelated, the analytical
expression of $p(R)$ is known~\cite{HINKLEY01121969}.  Since it is monotonly
rising till its maximum at $R^{\rm exp} \equiv \Br_1^{\rm exp}/\Br_2^{\rm exp}$
and then monotonly falling, the maximal value of the probability in the theory
interval can be easily found by evaluating $p(R)$ at
\begin{align}
  R & =   \left\{
  \begin{array}{ccl}
    \displaystyle
    R_i^{\rm th} + \Delta^+_i 
  & \mbox{if}
  & R^{\rm exp} \geq R_i^{\rm th} + \Delta^+_i
  \\[0.2cm]
    R_i^{\rm th} - \Delta^-_i
  & \mbox{if}
  & R^{\rm exp} \leq R_i^{\rm th} - \Delta^-_i \,.
  \\[0.2cm]
    \displaystyle
     R^{\rm exp}
  & \hskip 0.5cm \mbox{else} \hskip 0.5cm &
  \end{array}
  \right.
\end{align}
The probability value is converted to $\chi = -2 \log p(R)$.  Let us finally
note that the difference between the gaussian ratio distribution and a gaussian
distribution with central value $R$ and $\sigma(R)$ determined from simple
uncertainty propagation calculus, is numerically negligible unless large
deviations of experimental and theoretical values probe the tails of the
distributions, which are ``heavier'' for the gaussian ratio distribution.

Concerning the evaluation of the posterior probability, it is determined
numerically with the help of the Markov Chain Monte Carlo (MCMC) implementation
of the Bayesian Analysis Tool (BAT) \cite{Caldwell:2008fw}.  One and
two-dimensional posterior distributions are obtained in turn by marginalization
over the remaining parameters of interest. The best fit points are identified
with the help of Minuit that is initialized with the point of the highest
posterior found during the MCMC run.

%
%
\subsection{Pull value}

The deviation of a single measurement of an observable $O_{i}$ from its
prediction $O_i^{\rm th} \pm \Delta^\pm_i$ for a particular value of the
parameters of interest $\bftheta_{\boldsymbol*}$ will be given in terms of the
pull-value, accounting for theoretical uncertainties. Here, we define the pull
value, $\delta$, as the integral over those regions of the pdf[$O_{i}$], which
have higher probability as the maximal probability value $p^{\rm max}$ appearing
in the interval spanned by the theory prediction $[O_i^{\rm th} - \Delta^-_i,\,
O_i^{\rm th} + \Delta^+_i]$ evaluated at $\bftheta_{\boldsymbol*}$ and varying
the nuisance parameters, i.e.,
\begin{align}
  \delta & =
    \int_{-\infty}^{+\infty} d O_i\, p(O_i)\, \theta\big[p(O_i) - p^{\rm max}\big]
\end{align}
where $\theta(x)$ denotes the step function.  Consequently, the pull value is
zero if the maximum of the pdf is inside this theory interval. In the case of a
normally distributed pdf (with $\sigma^{+}_i = \sigma^{-}_i$), a non-zero pull
implies a symmetric integration interval around the central value $O_i^{\rm
  exp}$ of the distribution and the integrated fraction of probability can be
converted into the distance between the lower or upper boundary of the theory
uncertainty interval to $O_i^{\rm exp}$ in terms of its standard deviation
$\sigma_i$ depending on whether the theory prediction is above or below
$O_i^{\rm exp}$. In the case of non-gaussian pdf's, the pull value gives a
measure of the probability fraction that corresponds to those values of $O_i$
that have higher experimental probability than the ones contained in the
interval of the theory prediction\footnote{For very non-gaussian pdf's with
  several disconnected regions of probability, the pull value might give rise to
  misleading interpretations, however, all measurements at hand are gaussian or
  gaussian ratio distributed.}. The pull value is simply calculated by drawing
values for $O_i$ that are distributed according to the pdf[$O_i$] and taking the
ratio of the cases in which $p(O_i) > p^{\rm max}$ and the total number of
draws.

%
%
\subsection{$p$ Value}

As a measure of the goodness of fit, we will use $p$ values in order to compare
within the same theoretical model at some point $\bftheta_{\boldsymbol*}$ --
usually the best fit point(s) -- the quality of the fit for different sets of
data Set I and Set II. For this purpose we will assume the model with the
specific choice $\bftheta_{\boldsymbol*}$, allowing us to produce frequencies of
possible outcomes within the model. We will use two ways to calculate $p$
values.

The common definition is used as a first possibility, assuming the validity of
normal and all independent pdf's. It consists in the evaluation of the
cumulative of the $\chi^2$-distribution -- the latter denoted by $f(x, N_{\rm
  dof})$, with $N_{\rm dof}$ number of degrees of freedom -- starting from the
value $\chi^2_* = -2 \log {\cal L}(\bftheta_{\boldsymbol*})$
\begin{align}
  \label{eq:def:pValue}
  p = \int^{\infty}_{\chi^2_*} dx \, f(x, N_{\rm dof}) \,,
\end{align}
and corresponds to the probability of observing a test statistic at least as
extreme in a $\chi^2$ distribution with $N_{\rm dof}$. Values of $p < 5\%$ are
usually referred to as ``statistical significant'' deviation from the null
hypothesis, i.e., the validity of the model with parameters
$\bftheta_{\boldsymbol*}$. As usual, the number of degrees of freedom is given
as $N_{\rm dof} = (N_{\rm meas} - \mbox{dim} (\bftheta))$, with $N_{\rm meas}$
denoting the number of measurements.

As a second possibility we calculate the $p$ value defining a test statistics
based on the likelihood \cite{Beaujean:2012uj}. The according frequency
distribution is determined from $10^6$ pseudo experiments in the lack of raw
data and experimental efficiency corrections that require dedicated detector
simulations. For this purpose, the pdf of each observable $O_i$ is shifted such
that the position of it's maximum at $O_i = O_i^{\rm exp}$ coincides with the
prediction $O_i^{\rm th} (\bftheta_{\boldsymbol*})$ at the point
$\bftheta_{\boldsymbol*}$ of interest. In this way, the uncertainties of the
measurement with central value $O_i^{\rm exp}$ are adopted for $O_i^{\rm th}
(\bftheta_{\boldsymbol*})$, neglecting possibly different experimental
efficiency corrections. In each pseudo experiment, possible experimental
outcomes are drawn for all measurements in the data set from the shifted pdf's
and the likelihood value is compared to that of the observed data set,
determining this way the fraction of pseudo experiments with smaller likelihood
values. The $p$ value is identified with this fraction, however for the number
of degrees of freedom that corresponds to the number of measurements $N_{\rm
  meas}$ in the data set.  Subsequently, we correct the $p$ value by converting
it to a $\chi^2$ value with the help of the inverse cumulative distribution with
$N_{\rm meas}$ degrees of freedom and recalculate it for the actual $N_{\rm
  dof}$ \cite{Beaujean:2011zza} using \eqref{eq:def:pValue}.

%
%
\subsection{Probability distributions of observables}
\label{app:predictions}

If certain observables are not yet measured or despite an existing measurement
are not included in the data set $D$ of the fit, one might obtain a prediction
of its probability distribution given the data $D$ and model $M$
\cite{Beaujean:2012uj}.  We calculate the considered observables at each point
of the Markov Chain for the current value of $\bftheta$ and determine the
interval of the theory uncertainty $[O_i^{\rm th} - \Delta^-_i,\, O_i^{\rm th} +
\Delta^+_i]$ due to nuisance parameters as described in
\refapp{prob:region}. The obtained intervals are used to fill a histogram that
is normalized eventually to obtain a probability distribution.

%
%

\bibliographystyle{apsrev4-1}
\bibliography{draft_refs}

\end{document}